\documentclass[12pt,english]{book}
\usepackage{pslatex}
\usepackage[T1]{fontenc}
\usepackage[latin1]{inputenc}
\usepackage{geometry}
\usepackage{babel}
\usepackage{array}
\usepackage{graphics}
\usepackage{rotating}

\makeatletter

\providecommand{\LyX}{L\kern-.1667em\lower.25em\hbox{Y}\kern-.125emX\@}
\newlength{\LyXMinipageIndent}
\let\SF@@footnote\footnote
\def\footnote{\ifx\protect\@typeset@protect
    \expandafter\SF@@footnote
  \else
    \expandafter\SF@gobble@opt
  \fi
}
\expandafter\def\csname SF@gobble@opt \endcsname{\@ifnextchar[
  \SF@gobble@twobracket
  \@gobble
}
\edef\SF@gobble@opt{\noexpand\protect
  \expandafter\noexpand\csname SF@gobble@opt \endcsname}
\def\SF@gobble@twobracket[#1]#2{}

\makeatother
\begin{document}
\newcommand{\mW}{\ensuremath{m_{\scriptsize\mathrm{W}}}}
\newcommand{\mZ}{\ensuremath{m_{\scriptsize\mathrm{Z}}}}
\newcommand{\mh}{\ensuremath{m_{\scriptsize\mathrm{h}}}}
\newcommand{\Evis}{\ensuremath{\mathrm{E}_{\scriptsize\mathrm{vis}}}}
\newcommand{\Ebgo}{\ensuremath{\mathrm{E}_{\scriptsize\mathrm{BGO}}}} 
\newcommand{\Ehcal}{\ensuremath{\mathrm{E}_{\scriptsize\mathrm{HCAL}}}}
\newcommand{\Nww}{\ensuremath{{\cal N}_{\scriptsize \ \mathrm{WW}}}}
\newcommand{\Nzz}{\ensuremath{{\cal N}_{\scriptsize \ \mathrm{ZZ}}}}
\newcommand{\Nqq}{\ensuremath{{\cal N}_{\scriptsize \ \mathrm{q\bar{q}}}}}
\newcommand{\CLb}{\ensuremath{\mathrm{CL}_{\scriptsize \mathrm{b}}}}
\newcommand{\CLsb}{\ensuremath{\mathrm{CL}_{\scriptsize \mathrm{s+b}}}}
\newcommand{\CLs}{\ensuremath{\mathrm{CL}_{\scriptsize \mathrm{s}}}}
\newcommand{\BRgg}{\ensuremath{\mathrm{Br}_{\scriptsize\gamma\gamma}}}
\newcommand{\BRphobic}{\ensuremath{\mathrm{Br}_{\scriptsize\mathrm{phobic}}}}

\frontmatter

{\centering SEARCH FOR A HIGGS BOSON\par}

{\centering DECAYING TO MASSIVE VECTOR BOSON PAIRS\par}

{\centering AT LEP\par}

\vfill

{\centering Jeremiah M. Mans\par}

\vfill

{\centering A DISSERTATION\\
PRESENTED TO THE FACULTY\\
OF PRINCETON UNIVERSITY\\
IN CANDIDACY FOR THE DEGREE\\
OF DOCTOR OF PHILOSOPHY\par}

\vfill

{\centering RECOMMENDED FOR ACCEPTANCE\\
BY THE DEPARTMENT OF \\
PHYSICS\par}

\vfill

{\centering June 2002}

\newpage

\vfill

\vfill{}
{\centering \( \copyright  \) Copyright by Jeremiah Mans, 2002. All
rights reserved.\par}

\vfill\vfill\newpage\pagestyle{headings}
\section*{Abstract}

In the Standard Model, the Higgs boson is responsible for mass-generation
and stabilizing the electroweak interaction at high energies. The
boson has not been observed and the Standard Model does not predict
its mass. Direct searches have excluded the existence of a Higgs boson
with a mass less than 113 GeV. Searches to date have focussed on b-quark
decays of the Higgs, but the model predicts an increased branching
fraction to massive vector boson pairs for a heavier Higgs. In some
extensions of the Standard Model which predict multiple Higgs bosons,
the lightest Higgs boson couples primarily to bosons, not fermions.
Results excluding these {}``fermiophobic'' models have used the
two-photon decay to date, but for Higgs masses above 100 GeV, the
decay to massive vector boson pairs dominates. In this dissertation,
I present the first search for a Higgs boson decaying to massive vector
boson pairs. The search is based on data collected by the L3 experiment
at CERN during the 1999-2000 period.

The search uses the Higgsstrahlung production mode where the Higgs
is radiated from an off-shell Z boson, so the analysis must include
the decay of the Z boson as well as the decay of the two W or Z bosons
from the Higgs decay. The events will contain six final state fermions,
and the decays of the W and Z define nine different channels for the
\( \textrm{h}\rightarrow \textrm{WW} \) search. I present the details
and results of analyses for six of the channels. The combined analyses
exclude a fermiophobic Higgs decaying to massive vector boson pairs
for \( 83.8\textrm{ GeV }<\mh <104.2\textrm{ GeV} \) at a 95\% confidence
level with an unexcluded region between \( 88.8\textrm{ GeV}<\mh <89.6\textrm{ GeV} \).
Monte Carlo predictions of the analyses' performance predict an exclusion
range of \( 86.8\textrm{ GeV }<\mh <107.5\textrm{ GeV} \). I also
present model-independent branching ratio limits for the massive vector
boson search, as well as a scan of the fermiophobic plane combining
with the results of the LEP \( \textrm{h}\rightarrow \gamma \gamma  \)
search.

\newpage

\section*{Acknowledgements}

I want to thank my advisor Chris Tully for being an excellent advisor.
He put up with me when I was very new to the field of high energy
physics. I have learned most of what I know about how high energy
physics and high energy experiments work from Chris. His excitement
about the Higgs search and many other areas of high energy physics
is infectious. Chris has also given me the freedom to work on other
projects which I find interesting, and we have colloborated to modernize
the digital electronics class in the physics department.

I also want to thank Pierre Piroue, whom I consider my {}``advisor-emeritus.''
He offerred me a chance to visit Europe the summer after the end of
college and ended up with a graduate student. I have greatly enjoyed
our conversations on medicine, music, and natural disasters.

Wade Fisher has been a great fellow graduate student. He is responsible
for the \( \nu \nu \textrm{qql}\nu  \) and llqqqq analyses described
in this thesis, and he has been very patient with my many requests
for his time this spring. My wife and I greatly value the friendship
we have with him and Alana.

My time at Princeton has not just been about thesis research, but
also about teaching and working with great people. There are too many
great people at Princeton for me to list them all, but I particularly
want to acknowledge Stan Chidzik and Andrew Dutko. 

My family has been very supportative throughout my education, and
particularly interested in seeing me finish the dissertation. Above
everyone else, I want to thank my wife Tamara. She has been incredibly
patient with her always-busy husband this spring. She goads me when
I get too lazy and relaxes me when I get too stressed. Our wedding
was the greatest event, not just of my time at Princeton, but of my
whole life. 

\newpage

{\setlength\parindent{0pt}
\begin{minipage}[m]{1.00\columnwidth}
\setlength\parindent{\LyXMinipageIndent}
\centering \raisebox{-3in}{For Jesse, who would have loved this.}\end{minipage}
}

\tableofcontents{}

\listoffigures{}

\listoftables{}

\mainmatter

\chapter{LEP and the L3 Experiment}

\begin{quotation}
\noindent \textit{The machine does not isolate man from the great
problems of nature }\\
\textit{but plunges him more deeply into them.}

Antoine de Saint-Exup\'ery
\end{quotation}

\section{LEP}

\begin{figure}[tbp]
{\centering \resizebox*{0.7\textwidth}{!}{\includegraphics{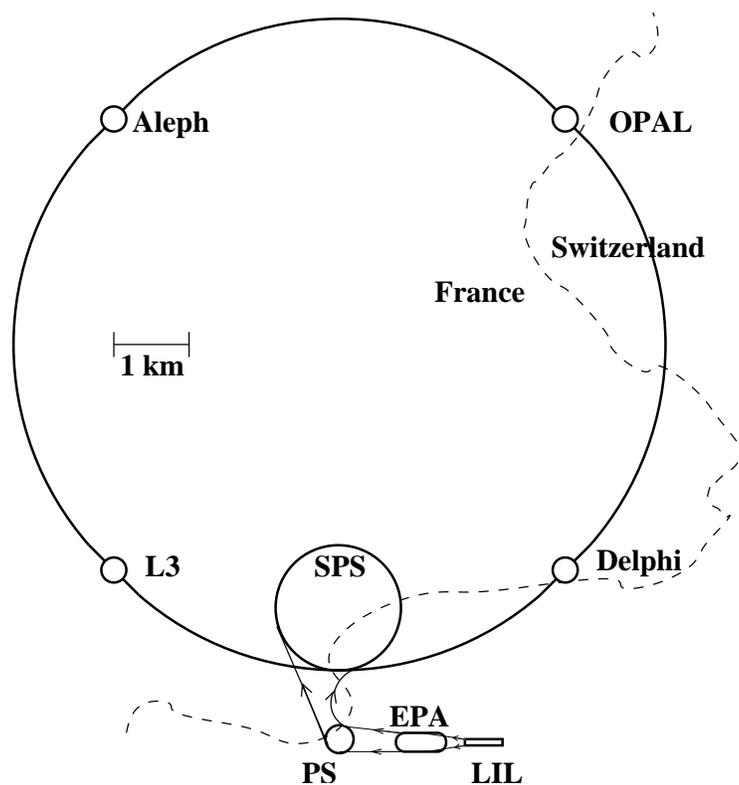}} \par}

\caption{Overhead sketch of LEP and the rest of the CERN accelerator complex.\label{fig:lepscheme}}
\end{figure}

LEP is the {}``Large Electron-Positron'' storage ring built 40 meters
under the countryside outside Geneva, Switzerland. The epithet {}``large''
was well chosen, since LEP is the largest accelerator in the world
at 27 km in circumference. Construction of the accelerator began in
1982 and the first collisions in the detectors were recorded on August
13,1989\cite{lep:history}. There are four large general-purpose detectors
equally spaced around the ring: L3, Aleph, Delphi, and Opal. The locations
of the four detectors are indicated in Figure \ref{fig:lepscheme}.

LEP sits at the end of a long chain of accelerators which work together
to accelerate electrons and positrons up to kinetic energies of 100
GeV and above. The electron and positron bunches are produced in the
Linear Injector for LEP (LIL) complex and stored at 600 MeV in the
Electron-Positron Accumulator (EPA). From the EPA, the bunches are
transfered to the CERN Proton Synchrotron (PS), where the magnets
are ramped down to accept the low-energy bunches. The PS accelerates
the bunches to 2.5 GeV and transfers them on to the Super Proton Synchrotron
(SPS) which provides the last pre-acceleration kick up to 22 GeV for
transfer into LEP. The LEP machine then accelerates the bunches up
to full energy, brings the beams into collision, and keeps them in
collision for several hours until the stored current drops to the
point where the operators decide to dump the beam.

Up until 1996, LEP ran at or near the Z pole (\( \sqrt{s}=91.2\textrm{ GeV} \)).
In 1996, CERN began adding superconducting cavities to LEP, which
allowed the beam energy to increase each year. At the beginning and
end of each year, a few \( \textrm{pb}^{-1} \) of data was taken
at the Z pole for calibration. Most of the data was taken at energies
near the upper limit of the machine. This limit increased as additional
accelerating cavities were added and the physicists and engineers
of the machine group tuned the machine for ever higher gradients and
beam energies. Between 1998 and 1999, CERN upgraded the cryogenic
systems as part of preparations for the LHC. The upgrade allowed the
machine group to push the accelerating gradient of the superconducting
cavities from their design of 6MV/m to more than 7.5MV/m. During the
1999 run, the beam energy was limited for several weeks by statute:
the original permit granted by the French nuclear authorities had
specified beam energies up to, but not exceeding 100 GeV, which limited
\( \sqrt{s} \) to 200 GeV. Once the authorities amended the permit,
the experiments collected a month's worth of data at 202 GeV, and
LEP even reached 204 GeV for one 15 minute run. The breakdown of luminosity
versus \( \sqrt{s} \) over the 1999-2000 running period is given
in Table \ref{tab:l3lumi}.
\begin{table}[tbp]
{\centering \begin{tabular}{|c|c|c|}
\hline 
&
\( \sqrt{s} \) (\( \pm 1\textrm{ GeV} \))&
Luminosity (\( \textrm{pb}^{-1}) \)\\
\hline
\hline 
1999&
191.6&
29.8\\
&
195.5&
83.7\\
&
199.5&
82.8\\
&
201.8&
37.0\\
\hline
2000&
203.1&
9.6\\
&
205.0&
68.9\\
&
206.5&
130.3\\
&
208.0&
8.5\\
\hline
\end{tabular}\par}

\caption{Luminosity collected by the L3 experiment over 1999-2000.\label{tab:l3lumi}}
\end{table}

In 2000, the machine operation was optimized for discoveries in the
Higgs and supersymmetric sectors, which required the maximum possible
beam energies. The machine group developed several improvements to
LEP operations which significantly improved the energy reach and integrated
luminosity collected in 2000 \cite{lep:op2000}.

\begin{figure}[tbp]
{\centering \resizebox*{0.5\textwidth}{!}{\includegraphics{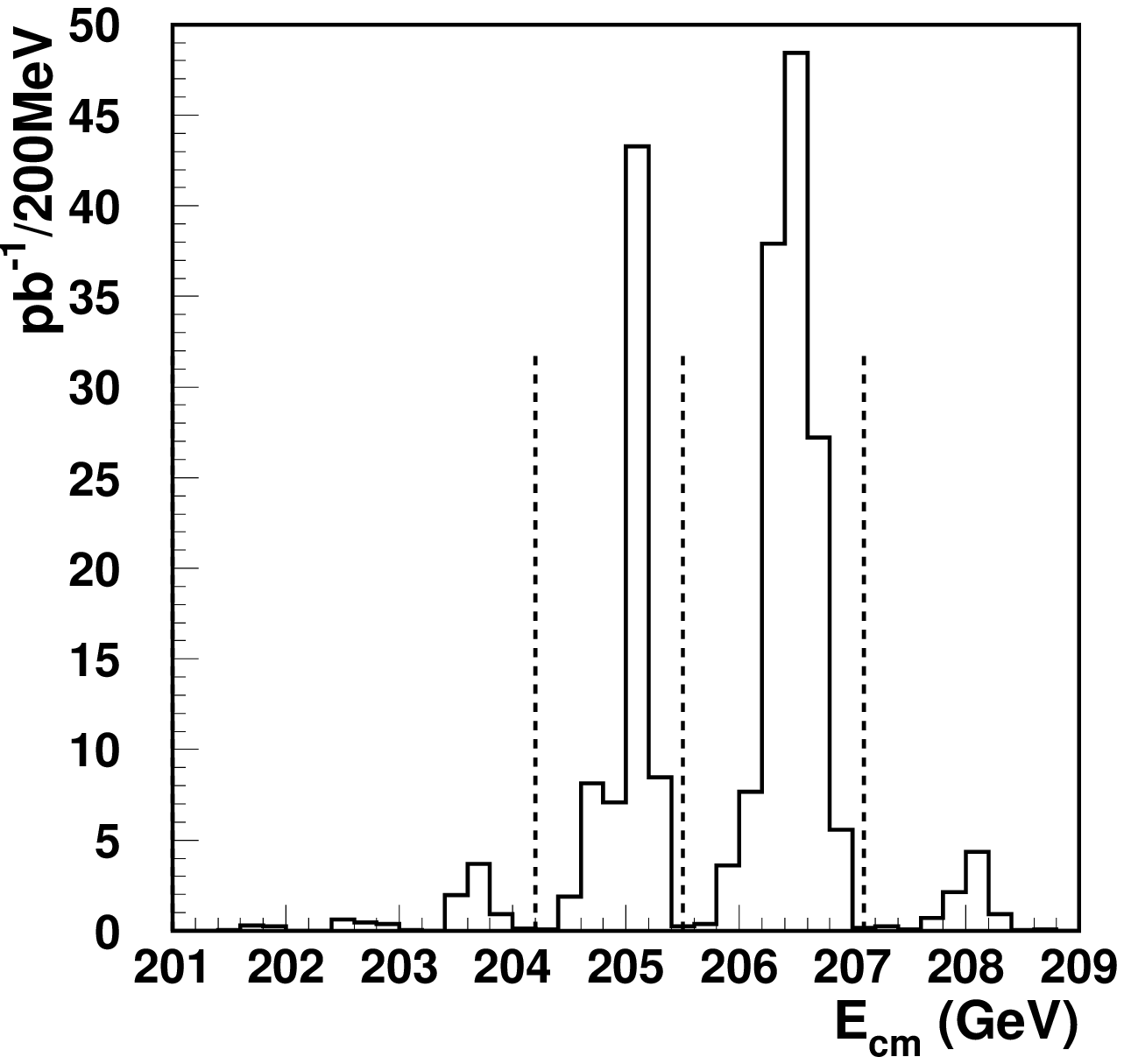}} \par}

\caption{Distribution of luminosity as a function of \protect\( \sqrt{s}\protect \)
in 2000.\label{fig:2000lumi}}

The four energy bins used for the \( \textrm{h}\rightarrow \textrm{WW}/\textrm{ZZ} \)
search are indicated by the dashed lines.
\end{figure}

\begin{itemize}
\item The upgraded cryogenics system increased the stability of the RF system,
which allowed the operations group to reduce the margin from two klystrons
to one. At full beam energy, the LEP RF system suffered a klystron
trip due to overheating about once an hour. With a one-klystron margin,
the RF system could absorb one trip, but a second occurring before
the first klystron could be restarted caused beam loss. The reduced
margin allowed an increase in \( \sqrt{s} \) of 1.5 GeV.
\item The machine group reduced the 350 MHz RF frequency driving the cavities
by 100 Hz to expand the orbit of the beams. The larger orbit reduced
the synchrotron radiation and allowed the dipolar component of the
quadrupole magnets to control the new orbit. The reduced frequency
also increased the RF margin slightly by reshaping the bunches. These
adjustments allowed an increase in \( \sqrt{s} \) of 1.4 GeV.
\item The machine group also enabled unused corrector magnets as additional
dipoles to further increase the effective LEP radius, which added
another 400 MeV.
\item Eight old LEP1 copper cavities were reinstalled, adding an additional
30MV in total accelerating gradient. This gradient increase translated
to an increase in \( \sqrt{s} \) of \( \sim 400\textrm{ MeV} \).
The increase in energy is larger than the gradient increase because
LEP does not have to accelerate the beams from rest each turn, but
rather just replace the energy lost to synchrotron radiation.
\item The machine's mode of operation was modified to add {}``miniramps.''
In previous years, once the machine reached its target energy and
the beams entered collision, the energy did not change. In 2000, the
operators would raise the energy several times during the physics
coast as the RF stabilized and current fell. Thus, a given fill would
generate data at several \( \sqrt{s} \) values.
\end{itemize}
The beam loss rate sharply increased with LEP operating at its limit.
Of the roughly 4000 fills made in the twelve years of LEP running,
1400 were made in the last year. To reduce the impact to physics beam
time, the machine group made special efforts to reduce the turnaround
time. The group was able to reduce the average turnaround time from
beam dump to stable collisions to less than an hour from the previous
average of about 2 hours. In the search for maximum energy, some of
the accelerating cavities were stressed beyond their limits and the
machine group had to reduce the maximum gradient of several during
the year. Some of these cavities recovered, but others did not. The
continual changes to the machine operating conditions meant that the
2000 dataset contained data from many different \( \sqrt{s} \) energies,
as shown in Figure \ref{fig:2000lumi}. For analysis purposes, we
grouped the data into the four energy bins indicated in the plot.

\section{The L3 Experiment}

\begin{figure}[tbp]
{\centering \resizebox*{0.8\textwidth}{!}{\includegraphics{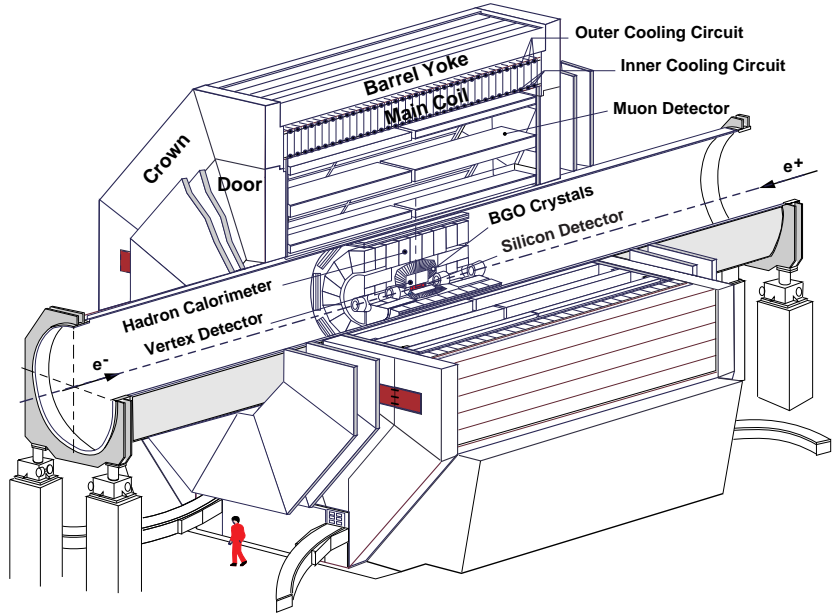}} \par}

\caption{Perspective view of the L3 experiment.\label{fig:l3persp}}
\end{figure}

The L3 experiment, located at point 2 on the LEP ring, is shown in
perspective view in Figure \ref{fig:l3persp}\cite{l3:construction}.
 The entire L3 experiment, the largest of the four LEP detectors,
is surrounded by a 7800 ton octagonal conventional solenoid electromagnet
which produces a 0.5 T field. Within the electromagnet are the muon
detection chambers, the calorimeters, and, closest to the beam pipe,
the inner tracking subdetectors. All of the detector elements are
mounted on a 281-ton steel support tube suspended along the central
axis of the detector. The muon chambers are mounted on the outside
of the support tube and the rest of the subdetectors are inside. While
the LEP machine experienced many changes in beam elements and operating
procedures during 1998-2000, the L3 experiment was extremely stable.
No major subdetectors were added during this time, and the calibration
procedures for the detector were perfected.

\subsection{The Inner Tracking Subdetectors}

The L3 experiment contains two major tracking subdetectors. The role
of these detectors is to measure the paths of charged particles through
the magnetic field with a minimum of disturbance to the particles'
paths and energies. The curvature of these tracks reveals the charge
and momentum of the particles. Very close to the beampipe there are
two layers of silicon strip detectors which are called the Silicon
Microvertex Detector (SMD). Around the SMD is a gas-filled drift chamber
called the Time-Expansion Chamber (TEC). 

The TEC consists of a long cylindrical tube filled with a mixture
of 80\% \( \textrm{CO}_{2} \) and 20\% isobutane \( \left( \textrm{iC}_{4}\textrm{H}_{10}\right)  \).
Charged particles passing through the tube ionize the gas molecules.
The TEC collects and times the arrival of the gas ions to determine
the path of the charged particles. The chamber is divided into two
rings -- an inner ring of 12 sectors and an outer ring of 24 sectors.
These sectors are defined by the arrangement of wires strung parallel
to the beam pipe. Most of the wires carry high voltages which supply
the drift and amplification electric fields, while the rest carry
the collected charge out to high speed analog-to-digital converters.
\begin{figure}[tbp]
{\centering \resizebox*{0.6\textwidth}{!}{\includegraphics{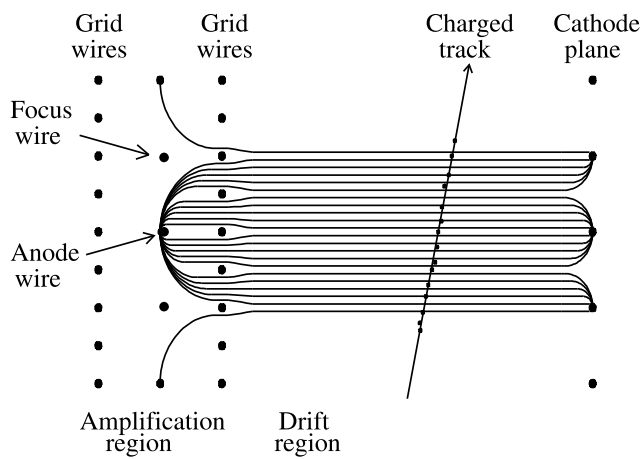}} \par}

\caption{The drift and amplification fields of the TEC. }

Charged particles ionize the gas, which drifts in a relatively low
field toward the grid wires. After passing the grid, the ions are
accelerated in a higher field and produce secondary ions which are
also collected at the anode.\label{fig:tecfield}
\end{figure}
 The fields set up in a TEC sector are shown in schematic view in
Figure \ref{fig:tecfield}. Each track is measured by up to 51 sense
wires to determine \( r-\phi  \) accurately. Additional charge-division
wires provide some information about the \( z \) position of the
track. There are also two cathode strip chambers mounted on the outside
of the TEC which measure the \( z \) position of the track with average
300 \( \mu \textrm{m} \) resolution. The analyses in this thesis
use the TEC, along with the SMD, for measuring tracks in jets and
identifying the isolated groups of odd-numbered tracks associated
with taus.

\begin{figure}[tbp]
{\centering \resizebox*{0.7\textwidth}{!}{\includegraphics{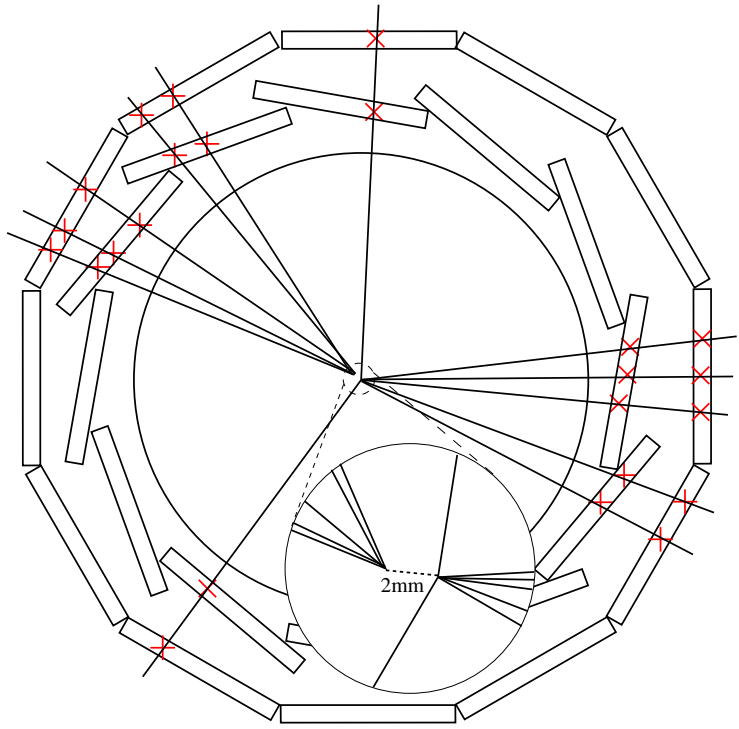}} \par}

\caption{\protect\( r-\phi \protect \) view of the Silicon Microvertex Detector.\label{fig:smd}}

The figure shows hits in the SMD and the tracks which might be expected
from a \( \textrm{ZZ}\rightarrow \textrm{e}^{+}\textrm{e}^{-}\textrm{b}\bar{\textrm{b}} \)
event. One of the \textbf{B} mesons has traveled 2mm in \( r-\phi  \)
before decaying at a displaced vertex.
\end{figure}

In 1991, the radius of the LEP beampipe was reduced from 8 cm to 5.5
\emph{}cm, which opened enough space to add a new silicon tracking
detector, the SMD \cite{l3:smd}. The SMD is a silicon strip detector,
composed of silicon wafers with metalized strips on both sides of
the wafer. The wafers are made of n-type silicon and have p-type strips
implanted on one side with a 25 \( \mu \textrm{m} \) pitch to measure
\( r-\phi  \). On the opposite side are \( \textrm{n}^{+} \)-type
strips with a wider readout pitch of 150 \( \mu \textrm{m} \) to
200 \( \mu \textrm{m} \) that measure \( r-z \). Charged particles
passing through the silicon wafer produce electron-hole pairs that
drift to a collection strip and the collected charge is read out.
The SMD improved the tracking resolution of the detector significantly.
The SMD is particularly important for reconstructing the primary vertex
(where the initial electron and position interaction occurred within
the beampipe) as well as for determining the location of secondary
vertices such as those from decays of \textbf{B} mesons. A schematic
\( r-\phi  \) view of the SMD along with some tracks which might
be expected from Z boson pair production are shown in Figure \ref{fig:smd}.

\subsection{Calorimetry}

In contrast to tracking, where the goal is to measure position and
momentum with very little disturbance of the particle, the goal in
calorimetry is to absorb the particle's energy completely and measure
it. Because of the differing interaction lengths of electrons/photons
versus pion/other hadrons, two types of high energy calorimeters are
required: electromagnetic calorimeters for electrons and photons and
hadron calorimeters for pions, kaons, and other hadrons. L3 has very
good calorimeters for both electromagnetic and hadronic showers. The
electromagnetic calorimeter(BGO) is a crystal calorimeter built out
of Bismuth Germanate, \( \textrm{Bi}_{4}(\textrm{GeO}_{4})_{3} \),
and the hadron calorimeter (HCAL) is built out of uranium plates with
interspersed proportional chambers.

The BGO electromagnetic calorimeter is a very important feature of
the L3 detector. The calorimeter is formed from 11,000 individual
2 cm \( \times  \) 2 cm \( \times  \) 24 cm crystals which point
at the interaction region. The heavy, high-charge nuclei in BGO cause
a strong electromagnetic cascade and eventually convert a fraction
of the electrons' and photons' energy into scintillation light, which
is measured using a photodiode. Crystal calorimeters did not originate
with the L3 detector, but L3 was the first large-scale detector to
use BGO as the crystal material%
\footnote{Since the development done for L3, BGO has found widespread use in
medical PET scanners.
}. The BGO has an average energy resolution of \( \frac{\sigma _{E}}{E}=\frac{1.57\%}{\sqrt{E}}+0.34\% \)
for electrons. The shower profile in the crystals surrounding the
peak crystal is also useful for separating hadrons, including \( \pi _{0} \)'s,
from electrons and photons. The very high resolution of the calorimeter
is important for measuring the electrons from Z decays and accurately
determining recoil masses.

As L3 was originally constructed, there was a small gap between the
barrel section of the BGO and the endcap. During the 1995-1996 shutdown,
a new subdetector was installed to fill the gap -- the so-called EGAP
detector\cite{l3:egap}. The detector is constructed of lead blocks
with scintillating fibers embedded longitudinally. Electromagnetic
showers in the lead generate light in the fibers. The light from the
fibers is coupled into plastic lightguides which are read out by phototriodes.
There are 24 blocks on each end of the BGO barrel to provide coverage
of the region \( 38^{\circ }<\theta <42^{\circ } \) and \( 138^{\circ }<\theta <142^{\circ } \).
The EGAP detector has poorer resolution than the BGO, at \( 12\%/\sqrt{E} \),
but the difference is relatively unimportant for searches: the increased
hermeticity of the detector is of greater importance.

The L3 HCAL, like most hadron calorimeters, is a sampling calorimeter.
The L3 HCAL consists of a series of depleted uranium plates interspersed
with gas proportional chambers. Most of the nuclear interactions occur
in the uranium and the gas chambers sample the developing shower. 
\begin{figure}[tbp]
{\centering \resizebox*{0.7\textwidth}{!}{\includegraphics{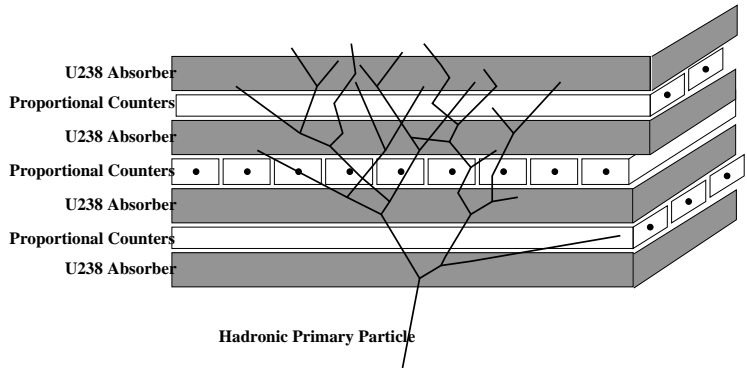}} \par}

\caption{Schematic view of the HCAL showing shower development.\label{fig:hcalshower}}
\end{figure}
A schematic view of a shower developing in a portion of the HCAL is
shown in Figure \ref{fig:hcalshower}. In the central barrel region,
there are 58 uranium layers in each module, while modules in the more
forward region have 51 layers. The HCAL is used in the analyses described
in this thesis to measure the energies of hadronic jets from Z and
W decays.

\begin{figure}[tbp]
{\centering \resizebox*{0.7\textwidth}{!}{\includegraphics{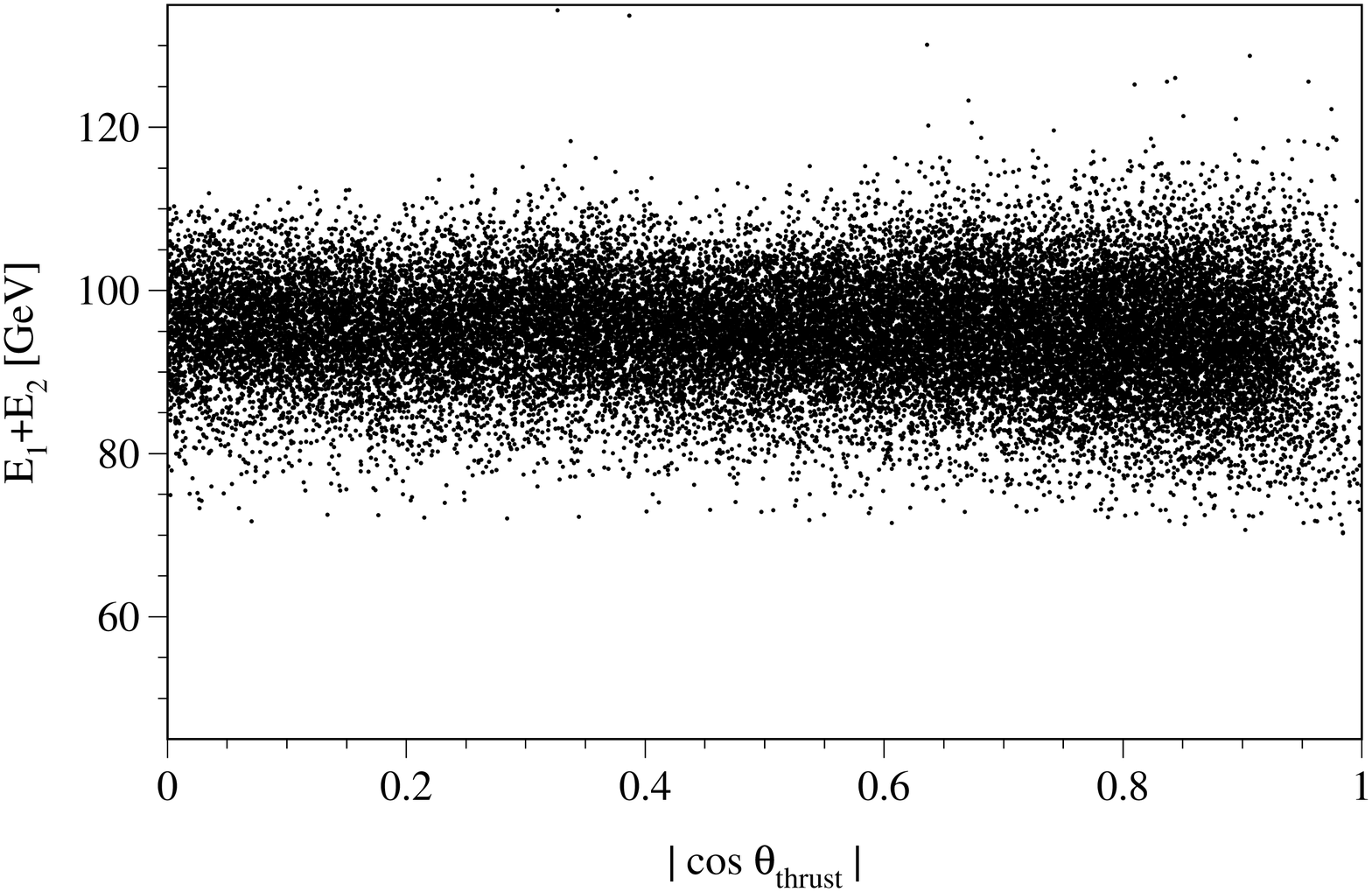}} \par}

\caption{Distribution of jet-energy as a function of the cosine of the thrust
angle.\label{fig:je_angle}}

This plot shows the sum of jet energies from Z-pole data. The resolution
is quite uniform across the entire detector, including the EGAP (\( 0.74<\cos \theta <0.81 \))\cite{l3:hnn_cand}.
\end{figure}

The L3 calorimeters work together to measure leptons and jets which
may be produced in any direction within the detector. The performance
of the calorimetry systems can be seen particularly well in Figure
\ref{fig:je_angle}. This figure shows the jet energy measured by
the calorimeter for Z-peak jets as a function of the angle of the
jet. The two jets produced from Z decay on the peak should sum to
\( \approx 91 \) GeV, as seen in the figure. The sum and resolution
are nearly constant for all jet production angles, despite the different
barrel, endcap, and EGAP calorimeters which are used to compute the
jet energies.

\subsection{Muon Chambers}

Besides the BGO electromagnetic calorimeter, the L3 muon chambers
are the most unique feature of the detector. They are unique not in
their construction or design but rather in their location. They are
located inside the magnet and flux-return yoke instead of outside
as is more common. This location provides the most uniform bending
field and reduces the multiple scattering of the muons on their way
out of the detector. Muons are the only particle, besides of course
the neutrinos, which emerge from the inner layers of BGO and uranium
with most of their energy intact. Thus they are the only particles
left to be measured by these tracking chambers. The muon system consists
of three layers of drift chambers spaced 145 cm apart, with each layer
forming an octagon centered on the interaction point. The chambers
are actively aligned using an LED-lens-quadrant photodiode system
and the alignment can be verified by ultraviolet laser shots that
simulate infinite-momentum muons produced at the interaction point.
Each chamber measures the position of passage of a muon with an average
accuracy of 168 \( \mu \textrm{m} \), which is sufficient to provide
a 2\% error measurement of muon momentum for 50 GeV muons.

The main muon chambers provide detector coverage in the barrel region
between \( 44^{\circ } \) and \( 136^{\circ } \). In 1995, an additional
set of chambers was added on the flux-return doors of the main solenoid
to provide measurements down to within \( 24^{\circ } \) of the beam
line \cite{l3:fbmuon}. These forward-backward muon chambers use a
1.24T toroidal field in the iron door to bend forward muons in the
region between the chambers. The coils to generate the toroidal field
were added as part of the detector upgrade, as well as resistive plate
chambers for triggering on forward muons.

Despite the depth of the experiment underground, cosmic ray muons
do penetrate down to L3%
\footnote{In fact, another experiment called L3 Cosmics used the L3 muon chambers
to study these very high energy muons from cosmic rays.
}. With very precise arrival-time information about these muons, it
is possible to reject those which do not occur in time with a beam
crossing. Since the muon chambers cannot provide this information,
there is a layer of scintillator panels located between the BGO and
the HCAL. These scintillation panels are read out by high speed photomultipliers
and time-to-digital converters. The scintillator system has a timing
resolution of about 1 ns, which allows the separation of cosmic ray
muons from muon pairs produced at the interaction point. A cosmic
ray muon would require 5.8 ns to travel across from top scintillator
panel to bottom scintillator panel, while a di-muon pair would arrive
on the two sides simultaneously.

\subsection{Monte Carlo}

There are very few background-free event signatures for a Higgs produced
at LEP, so it is important to understand the behavior of the detector
for various Standard Model processes known to be present as well as
for the predicted process. For processes which are well-understood
theoretically, example events can be generated by sampling the theoretical
distributions in a random manner. This technique is called Monte Carlo
(MC) and is widely used in high energy physics and in many other fields. 

A primary process such as \( \textrm{e}^{+}\textrm{e}^{-}\rightarrow \textrm{WW}\rightarrow \textrm{c}\bar{\textrm{s}}\mu \nu  \)
is specified by the user, and MC generator creates events of this
process using a level of accuracy defined by the number of Feynman
diagrams included in the generator. The generator carries out the
processes of hadronization and quark decay. The generator produces
a list of four-vectors representing the {}``stable'' mesons, baryons,
photons, and leptons produced in the event. L3 uses several different
MC generator programs depending on the processes which are under study.
Table \ref{tab:mcgen} lists several of the important generators and
what processes are generated using them for the work presented in
this thesis. 

Of course, the detector does not produce a list of four-vectors; it
reports energies in calorimeter cells and hits in trackers. In order
to match the Monte Carlo with the data, the list of four-vectors must
be converted to the same form as data from the detector. This difficult
task is carried out by a simulation program based on GEANT 3.15\cite{mc:geant3}.
This program simulates the response of the entire detector to this
event. The simulation includes the full complex geometry of the detector,
with material-specific properties for both the active regions and
for structural elements that may cause scattering or shower initiation.
The interaction of hadrons inside the detector is handled by a package
called GEISHA\cite{mc:geisha}. The result of the simulation is stored
in the same format as that used for the data, and from this point
the same reconstruction and analysis techniques can be applied identically
to data and Monte Carlo.
\begin{table}[tbp]
{\centering \begin{tabular}{|c|c|}
\hline 
Generator&
Processes\\
\hline
\hline 
PYTHIA\cite{mc:pythia}&
\( \begin{array}{c}
\textrm{e}^{+}\textrm{e}^{-}\rightarrow \textrm{ZZ}\\
\textrm{e}^{+}\textrm{e}^{-}\rightarrow \textrm{Ze}^{+}\textrm{e}^{-}\\
\textrm{e}^{+}\textrm{e}^{-}\rightarrow \textrm{ZH}
\end{array} \)\\
\hline 
KK2F\cite{mc:kk2f}&
\( \begin{array}{l}
\textrm{e}^{+}\textrm{e}^{-}\rightarrow \textrm{q}\bar{\textrm{q}}(\gamma )\\
\textrm{e}^{+}\textrm{e}^{-}\rightarrow \mu ^{+}\mu ^{-}(\gamma )\\
\textrm{e}^{+}\textrm{e}^{-}\rightarrow \tau ^{+}\tau ^{-}(\gamma )
\end{array} \)\\
\hline 
KORALW\cite{mc:koralw}&
\( \textrm{e}^{+}\textrm{e}^{-}\rightarrow \textrm{WW} \)\\
\hline 
EXCALIBUR\cite{mc:excalibur}&
\( \begin{array}{l}
\textrm{e}^{+}\textrm{e}^{-}\rightarrow \textrm{q}\bar{\textrm{q}}'\textrm{e}\nu \\
\textrm{e}^{+}\textrm{e}^{-}\rightarrow \textrm{f}_{1}\bar{\textrm{f}}'_{1}\textrm{f}_{2}\bar{\textrm{f}}'_{2}
\end{array} \)\\
\hline 
PHOJET\cite{mc:phojet}&
\( \textrm{e}^{+}\textrm{e}^{-}\rightarrow \textrm{e}^{+}\textrm{e}^{-}\textrm{q}\bar{\textrm{q}} \)\\
\hline
\end{tabular}\par}

\caption{Partial list of Monte Carlo Generators used by L3.\label{tab:mcgen}}
\end{table}

Generation, production, and reconstruction of Monte Carlo is a complicated
and CPU-intensive process which is managed from CERN, but carried
out at multiple institutes around the world. Farms of PCs are used
as well as idle workstations around the experiment during evenings
and weekends.

\newcommand{\eplusminus}{\ensuremath{\mathrm{e^+e^-}}}

\chapter{Theory of the Higgs Boson}

\begin{quotation}
\noindent \textit{{[}In a system of physics{]} we adopt, at least
insofar as we are reasonable, the simplest conceptual scheme into
which the disordered fragments of raw experience can be fitted and
arranged.}

Willard Van Orman Quine
\end{quotation}

\section{Review of the Standard Model}

The Standard Model (SM) of particle physics, developed by Weinberg,
Glashow, and Salam \cite{hep:sm1,hep:sm2,hep:sm3}, has proven to
be an extremely effective theory for predicting the results of high-energy
physics experiments over the last 25 years. The model was developed
in the 1960s and 1970s to bring together the results of many different
experiments and ad hoc theories. The theory describes the behavior
of three forces: the electromagnetic force which acts between charges,
the weak force which is responsible for beta decay, and the strong
force felt only by quarks and which binds the nuclei of atoms. The
theory is silent about gravity, which is too weak at these scales
to be felt. The forces are carried by gauge bosons: the \( \textrm{W}^{+} \),\( \textrm{ W}^{-} \),
and Z of the weak interaction, the photon (\( \gamma  \)) of the
electromagnetic interaction, and the gluons of the strong interaction.
The matter constituents of the theory are twelve particles that are
organized into three generations of quarks and three generations of
leptons and neutrino partners. The two lightest quarks, the up and
down quarks, combine to form protons and neutrons in normal matter,
while the lightest charged lepton is the familiar electron. All the
particles of the Standard Model are listed in Table \ref{tab:smparticles}.
The discoveries of the gluon in 1979\cite{hep:gluon_discovery} and
the W and Z particles in 1983\cite{hep:W_discovery,hep:Z_discovery}
were major triumphs for the Standard Model. 
\begin{table}
\begin{center}\begin{tabular}{ccc}
\( \left( \begin{array}{cc}
\mbox {\huge u} & \begin{array}{c}
+2/3\\
1-5
\end{array}\\
\mbox {\huge d} & \begin{array}{c}
-1/3\\
3-9
\end{array}
\end{array}\right)  \)&
\( \left( \begin{array}{cc}
\mbox {\huge c} & \begin{array}{c}
+2/3\\
1150-1350
\end{array}\\
\mbox {\huge s} & \begin{array}{c}
-1/3\\
75-170
\end{array}
\end{array}\right)  \)&
\( \left( \begin{array}{cc}
\mbox {\huge t} & \begin{array}{c}
+2/3\\
174300
\end{array}\\
\mbox {\huge b} & \begin{array}{c}
-1/3\\
4000-4400
\end{array}
\end{array}\right)  \)\\
&
&
\\
\( \left( \begin{array}{cc}
\mbox {\huge \ensuremath {\nu _{\textrm{e}}}} & \begin{array}{c}
0\\
1-5
\end{array}\\
\mbox {\huge e} & \begin{array}{c}
-1\\
0.51
\end{array}
\end{array}\right)  \)&
\( \left( \begin{array}{cc}
\mbox {\huge \ensuremath {\nu _{\mu }}} & \begin{array}{c}
0\\
<0.2
\end{array}\\
\mbox {\huge \ensuremath {\mu }} & \begin{array}{c}
-1\\
106
\end{array}
\end{array}\right)  \)&
\( \left( \begin{array}{cc}
\mbox {\huge \ensuremath {\nu _{\tau }}} & \begin{array}{c}
0\\
<18
\end{array}\\
\mbox {\huge \ensuremath {\tau }} & \begin{array}{c}
-1\\
1777
\end{array}
\end{array}\right)  \)\\
\multicolumn{1}{l}{Fermions}&
&
\\
\hline
\hline 
\multicolumn{1}{l}{Bosons}&
&
\\
{\LARGE \( \gamma  \)}\( \begin{array}{c}
0\\
0
\end{array} \)&
{\LARGE g}\( \begin{array}{c}
0\\
0
\end{array} \)&
{\LARGE H}\( \begin{array}{c}
0\\
?
\end{array} \)\\
&
&
\\
{\LARGE \( \textrm{W}^{\pm } \)}\( \begin{array}{c}
\pm 1\\
80419
\end{array} \)&
{\LARGE Z}\( \begin{array}{c}
0\\
91118
\end{array} \)&
\\
\end{tabular}

\end{center}

\caption{Constituent particles of the Standard Model.\label{tab:smparticles} }

Each particle is listed with its charge and the particle's mass in
MeV as listed in the Particle Data Book \cite{hep:pdb2000}. For reference,
recall that the mass of the proton is 938 MeV. The d,s, and b quarks
and the charged leptons are collectively referred to as {}``down-type''
particles, while u,c,t, and the neutrinos are {}``up-type''.
\end{table}

The Standard Model is a quantum field theory, where all particles
appear as fields and their behavior and interaction can be described
by a Lagrangian. For example, a massless fermion field \( \psi  \)
freely propagating through space has a Lagrangian of the form \[
{\cal L}=i\, \hbar c\, \bar{\psi }\, \gamma ^{\mu }\, \partial _{\mu }\, \psi \]
while a massless vector (spin-1) boson field \( A^{\mu } \) has a
free Lagrangian of the form\[
{\cal L}=-\frac{1}{16\pi }(\partial ^{\mu }A^{\nu }-\partial ^{\nu }A^{\mu })(\partial _{\mu }A_{\nu }-\partial _{\nu }A_{\mu })=-\frac{1}{16\pi }F^{\mu \nu }F_{\mu \nu }.\]
 Interactions between bosons and fermions are written as terms like
\[
{\cal L}_{\textrm{int}}=-(q\, \bar{\psi }\, \gamma ^{\mu }\, \psi )A_{\mu },\]
involving three fields. There are also terms which describe the interaction
of four boson fields. The full Standard Model Lagrangian is quite
large, but all of the terms have one of these basic forms. In the
case of the photon and Z, the two fermions involved are the same flavor,
while the W couples to a weak isospin doublet%
\footnote{To be accurate, the W can couple across quark generations with reduced
probabilities given by the squares of the off-diagonal terms of the
CKM matrix \( V_{ij} \).
} such as \( \mu  \) and \( \nu _{\mu } \) or charm and strange quarks.

\section{Motivation for a Higgs}

In the above discussion of fields, the bosons were explicitly massless,
but the physical W and Z are indeed quite massive. The simplest way
to add a boson mass term to the Lagrangian is to append \[
m^{2}B_{\mu }B^{\mu }.\]
However, this term is not invariant under transformations which take
\( B_{\mu }\rightarrow B_{\mu }-\partial _{\mu }\chi  \). Therefore,
some other gauge-invariant technique is needed to provide masses.
Gauge-invariance is a very important principle in quantum field theories
because it guarantees a theory to be renormalizable \cite{higgs:renorm1}.
Renormalization is a process of canceling the many infinities which
can appear in the field theory allowing reasonable calculations to
be performed.

The gauge-invariant solution used in the Standard Model is the Higgs
mechanism. To understand the SM's Higgs mechanism, consider first
the simpler situation of a theory which contains only a massless gauge
boson \( A^{\mu } \) to which we add a massless complex scalar field
\( \phi  \) \cite{higgs:hhg}. For this situation, the Lagrangian
has the form\[
{\cal L}=(D_{\mu }\phi )^{*}(D^{\mu }\phi )+\mu ^{2}\phi ^{*}\phi -\lambda (\phi ^{*}\phi )^{2}-\frac{1}{4}F^{\mu \nu }F_{\mu \nu },\]
with the covariant derivative \( D^{\mu }=\partial ^{\mu }+igA^{\mu } \)
to achieve invariance under a local gauge transformation. We see that
the scalar field has its minimum at \( \phi =\sqrt{\mu ^{2}/2\lambda }=v/\sqrt{2} \).
If we expand the field near the minimum as \( \phi =\left( v+h(x)\right) /\sqrt{2} \)
we obtain\begin{eqnarray*}
{\cal L} & = & \frac{1}{2}\left[ (\partial _{\mu }-igA_{\mu })(v+h)(\partial ^{\mu }+igA^{\mu })(v+h)\right] \\
 &  & +\frac{1}{2}\mu ^{2}(v+h)^{2}-\frac{1}{4}\lambda (v+h)^{4}-\frac{1}{4}F^{\mu \nu }F_{\mu \nu }.
\end{eqnarray*}
This Lagrangian contains the term \( \frac{g^{2}v^{2}}{2}A_{\mu }A^{\mu } \),
which is a mass term for the (previously massless) gauge boson, obtained
in a gauge-invariant manner. The term \( \lambda v^{2}h^{2} \) is
a mass term for the quantum excitation of the scalar field -- a new
massive scalar boson. In addition, there are \( hAA \), \( h^{3} \),
and \( h^{4} \) interaction terms. The mass of the \( A \) boson
fixes \( v^{2} \) but \( \lambda  \) is not predicted by the model,
and the mass of the scalar is a free parameter.

In review, the algebra above converted an apparently massless complex
scalar field with two degrees of freedom into a real massive field
and the longitudinal polarization state of the gauge boson, again
two total degrees of freedom. The SM, with \( \textrm{W}^{+} \),
\( \textrm{W}^{-} \), and Z to provide mass for, must have at least
an SU(2) doublet of complex scalar fields, \( \phi =\left( \begin{array}{c}
\phi ^{+}\\
\phi ^{0}
\end{array}\right)  \). Symmetry-breaking is initiated by giving a vacuum expectation only
to the real part of the neutral field \( \left\langle \phi ^{0}\right\rangle =v/\sqrt{2} \).
Three of the degrees of freedom become the longitudinal polarizations
of the massive weak bosons, and the fourth remains as a real observable
scalar boson, the Higgs boson.

\begin{figure}
{\centering \resizebox*{0.3\textwidth}{!}{\includegraphics{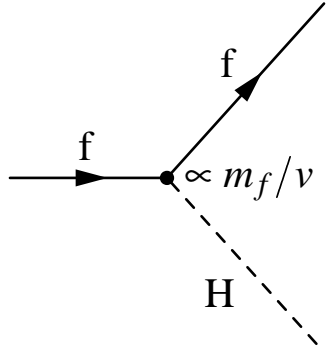}} \par}

\caption{Higgs-fermion vertex.\label{fig:higgsfermion}}
\end{figure}

With the Higgs mechanism it is also possible to add masses for the
fermions in the theory using Yukawa-type terms. Since the left-handed
fermions in the SM are SU(2) doublets and the right-handed fermions
are SU(2) singlets, a mass term such as \[
m\bar{f}\, f=m(\bar{f}_{L}f_{R}+\bar{f}_{R}f_{L})\]
is not SU(2) invariant. With the Higgs SU(2) doublet, we may write
an interaction Lagrangian \[
{L}_{{\scriptsize \textrm{int}}}=g_{f}\left[ (\bar{f}_{L}\phi )f_{R}+(\phi ^{\dagger }\bar{f}_{R})f_{L}\right] ,\]
where \( g_{f} \) is different for each fermion. This interaction
Lagrangian transforms satisfactorily under SU(2), although it is an
unusual Lagrangian since it explicitly contains the conjugate of the
Higgs field. When the Higgs acquires a vacuum expectation, \[
\phi \rightarrow \left( \begin{array}{c}
0\\
\frac{v+h}{\sqrt{2}}
\end{array}\right) ,\]
the fermion interaction becomes\[
{L}_{{\scriptsize \textrm{int}}}=\frac{g_{f}v}{\sqrt{2}}\bar{f}f+\frac{g_{f}}{\sqrt{2}}\bar{f}fh.\]
 The first part of the interaction Lagrangian is a mass term for the
fermion, where \( m_{f}=\frac{g_{f}v}{\sqrt{2}} \). The values \( g_{f} \)
are free parameters, so the model does not predict the masses of the
fermions. Instead, one measures the mass of the fermion experimentally
and uses \( m_{f} \) equation as a definition of \( g_{f} \), so
\( g_{f}\equiv \frac{\sqrt{2}m_{f}}{v} \). With this substitution,
the second part of the interaction Lagrangian becomes \[
\frac{\sqrt{2}m_{f}}{v}\bar{f}fh.\]
This is an interaction term between the fermion and the Higgs particle,
describing the vertex in Figure \ref{fig:higgsfermion} which has
a coupling proportional to the mass of the fermion.

\section{Production of a Higgs Boson}

In order to search for the Higgs at an accelerator, the experimental
production and decay of the Higgs must be considered. The production
mechanism for the Higgs is very dependent on the collider used to
produce it. In the case of LEP, only the diagrams beginning with an
\eplusminus\ pair are relevant. The direct coupling of the Higgs to
\eplusminus\ is very small since the coupling is proportional to the
fermion mass, which is extremely small for the electron. Therefore,
the direct \( \eplusminus \rightarrow \textrm{H} \) production rate
is very small, and indirect processes dominate. There are two classes
of indirect production which are important at LEP: Higgsstrahlung
and vector boson fusion. In these indirect processes, additional particles
are produced along with the Higgs, so their presence and possible
decays must be taken into account when describing the physical signature
of a Higgs-containing event.

\begin{figure}
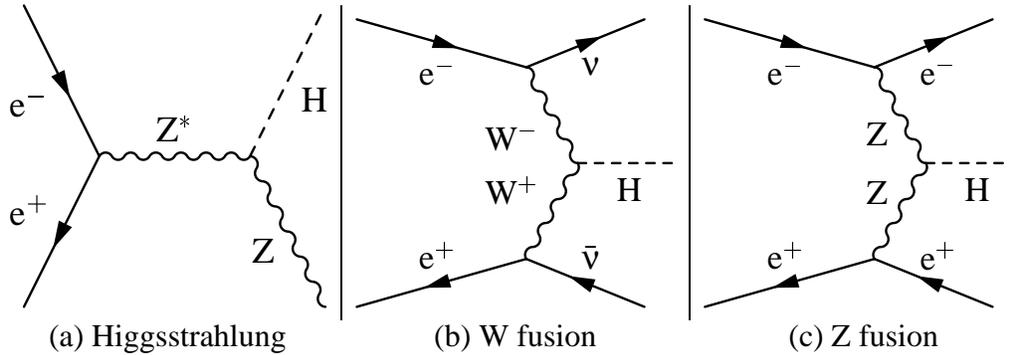

{\centering \begin{tabular}{ccc}
\resizebox*{0.3\textwidth}{!}{\includegraphics{higgsstrahlung.epsi}} &
\multicolumn{1}{|c|}{\resizebox*{0.3\textwidth}{!}{\includegraphics{wfusion.epsi}} }&
\resizebox*{0.3\textwidth}{!}{\includegraphics{zfusion.epsi}} \\
(a) Higgsstrahlung&
(b) W fusion&
(c) Z fusion\\
\end{tabular}\par}

\caption{Production diagrams for the Higgs at LEP.\label{fig:higgsprod}}
\end{figure}

The first type of indirect production is the so-called Higgsstrahlung
process, where the electron and positron annihilate to produce an
off-shell Z (Z{*}). The Z{*} decays to its mass-shell by emitting
a Higgs in a manner quite similar to Bremsstrahlung. The Feynman diagram
for this process is shown in Figure \ref{fig:higgsprod}a. The physical
signature of the event includes the decay products of a Z boson as
well as the Higgs. This associated Z can be used to tag the Higgs
events. Also, since the Z and Higgs are produced in a two-body decay
of the Z{*} produced at rest, the momentum of the Z should be equal
to that of the Higgs; thus the event should be balanced in the detector.
For a high rate, the final Z should be near its mass shell, which
implies a production mass limit of \( m_{{\scriptsize \textrm{H}}}\widetilde{<}\sqrt{s}-\mZ  \).
Thus, if the Higgs's mass is less than 116 GeV, LEP should be able
to produce it by Higgsstrahlung with a center-of-mass energy of 207
GeV.

The second class of diagrams is the vector boson fusion diagrams,
both W fusion and Z fusion. In W fusion, the incoming electron and
positron emit W bosons which combine to form a Higgs. The emission
of the W boson converts the electron and positron into a neutrino
and antineutrino respectively, as shown in Figure \ref{fig:higgsprod}b.
Z fusion is similar except the scattered electron and positrons remain
in the final state (Figure \ref{fig:higgsprod}c). Since fusion diagrams
are three-body processes which involve a t-channel diagram, the Higgs
produced in the fusion process will have an arbitrary boost relative
to the experiment. Also, the missing mass (for W fusion) or \eplusminus\
invariant mass (for Z fusion) will not have any particular value since
neither corresponds to any resonance. 

The greater cross-section and kinematic advantages of Higgsstrahlung
over the fusion diagrams mean that the searches are tuned for the
HZ process, despite the resulting search limit at \( m_{{\scriptsize \textrm{H}}}=\sqrt{s}-\mZ  \).
Above that limit, the fusion diagrams dominate, but the absolute rate
is too small for an effective search at LEP. As a result, we will
consider only HZ production for this thesis, which means we must take
into account the decay products of the Z in our analysis.
\begin{figure}
\begin{minipage}{0.48\linewidth}

\resizebox*{1\textwidth}{!}{\includegraphics{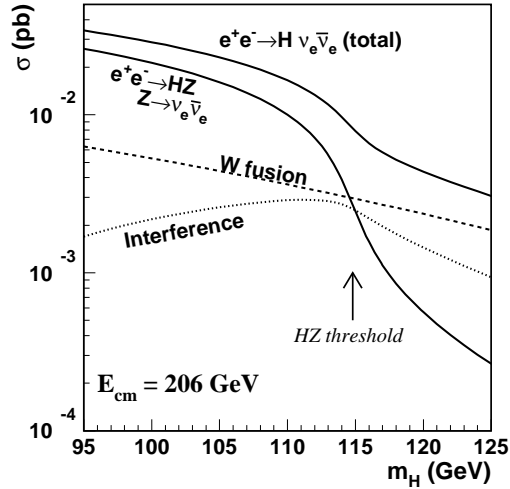}} 

\end{minipage}\hfill\begin{minipage}{0.48\linewidth}

\caption{Cross-sections for different production diagrams for the process
\protect\( \eplusminus \rightarrow \textrm{H}\nu _{e}\bar{\nu }_{e}\protect \). }

The plot is shown for \( \sqrt{s}=206\textrm{ GeV} \). For \( m_{{\scriptsize \textrm{H}}}<115\textrm{ GeV} \),
the Higgsstrahlung process dominates the cross-section. For higher
Higgs masses, the W fusion process becomes dominant, but the total
cross-section becomes small. There is also an interference term which
is important when the two diagrams have similar strength near 115
GeV.

\label{fig:prodrate}\end{minipage}
\end{figure}

\section{Decays of the Higgs}

The decay of the Higgs is independent of how it is produced, so the
same decay channels will occur in the same ratio at any collider.
However, the relative usefulness of different decay modes of the Higgs
(and associated particles) will vary depending on the background processes
at different kinds of colliders. For example, at hadron machines there
are many jets arising from soft QCD interactions, so the decays of
the Higgs and associated particles into jets would be less useful
than those with leptons or photons in the final state. At LEP, the
jet-like backgrounds are easier to control through momentum and mass
constraints, so the importance of a channel depends more on its branching
ratio.

\begin{figure}
{\centering \resizebox*{0.5\textwidth}{!}{\includegraphics{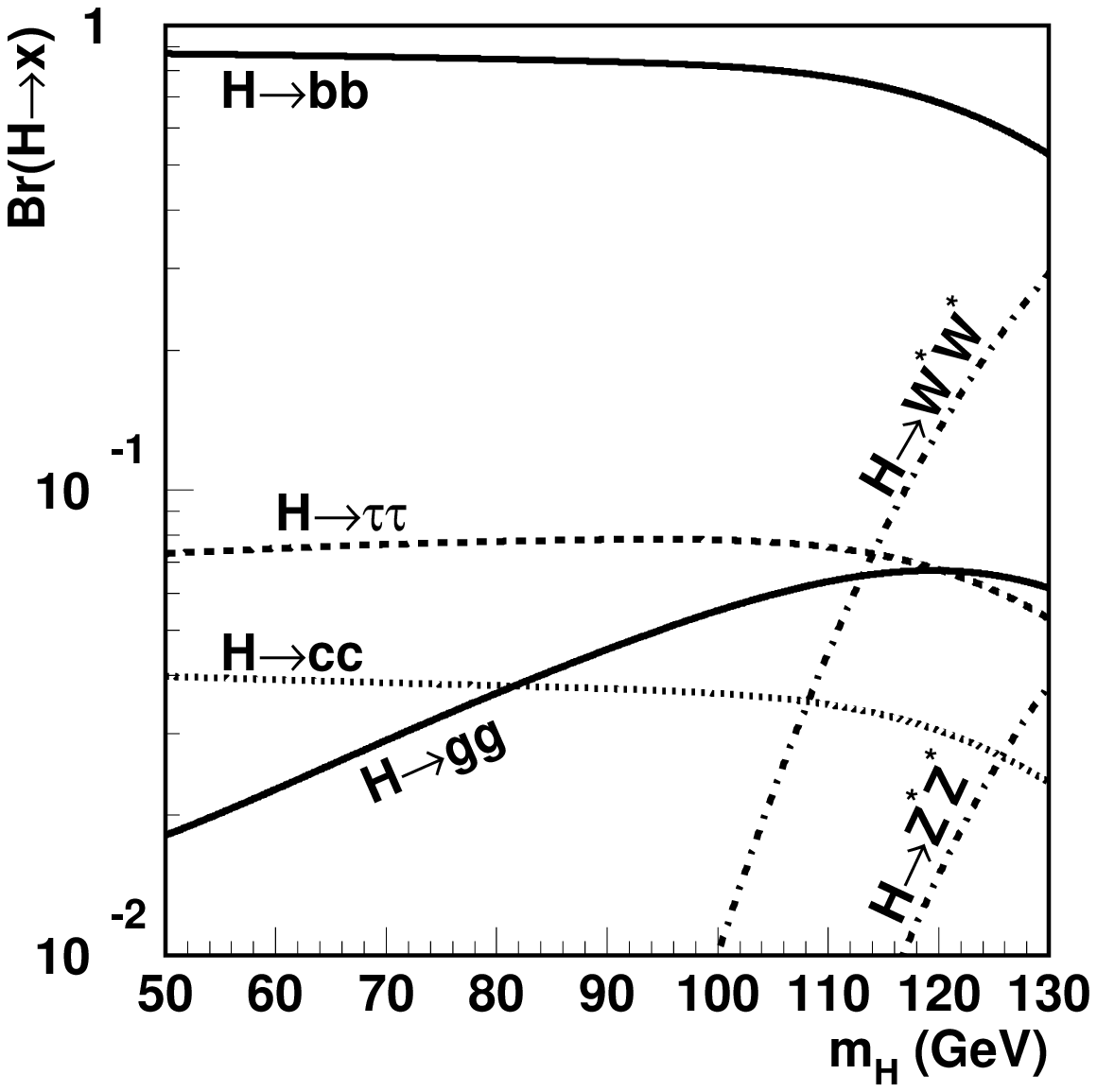}} \par}

\caption{Predicted branching fractions for a minimal Standard Model Higgs.\label{fig:smhiggs}}

The branching fractions are from the standard LHWG database and were
calculated using the HZHA program \cite{higgs:hzha}.
\end{figure}

The decay of the Higgs is very dependent on the detailed physics of
the Higgs, which implies strong model-dependence. The structure of
the electroweak symmetry-breaking puts strong limits on the Z-H-Z
vertex, which means that for most models the production rate is similar.
The models are primarily distinguished by their decays. In the Standard
Model, the expected decays of the Higgs depend on the mass of the
Higgs. The general rule is that the Higgs will primarily decay to
the heaviest particles kinematically available, since the Higgs coupling
is proportional to mass. Thus, for a Higgs with mass between 12 GeV
and \textasciitilde{}150 GeV, the SM predicts the Higgs will decay
primarily to \( \textrm{b}\bar{\textrm{b}} \), with small branching
fractions into \( \tau ^{+}\tau ^{-} \) and \( \textrm{c}\bar{\textrm{c}} \)
at lower mass and a rising branching fraction to gauge boson pairs
at higher mass. Although the gluon is massless, there is also a substantial
Higgs branching fraction to two gluons through top-quark loops. For
Higgsstrahlung searches at LEP, \( \textrm{b}\bar{\textrm{b}} \)
is the most important channel, as seen in Figure \ref{fig:smhiggs}.

\section{Two Higgs Doublet Models}

In the minimal SM, after providing longitudinal polarizations to the
massive weak bosons there is one degree of freedom left which becomes
the Higgs boson. A more general assumption is that there are two doublets
of complex scalar fields, \( \phi _{1} \) and \( \phi _{2} \). Many
theorists find advantages in this more extensive Higgs sector. As
a group, these models are called {}``Two-Higgs-Doublet Models''
or 2HDMs. The most general 2HDM potential\cite{higgs:hhg} is quite
extensive:\begin{eqnarray*}
V & = & \lambda _{1}(\phi ^{\dagger }_{1}\phi _{1}-v_{1}^{2})^{2}+\lambda _{2}(\phi ^{\dagger }_{2}\phi _{2}-v_{2}^{2})^{2}\\
 &  & +\lambda _{3}\left[ (\phi ^{\dagger }_{1}\phi _{1}-v_{1}^{2})+(\phi ^{\dagger }_{2}\phi _{2}-v_{2}^{2})\right] \\
 &  & +\lambda _{4}\left[ (\phi ^{\dagger }_{1}\phi _{1})(\phi ^{\dagger }_{2}\phi _{2})-(\phi ^{\dagger }_{1}\phi _{2})(\phi ^{\dagger }_{2}\phi _{1})\right] \\
 &  & +\lambda _{5}\left[ \textrm{Re}(\phi ^{\dagger }_{1}\phi _{2})-v_{1}v_{2}\cos \xi \right] ^{2}\\
 &  & +\lambda _{6}\left[ \textrm{Im}(\phi ^{\dagger }_{1}\phi _{2})-v_{1}v_{2}\sin \xi \right] ^{2}.
\end{eqnarray*}
If \( \sin \xi \ne 0 \), then the theory will break CP explicitly,
so we will set \( \xi =0 \) which makes the potential minimum\[
\left\langle \phi _{1}\right\rangle =\left( \begin{array}{c}
0\\
v_{1}
\end{array}\right) ,\; \left\langle \phi _{2}\right\rangle =\left( \begin{array}{c}
0\\
v_{2}
\end{array}\right) .\]
After significant algebra to remove the Goldstone bosons, we are left
with two charged Higgs bosons: \( \textrm{H}^{\pm } \), one CP-odd
scalar: A\( ^{0} \), and two CP-even physical scalars: H and h. These
last two physical scalars are constructed from a linear combination
of the \( \phi _{1} \) and \( \phi _{2} \) fields as\begin{eqnarray*}
\textrm{H} & = & \sqrt{2}\left[ \left( \Re (\phi ^{0}_{1})-v_{1}\right) \cos \alpha +\left( \Re (\phi ^{0}_{2})-v_{2}\right) \sin \alpha \right] \\
\textrm{h} & = & \sqrt{2}\left[ -\left( \Re (\phi ^{0}_{1})-v_{1}\right) \sin \alpha +\left( \Re (\phi ^{0}_{2})-v_{2}\right) \cos \alpha \right] ,
\end{eqnarray*}
where \( \alpha  \) is the mixing angle between the doublets and
the two CP-even scalars. By convention, the h is the less-massive
of the two CP-even bosons. The sum \( v^{2}_{1}+v^{2}_{2} \) is set
by the mass of the W boson, so there are six free parameters: four
Higgs boson masses, \( \alpha  \), and \( \tan \beta \equiv v_{2}/v_{1} \).

When one wishes to couple the Higgs fields of the 2HDM model to the
particles of the Standard Model, there are several different strategies.
In one type of model, one doublet couples to the up-type quarks and
leptons and the other to the down-type fermions. This type of model
is referred to as a {}``Type II 2HDM.'' The most well-known Type
II model is the Minimal Supersymmetric Standard Model (MSSM) \cite{higgs:mssm1,higgs:mssm2,higgs:mssm3}.
Alternatively, one can construct a model where one doublet couples
to bosons and the other to fermions, which is a Type I model. In this
type of model, the coupling of the lightest Higgs to fermions is proportional
to \( \cos \alpha  \). Thus, for values of \( \alpha \rightarrow \frac{\pi }{2} \),
the couplings of the light Higgs to fermions tend toward zero. The
model is generally referred to as a {}``fermiophobic'' model, since
the light Higgs does not couple to the fermions \cite{fp:renorm2hdm}.

\begin{figure}
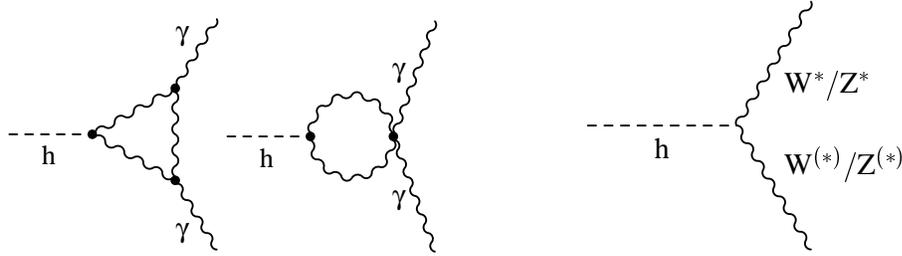

\begin{tabular}{cc}
\resizebox*{0.2\textwidth}{!}{\includegraphics{h2gg.epsi}} \resizebox*{0.2\textwidth}{!}{\includegraphics{h2gg_ii.epsi}} &
\resizebox*{0.3\textwidth}{!}{\includegraphics{HtoVV.epsi}} \\
(a) Loop diagrams for \( \textrm{h}\rightarrow \gamma \gamma  \).&
(b) Higgs decay to a pair of Z or W bosons.\\
\end{tabular}

\caption{Leading decay diagrams for a fermiophobic Higgs.\label{fig:phobicfeyn}}
\end{figure}

Since the fermiophobic Higgs does not couple directly to \( \textrm{b}\bar{\textrm{b}} \)
or \( \tau ^{+}\tau ^{-} \) as in the Standard Model, what decay
channels does this leave? Somewhat surprisingly, a low-mass fermiophobic
Higgs decays primarily to two photons. The Higgs does not couple directly
to the photon, but it can decay through a W loop or a charged Higgs
loop, as shown in the \ref{fig:phobicfeyn}a. For a low mass fermiophobic
Higgs boson, the two photon decay is expected to be dominant, and
all the LEP experiments have carried out searches for it. These analyses
are reviewed in Chapter \ref{chap:results} before their results are
combined with the \( \textrm{h}\rightarrow \textrm{WW}/\textrm{ZZ} \)
channels to fully cover the fermiophobic search.

\begin{figure}
\begin{minipage}{0.45\linewidth}

{\centering \resizebox*{1\textwidth}{!}{\includegraphics{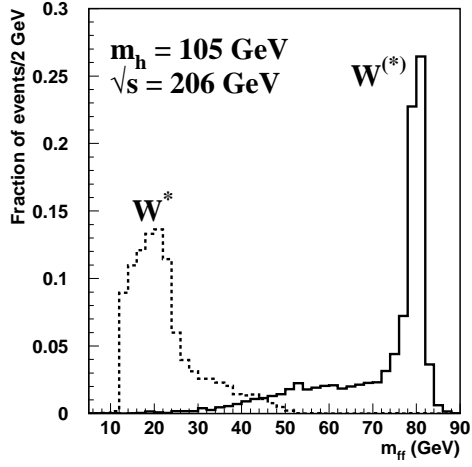}} \par}

\end{minipage}\hfill\begin{minipage}{0.45\linewidth}

\caption{Generated mass of W bosons in \protect\( \textrm{h}\rightarrow \textrm{WW}\protect \).}

This plot shows the masses of the W bosons produced by the PYTHIA
Monte Carlo generator with \( m_{\textrm{h}}=105\textrm{ GeV} \)
and \( \sqrt{s}=206 \) GeV.\label{fig:genmass} The solid curve represents
the more massive W produced while the dashed curve which represents
the lighter W. The heavier W has an average mass of 71.7 GeV with
a pronounced peak at 80 GeV, while the lighter W has an average mass
of 22.6 GeV.

\end{minipage}
\end{figure}

The fermiophobic Higgs can also decay to a pair of weak gauge bosons.
At the masses which LEP can reach, the Higgs cannot decay to two real
W's or Z's, so the decay is \( \textrm{h}\rightarrow V^{*}V^{*} \),
where the star indicates that the vector boson is off its mass shell.
How far off mass shell? Consider the differential width for \( \textrm{h}\rightarrow \textrm{W}^{(*)}\textrm{W}^{*}\rightarrow \textrm{f}_{1}\bar{\textrm{f}}'_{1}\textrm{f}_{2}\bar{\textrm{f}}'_{2} \):
\cite{fp:width} \begin{eqnarray*}
d\Gamma _{h} & = & dm^{2}_{(*)}dm^{2}_{*}\frac{g^{6}m_{W}^{2}\sqrt{\lambda (m^{2}_{(*)},m^{2}_{*},m^{2}_{h})}}{16(4\pi )^{8}m_{h}}\cdot \\
 &  & \left| \frac{{\cal F}_{1}(\theta ^{(*)}_{\textrm{f}_{1}},\theta ^{*}_{\textrm{f}_{2}})}{(m^{2}_{(*)}-m^{2}_{W}+im_{W}\Gamma _{W})(m^{2}_{*}-m^{2}_{W}+im_{W}\Gamma _{W})}\right| ^{2}{\cal F}_{2}(\textrm{f}_{1}){\cal F}_{2}(\textrm{f}_{2}),
\end{eqnarray*}
where \( m_{(*)} \) is the mass of the W boson closer to its mass
shell, and \( m_{*} \) that of the lighter W. The angles \( \theta ^{(*)}_{\textrm{f}_{1}} \)
and \( \theta ^{*}_{\textrm{f}_{2}} \) are measured in the rest frame
of the appropriate W boson. The functions are defined as \begin{eqnarray*}
{\cal F}_{1} & = & 2m_{(*)}m_{*}(1-\cos \theta ^{(*)}_{\textrm{f}_{1}}\cos \theta ^{*}_{\textrm{f}_{2}})+(m^{2}_{h}-m^{2}_{(*)}-m^{2}_{*})\sin \theta ^{(*)}_{\textrm{f}_{1}}\sin \theta ^{*}_{\textrm{f}_{2}}\\
{\cal F}_{2}(\textrm{f}_{\textrm{j}}) & = & \left\{ \begin{array}{ll}
1 & ,\textrm{ for leptons}\\
3\left| V_{\textrm{f}_{\textrm{j}}\bar{\textrm{f}}'_{\textrm{j}}}\right| ^{2} & ,\textrm{ for quarks}
\end{array}\right. \\
\lambda (a,b,c) & = & \left( 1-\frac{a}{c}-\frac{b}{c}\right) ^{2}-\frac{4ab}{c^{2}}.
\end{eqnarray*}
Examining the denominator of the differential width, it is clear that
the width is maximized for \( m_{(*)}\approx \mW  \) and \( m_{*}\approx \sqrt{s}-\mZ -\mW  \)
. This effect can be seen clearly by plotting the invariant masses
for the W's produced by the Pythia MC generator in Figure \ref{fig:genmass}.
Thus, at LEP fermiophobic Higgs decays should have one vector boson
near its mass-shell and the other far off it. This feature strongly
influenced the design of analyses intended to search out the Higgs
in this channel.

\begin{figure}
{\centering \resizebox*{0.55\textwidth}{!}{\includegraphics{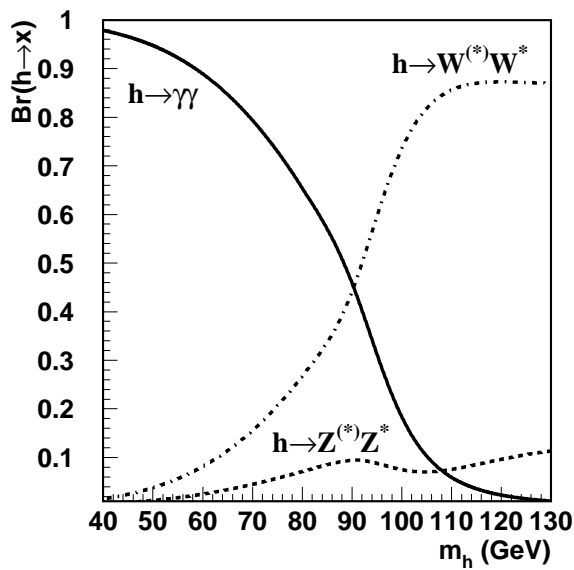}} \par}

\caption{Branching fractions of the benchmark fermiophobic model.\label{fig:phobicbr}}
\end{figure}

The relative rates of the \( \textrm{h}\rightarrow \gamma \gamma  \)
channel and the \( \textrm{h}\rightarrow \textrm{WW}/\textrm{ZZ} \)
channels within a Higgs model depend on the details of the model.
The partial widths of \( \textrm{h}\rightarrow \textrm{WW} \) and
\( \textrm{h}\rightarrow \textrm{ZZ} \) are dominated by direct coupling
terms which are strongly constrained by the Higgs's role in generating
the masses of the W and Z bosons. We may thus assume the rates of
\( \textrm{h}\rightarrow \textrm{ZZ} \) and \( \textrm{h}\rightarrow \textrm{WW} \)
to be in constant proportion. Conversely, the \( \textrm{h}\rightarrow \gamma \gamma  \)
decay is entirely dependent on loops, which makes it more sensitive
to the details of the theory and thus more model-dependent. As a baseline
of comparison, the LEP Higgs Working Group settled on a benchmark
model which produces the branching ratios plotted in Figure \ref{fig:phobicbr}.
All results use these branching ratios unless otherwise stated.

\section{Indirect Measurements of the Higgs Mass}

\begin{figure}
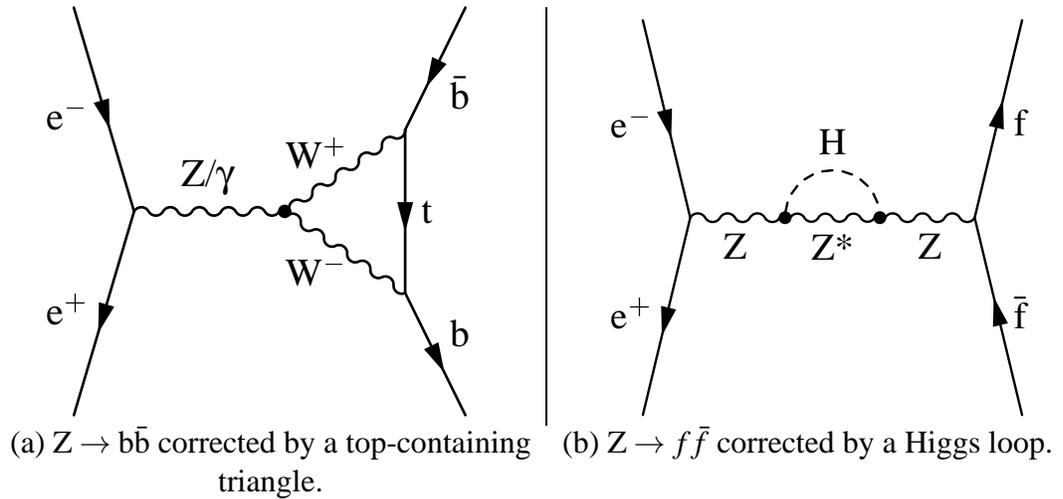

{\centering \begin{tabular}{c|c}
\resizebox*{0.4\textwidth}{!}{\includegraphics{topbb.epsi}} \hspace{6pt} &
\hspace{6pt} \resizebox*{0.4\textwidth}{!}{\includegraphics{electroweak2.epsi}} \\
\multicolumn{1}{c}{(a) \( \textrm{Z}\rightarrow \textrm{b}\bar{\textrm{b}} \) corrected
by a top-containing }&
\multicolumn{1}{c}{(b) \( \textrm{Z}\rightarrow f\bar{f} \) corrected by a Higgs loop.}\\
\multicolumn{1}{c}{triangle.}&
\multicolumn{1}{c}{}\\
\end{tabular}\par}

\caption{Example corrections to basic LEP I electroweak diagrams from top
and Higgs loops.\label{fig:electroweak}}
\end{figure}

Although the Higgs has not been observed, its presence can potentially
be deduced by a careful study of standard electroweak processes. The
Standard Model predicts higher-order corrections to many processes
which are sensitive to the masses of the bosons and the heavier quarks.
For example, the major process studied at LEP for the first five years
was \( \eplusminus \rightarrow f\bar{f} \) via the Z resonance, including
\( \eplusminus \rightarrow \textrm{Z}\rightarrow \textrm{b}\bar{\textrm{b}} \).
This process has significant corrections from top quarks, including
the diagram in Figure \ref{fig:electroweak}a. The presence of the
top quark in these loop diagrams allowed the LEP Electroweak Working
Group to predict \( m_{{\scriptsize \textrm{t}}}=173^{+12\: +18}_{-13\: -20}\textrm{ GeV} \)
in 1994 \cite{hep:electroweak94}. The CDF and D0 collaborations published
the first direct observation of the top quark in 1995 with the mass
values of \( m_{{\scriptsize \textrm{t}}}=176\pm 8(\textrm{stat}.)\pm 10(\textrm{syst}.)\textrm{ GeV} \)
\cite{hep:cdf_top_mass} and \( m_{{\scriptsize \textrm{t}}}=199_{-21}^{+19}(\textrm{stat}.)\pm 22(\textrm{syst}.)\textrm{ GeV} \)
\cite{hep:d0_top_mass} respectively. The agreement between the indirect
prediction and the observation is quite remarkable.

The electroweak observables are sensitive to \( m_{{\scriptsize \textrm{t}}} \)
to order \( m^{2}_{{\scriptsize \textrm{t}}} \) , but they are also
sensitive to \( \log \mh  \), which allows indirect limits on \mh\
to be set with sufficient data. This dependency arises from diagrams
such as \ref{fig:electroweak}b, which is essentially the virtual
form of the Higgsstrahlung diagram. The success of the LEP electroweak
fits encouraged the combination of the original LEP I results with
data from Stanford Linear Detector (SLD), W mass measurements from
the Tevatron and LEP II, and even results from neutrino-nucleon scattering
and atomic parity violation in Cesium atoms. When the data from all
these sources are combined, the electroweak fit establishes a favored
region for the Standard Model Higgs. With enough independent data,
the electroweak fit becomes a strong test of the internal consistency
of the Standard Model.

\begin{figure}
{\centering \begin{tabular}{lc}
\resizebox*{0.49\textwidth}{!}{\includegraphics{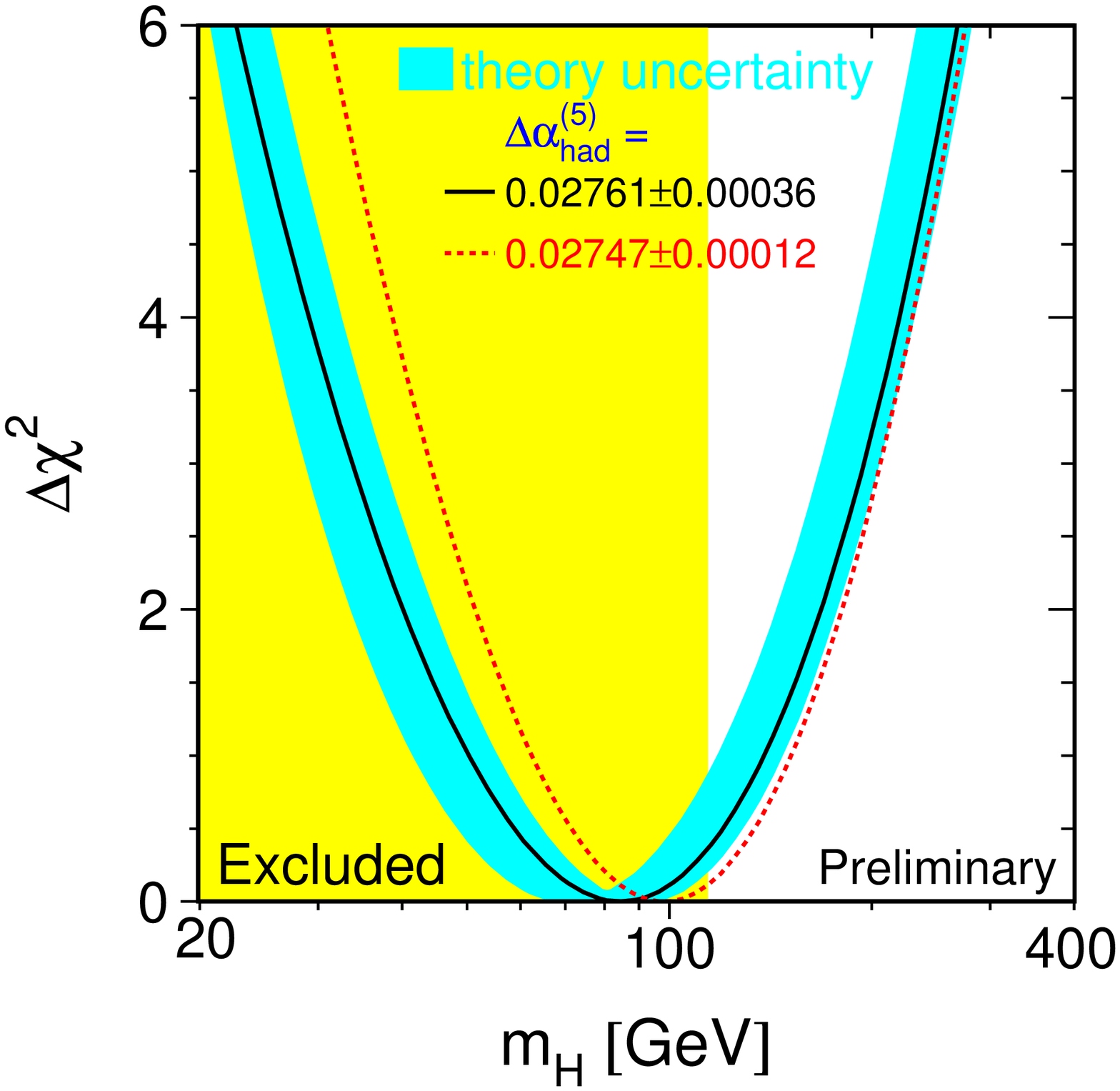}} &
\hspace{-6pt}\resizebox*{0.49\textwidth}{!}{\includegraphics{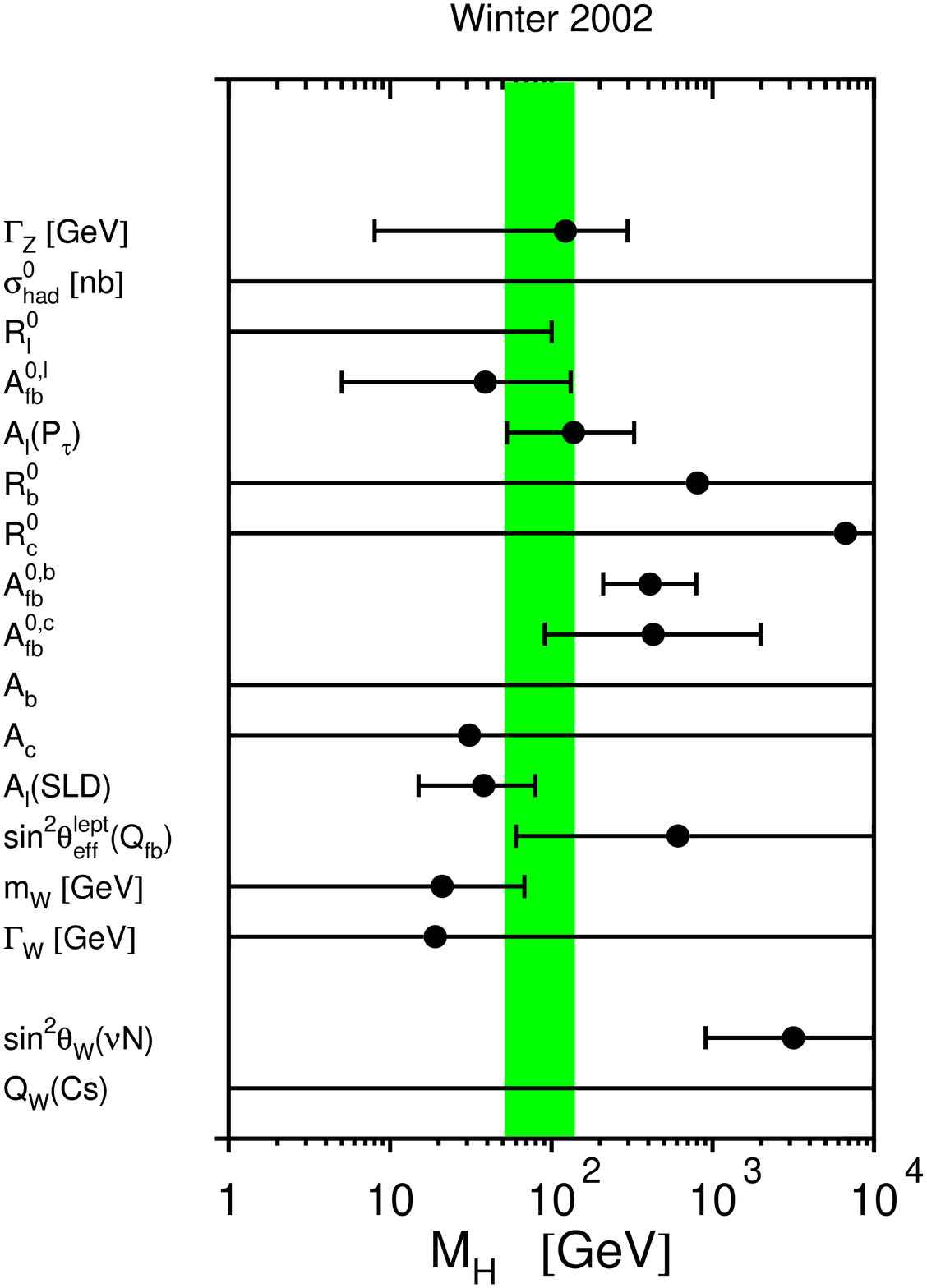}} \\
\end{tabular}\par}

\caption{Results of the most recent electroweak fit for the Higgs mass. \label{fig:ew_results}}

Several of the results used in this fit are preliminary and the fit
itself should also be considered preliminary at this time \cite{hep:electroweak02}.
\end{figure}

The most recent results of the electroweak fit for the Higgs mass
are given in Figure \ref{fig:ew_results} \cite{hep:electroweak02}. 
Figure \ref{fig:ew_results}a shows the result of the Standard Model
fit for the Higgs combining all the observations. The fit favors a
low mass for the Higgs, around 85 or 95 GeV depending on the value
of \( \Delta \alpha ^{(5)}_{{\scriptsize \textrm{had}}} \) chosen.
The presence of the Higgs has been excluded by the direct Standard
Model search up to \( \approx 113 \) GeV. The results of the fit
suggest that the Higgs might be within the reach of the LEP experimental
data. 

Figure \ref{fig:ew_results}b gives the favored Higgs mass region
for particular sets of the input data to the fit. Many of the input
parameters do not have sufficient sensitivity to the Higgs mass to
set useful limits, and several of the sensitive parameters are not
in agreement. The left-right asymmetry measured at the SLD and the
W mass measurement favor a low Higgs mass (around 40 GeV) while the
forward-backward asymmetry measured at LEP and other measurements
favor a much heavier Higgs mass above 200 GeV. The overall \( \chi ^{2} \)
of the electroweak fit fairly poor at 29 for 15 degrees of freedom.
The electroweak fit indicates that all is not well with the Standard
Model and suggests that the problem may be in the Higgs sector. Unfortunately,
the electroweak fit is sufficiently complicated that no work has been
done to determine if a Type I 2HDM might better fit the electroweak
data. A direct search is certainly worthwhile given the indications
of the electroweak data.

\chapter{The Search Process for \protect\( \textrm{h}\rightarrow \textrm{V}^{(*)}\textrm{V}^{*}\protect \)}

\textit{Algebra is a jolly game; we go searching for} \textit{\emph{x}}\textit{,
only we don't know what it is...}

\begin{quotation}
Hermann Einstein
\end{quotation}
Prior to 1999, neither the Standard Model nor the benchmark fermiophobic
model predicted any success in a search for a Higgs decaying to massive
boson pairs. The maximum energy achieved by the LEP accelerator was
189 GeV, implying a Higgsstrahlung reach to 99 GeV, a mass too low
for a significant rate in most models. However, the significant beam
energy increases of late 1999 and 2000 extended the search range above
110 GeV : the point in the Standard Model where \( \textrm{BR}(\textrm{H}\rightarrow \textrm{W}^{*}\textrm{W}^{*}) \)
becomes larger than \( \textrm{BR}(\textrm{H}\rightarrow \tau ^{+}\tau ^{-}) \),
and thus becomes the second-largest decay. At such masses, the decay
to massive boson pairs also dominates the benchmark fermiophobic model.

Any Higgs decay to massive boson pairs within the reach of the Higgsstrahlung
process at LEP necessarily involved a virtual boson, which incurs
a penalty in the rate. The W is lighter than the Z, so it naturally
dominates the branching fraction for the LEP mass range. In fact,
there is an additional exchange term arising from the distinguishablity
of \( \textrm{W}^{+} \)and \( \textrm{W}^{-} \) versus Z and Z,
so the \( \textrm{h}\rightarrow \textrm{WW} \) is expected to dominate
for all Higgs masses. Accordingly, we focused on \( \textrm{h}\rightarrow \textrm{WW} \),
but supplemented the search with \( \textrm{h}\rightarrow \textrm{ZZ} \)
where possible.

In the \( \textrm{e}^{+}\textrm{e}^{-}\rightarrow \textrm{Zh}\rightarrow \textrm{f}\bar{\textrm{f}}\textrm{W}^{(*)}\textrm{W}^{*} \)
search, there are nine different channels, defined by the decays of
the Z and the two W bosons. These nine channels are listed in Table
\ref{tab:subchan}, along with their theoretical branching fractions. 
\begin{table}
{\centering \begin{tabular}{|c|cc|cc|cc|cc|}
\cline{4-9} 
\multicolumn{3}{c|}{}&
\multicolumn{6}{c|}{\( \textrm{WW}\rightarrow  \)}\\
\cline{4-4} \cline{5-5} \cline{6-6} \cline{7-7} \cline{8-8} \cline{9-9} 
\multicolumn{3}{c|}{}&
qqqq&
(47\%)&
qql\( \nu  \)&
(43\%)&
l\( \nu  \)l\( \nu  \)&
(10\%)\\
\cline{4-4} \cline{5-5} \cline{6-6} \cline{7-7} \cline{8-8} \cline{9-9} 
\hline 
\multicolumn{1}{|c|}{}&
qq&
(70\%)&
qqqqqq&
(32.8\%)&
qqqql\( \nu  \)&
(30.2\%)&
qq\( \textrm{l}\nu \textrm{l}\nu  \)&
(7.0\%)\\
\cline{2-2} \cline{3-3} \cline{4-4} \cline{5-5} \cline{6-6} \cline{7-7} \cline{8-8} \cline{9-9} 
\multicolumn{1}{|c|}{\( \textrm{Z}\rightarrow  \)}&
\( \nu \nu  \)&
(20\%)&
\( \nu \nu \textrm{qqqq} \)&
(9.4\%)&
\( \nu \nu \textrm{qql}\nu  \)&
(8.7\%)&
\( \nu \nu \textrm{l}\nu \textrm{l}\nu  \)&
(2.0\%)\\
\cline{2-2} \cline{3-3} \cline{4-4} \cline{5-5} \cline{6-6} \cline{7-7} \cline{8-8} \cline{9-9} 
\multicolumn{1}{|c|}{}&
ll&
(10\%)&
llqqqq&
(4.7\%)&
llqql\( \nu  \)&
(4.4\%)&
lll\( \nu  \)l\( \nu  \)&
(1.0\%)\\
\hline
\end{tabular}\par}

\caption{Channels of the \protect\( \textrm{h}\rightarrow \textrm{WW}\protect \)
search and theoretical branching fractions.\label{tab:subchan}}

Branching fractions are calculated using the most recent results from
the Particle Data Group \cite{hep:pdb2000}.
\end{table}
 We constructed channel names by first listing the decay products
of the Z boson and then listing the four decay products from the two
W bosons. For example, in the \( \textrm{qqqql}\nu  \) channel, the
Z decays to \( \textrm{q}\bar{\textrm{q}} \), one W decays to \( \textrm{q}\bar{\textrm{q}}' \)
and the other W decays to \( \textrm{l}\nu  \).

The potential search significance of a channel depends on both the
expected background level after selection and the intrinsic physical
branching fraction of the channel. To estimate the number of expected
signal events, assume that all 217 \( \textrm{pb}^{-1} \) of 2000
data was taken at \( \sqrt{s}=206\textrm{ GeV} \), where the \( \textrm{e}^{+}\textrm{e}^{-}\rightarrow \textrm{Zh} \)
cross-section for \mh=110 GeV is 0.15 pb and the benchmark branching
ratio \( \textrm{h}\rightarrow \textrm{WW} \) is 86\%. The total
number of events expected would be 28 events in all channels, assuming
100\% efficiency. Applying the branching ratios yields an expectation
of less than one event in the \( \nu \nu \textrm{l}\nu \textrm{l}\nu  \)
and \( \textrm{lll}\nu \textrm{l}\nu  \) channels; these channels
are likely to be unimportant for the search. On the other hand, a
qqqqqq analysis would provide \textasciitilde{}9 events, which could
be a significant number depending on the signal-background separation
which is possible in the analysis.

\begin{figure}
{\centering \resizebox*{0.9\textwidth}{!}{\includegraphics{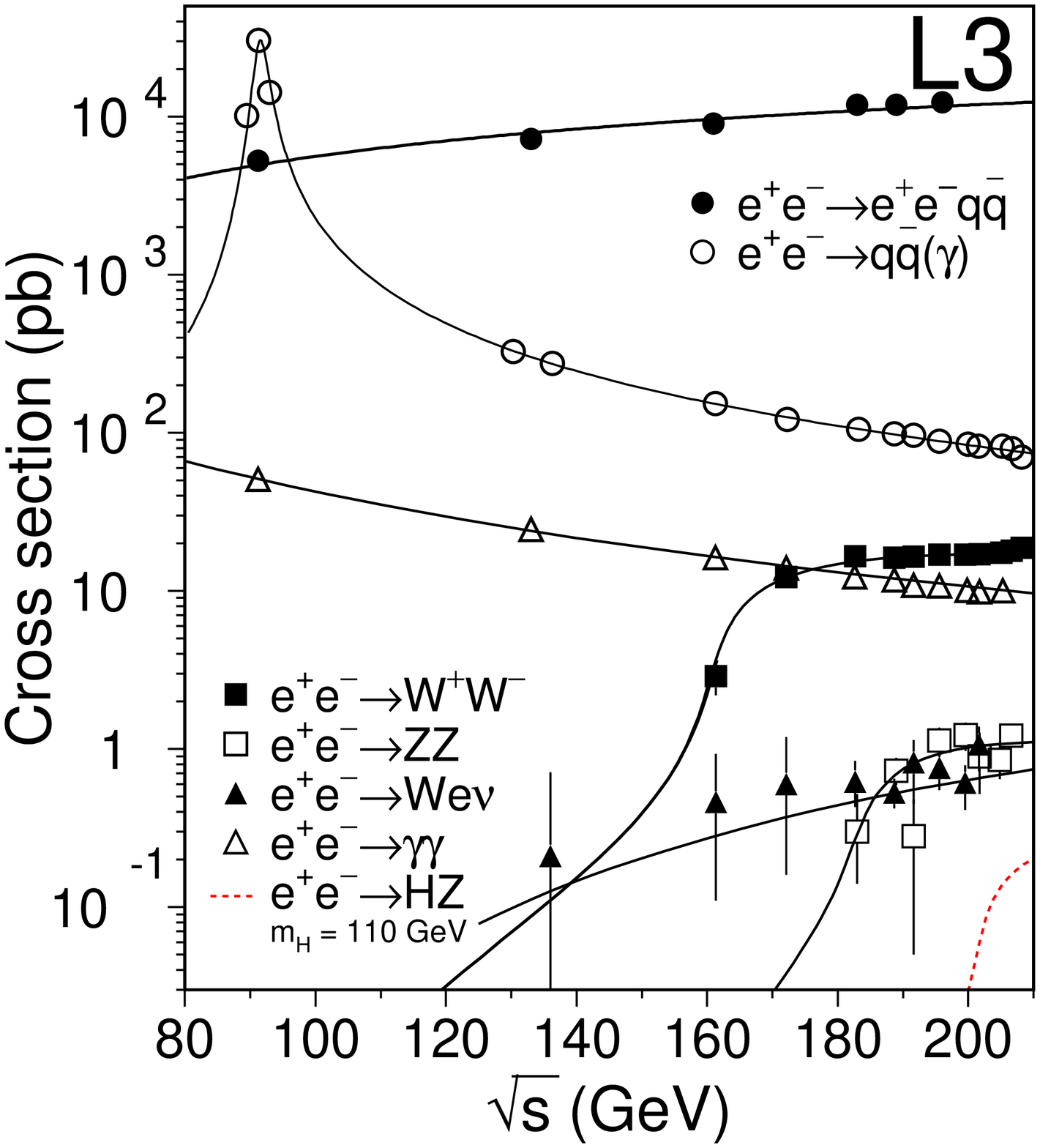}} \par}

\caption{Cross-sections for Higgs production and background processes.\label{fig:xsecs}}

The data points are background cross-section measurements from L3.
\end{figure}

Of the nine channels, we have analyzed six of them: qqqqqq, qqqql\( \nu  \),
\( \nu \nu  \)qqqq, \( \nu \nu \textrm{qql}\nu  \), llqqqq and qql\( \nu  \)l\( \nu  \).
The analyses of each channel follow the same pattern:

\begin{enumerate}
\item A set of preselection cuts removed the {}``obvious'' background
events. These preselection cuts remove classes of events which appear
in the data, but are not well covered by Monte Carlo. These include
cosmic muon events, beam-gas events, and some types of two-photon
events.
\item More difficult backgrounds were removed at the final selection step
using one or more neural networks. All the analyses used a common
class of neural network techniques described in Appendix \ref{sec:neuralnets}.
These networks were trained to produce an output of zero for background
events and one for signal events.
\item Final distributions of the selected signal, background, and data were
produced, generally using a discriminant combination of the neural
networks and a reconstructed Higgs mass as described in Appendix \ref{sec:discriminant}.
\end{enumerate}
How much background needed to be removed? Figure \ref{fig:xsecs}
is a plot of cross-section measurements made at L3 of different Standard
Model processes as a function of center-of-mass energy. Near \( \sqrt{s}=91\textrm{ GeV} \),
the Z pole is clearly visible in the \( \textrm{e}^{+}\textrm{e}^{-}\rightarrow \textrm{q}\bar{\textrm{q}}(\gamma ) \)
cross-section. At 160 GeV, W pair production crosses threshold, and
Z pair production turns on around 183 GeV. Down in the lower right
corner is the predicted cross-section for a \( \mh =110\textrm{ GeV} \)
Higgs. Thus, we attempted to detect a process which is predicted to
occur at a rate five orders of magnitude smaller than the many Standard
Model background processes, so significant data analysis efforts were
required to remove background and isolate any candidate signal events.

The rest of this chapter discusses the analyses of the various channels
in detail. The results of the search are given in Chapter \ref{chap:results}.

\newpage\begin{minipage}{0.60\linewidth}\vfill\section{The qqqqqq Channel}\end{minipage}\hfill\begin{minipage}{0.35\linewidth}

\vspace{0.3cm}
{\centering \resizebox*{0.9\textwidth}{!}{\includegraphics{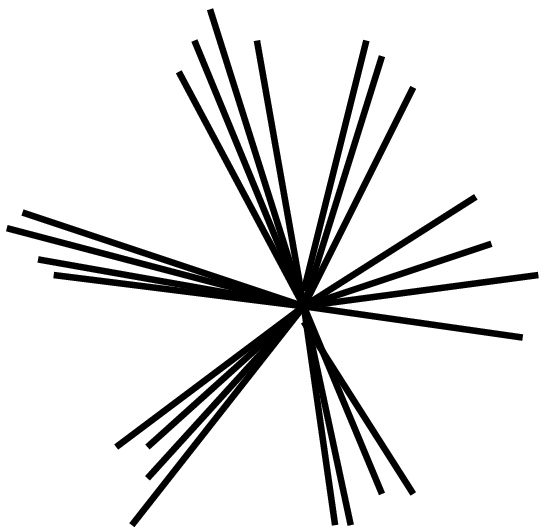}} \par}
\vspace{0.3cm}

\end{minipage}

The most spectacular events in the \( \textrm{h}\rightarrow \textrm{WW} \)
search are the qqqqqq events. In this channel, the Z and both W's
decay hadronically, so the physical signature is six jets with many
charged tracks and the full collision energy spread around the detector.
One pair of jets should have the Z mass, and another should have an
invariant mass near \mW. The last two jets should be fairly low mass
and low energy. 

The major backgrounds to this search are \( \textrm{e}^{+}\textrm{e}^{-}\rightarrow \textrm{WW}\rightarrow \textrm{qqqq} \),
\( \textrm{e}^{+}\textrm{e}^{-}\rightarrow \textrm{ZZ}\rightarrow \textrm{qqqq} \),
and \( \textrm{e}^{+}\textrm{e}^{-}\rightarrow \textrm{q}\bar{\textrm{q}}(\gamma ) \)
processes. These processes are backgrounds even though at first glance
there should be only two or four jets in the event. Any of the quarks
in these events may radiate one or more hard gluons, which will hadronize
into another jet. This jet will typically have less energy and fewer
charged tracks than a primary quark jet. Gluon jets may easily mimic
the two weak jets expected from the \( \textrm{W}^{*} \) decay. The
inner tracking volume of the L3 detector is quite small, so fluctuations
in a single jet can be hard to distinguish from two separate jets.
The goal of the analysis is to remove the events with poorly reconstructed
jets or radiated gluons.

At preselection,we required \( 0.85<\frac{\Evis }{\sqrt{s}}<1.15 \)
to eliminate two photon processes and other low energy background
processes. We selected hadronic events by requiring at least 30 calorimetric
clusters and 30 tracks, as well as \( \Ebgo >70\textrm{ GeV} \) and
\( \Ehcal >25\textrm{ GeV} \). We reduced the contamination from
\( \textrm{e}^{+}\textrm{e}^{-}\rightarrow \textrm{q}\bar{\textrm{q}}(\gamma ) \)
by requiring the event thrust to be less than 0.9. The thrust variable
measures the extent to which all the particles point along a single
direction, as would be the case for \( \textrm{q}\bar{\textrm{q}}(\gamma ) \).
Finally, we forced the event to six jets using the Durham algorithm\cite{algo:durham}
and required each of the six resulting jets to contain at least one
charged track. The Durham algorithm is a jet-building algorithm which
iteratively combines the two calorimeter clusters with the smallest
\( Y(i,j)\equiv 2E_{i}E_{j}\frac{1-\cos \theta _{ij}}{s} \) to produce
proto-jets. As the number of proto-jets decreases, the \( Y \) value
for combining the remaining proto-jets rises. We required events to
have a minimum \( Y \) value for combining the six jets to produce
five using the requirement \( \textrm{log}_{10}Y_{56}>-4.1 \).

After preselection, we applied a constrained fit to the six jets requiring
momentum and energy balance. The details of the constrained fit algorithm
are available in Appendix C of \cite{l3:constrainedfit}. Next, we
chose the pair of jets with invariant mass closest to \mZ\ after the
fit. This pair was the Higgsstrahlung Z candidate, and the four remaining
jets became the Higgs candidate. Of the remaining four jets, we made
the pair with invariant mass closest to \mW\ the \( \textrm{W}^{(*)} \)
candidate. For selection, we prepared three neural networks with the
structure of eleven inputs, twenty-five hidden nodes, and one output
node. The eleven inputs are listed in Table \ref{tab:6jetnn}, along
with descriptions of the general event features which each variable
used.
\begin{table}
\begin{tabular}{|c|p{2.1in}|p{2.5in}|}
\hline 
\emph{Variable}&
\emph{Description}&
\emph{Trend}\\
\hline
\hline 
\( \textrm{E}^{\textrm{max}}_{\textrm{jet}6} \)&
Energy of the most energetic jet from the 6 jet fit.&
\multicolumn{1}{p{2.5in}|}{}\\
\cline{1-1} \cline{2-2} 
\( \textrm{E}^{\textrm{min}}_{\textrm{jet}6} \)&
Energy of the least energetic jet from the 6 jet fit.&
\multicolumn{1}{p{2.5in}|}{\vspace{-40pt}Signal events should have six reasonably equal jets,
while many backgrounds have several high energy jets and several very
low energy gluon jets.}\\
\hline 
\( n^{\textrm{min}}_{\textrm{jet}6} \)&
Minimum number of charge tracks in any of the jets from the 6 jet
fit.&
Gluon jets and other {}``reconstruction{}`` jets will have fewer
charge tracks than signal jets.\\
\hline 
\( \theta ^{\textrm{min}}_{\textrm{jet}6} \)&
Minimum angle between any two of the six jets.&
Gluon-radiation jets will tend to have a relatively small angle with
respect to other jets.\\
\hline 
\( \log Y_{45} \)&
Durham Y value where the fit changes from four jets to five jets.&
\multicolumn{1}{m{2.5in}|}{ }\\
\cline{1-1} \cline{2-2} 
\( \log Y_{56} \)&
Durham Y value where the fit changes from five jets to six jets.&
\multicolumn{1}{p{2.5in}|}{\vspace{-0.28in}

True six jet events should have larger values of the Durham cut values.}\\
\hline 
\( m_{\textrm{eq}} \)&
Mass determined by a 5C fit assuming four jets and two equal mass
dijets.&
WW and ZZ background processes should have \( m_{\textrm{eq}}=\mW  \)
and \( m_{\textrm{eq}}=\mZ  \) respectively.\\
\hline 
\( \chi _{\textrm{WW}}^{2} \)&
\( \chi ^{2} \) of a 5C fit to \( \textrm{e}^{+}\textrm{e}^{-}\rightarrow \textrm{WW}\rightarrow \textrm{qqqq} \)&
WW background events should have a good \( \chi ^{2} \) for this
fit, while signal events should not fit as well.\\
\hline 
\( m^{4c}_{\textrm{Z}} \)&
Mass of the Z candidate from the 4C fit.&
For signal, this should be close to \mZ.\\
\hline 
\( m^{4c}_{\textrm{W}} \)&
Mass of the W candidate from the 4C fit.&
For signal, this should be close to \mW.\\
\hline 
\( \alpha _{\textrm{W}^{(*)}\textrm{W}^{*}} \)&
Angle between the decay planes of the W candidate and \( \textrm{W}^{*} \)
candidate.&
This angle is likely to be smaller for gluon jets which fake the \( \textrm{W}^{*} \).\\
\hline
\end{tabular}

\caption{Neural network variables for the qqqqqq selection networks.\label{tab:6jetnn}}
\end{table}
 We trained the three networks using the same set of Higgs signal
events but using a different type of background: either WW, ZZ, or
\( \textrm{q}\bar{\textrm{q}} \). We cut independently on all three
networks, requiring \( \Nww >0.3 \), \( \Nzz >0.3 \), and \( \Nqq >0.7 \).
Table \ref{tab:qqqqqq_numbers} gives the numbers of signal and background
events expected and data observed after preselection and selection.
\begin{table}
{\centering \begin{tabular}{|c|cc|cc|}
\hline 
&
\multicolumn{2}{c|}{1999 }&
\multicolumn{2}{c|}{2000}\\
&
Preselection&
Selection&
Preselection&
Selection\\
\hline
\hline 
WW background&
451.1&
119.3&
823.3&
228.4\\
\hline 
ZZ background&
34.9&
14.7&
69.8&
29.4\\
\hline 
\( \textrm{q}\bar{\textrm{q}}(\gamma ) \) background&
184.4&
18.9&
304.5&
35.1\\
\hline 
Total MC Background\( ^{a} \)&
671.0&
153.0&
1199.1&
293.0\\
\hline 
Data&
652&
155&
1234&
288\\
\hline 
Signal for \( m_{\textrm{h}}=110 \) GeV&
1.02&
0.94&
8.0&
7.1\\
\hline
\end{tabular}\par}

{\centering \( ^{a} \) Includes very small contributions from Zee
and \( \textrm{e}\nu \textrm{qq} \) processes.\par}

\caption{Preselection and selection totals for the qqqqqq channel.\label{tab:qqqqqq_numbers}}
\end{table}
 The final variable was a discriminant combining the three network
outputs and the reconstructed Higgs mass from the 4C fit, as described
in Appendix \ref{sec:discriminant}.

Besides production from \( \textrm{h}\rightarrow \textrm{WW} \),
the six-jet signature can also be produced by the \( \textrm{e}^{+}\textrm{e}^{-}\rightarrow \textrm{Zh}\rightarrow \textrm{ZZ}^{(*)}\textrm{Z}^{*}\rightarrow \textrm{qqqqqq} \)
process. In fact, the same analysis efficiently selects both channels,
since the mass reconstruction is sufficiently broad to accept either
W or Z di-jet pairs. Therefore we included the \( \textrm{h}\rightarrow \textrm{ZZ} \)
signal in the analysis, which effectively added 15\% to the expected
rate relative to using only \( \textrm{h}\rightarrow \textrm{WW} \)
for the six-jet channel.

\newpage\begin{minipage}{0.60\linewidth}\vfill\section{The \ensuremath{\mathrm{qqqql}\nu} Channel}\end{minipage}\hfill\begin{minipage}{0.35\linewidth}

\vspace{0.3cm}
{\centering \resizebox*{0.9\textwidth}{!}{\includegraphics{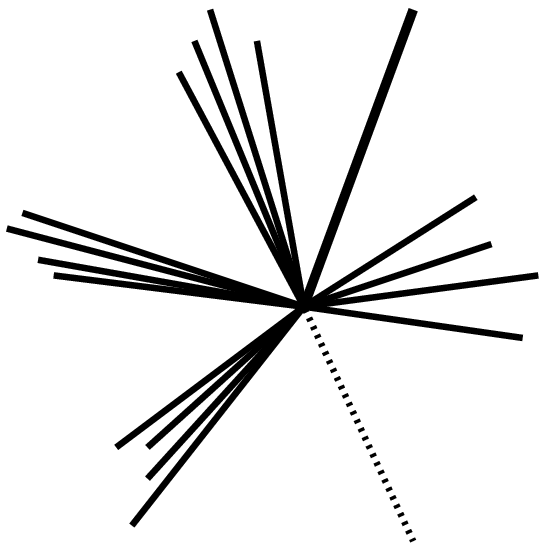}} \par}
\vspace{0.3cm}

\end{minipage}

In this channel, the Z decays hadronically, while one W decays hadronically
and the other decays leptonically. The different lepton flavors naturally
define three different subchannels: \( \textrm{qqqqe}\nu  \), \( \textrm{qqqq}\mu \nu  \),
and \( \textrm{qqqq}\tau \nu  \). Further, the difference between
leptons coming from the \( \textrm{W}^{(*)} \) and from the \( \textrm{W}^{*} \)
doubles the number of subchannels. In one set of signatures, the \( \textrm{W}^{(*)} \)
decays hadronically and the \( \textrm{W}^{*} \) decays leptonically,
which means the lepton energy is small and the neutrino energy is
also small, so the missing energy in the event should be small. In
the other set, the \( \textrm{W}^{(*)} \) decays to \( \textrm{l}\nu  \)
and the \( \textrm{W}^{*} \) decays hadronically, leading to a high-energy
lepton and a good deal of missing energy. Since the kinematics of
the two cases are quite different, we considered the \( \textrm{qqqql}\nu  \)
channel to have six subchannels. For brevity, we will refer to events
where the lepton is produced from the decay of the \( \textrm{W}^{(*)} \)
as \( (\textrm{l}\nu ) \) events and events where the lepton comes
from the \( \textrm{W}^{*} \) as \( (\textrm{l}\nu )^{*} \) events.

The major backgrounds to this channel differ somewhat depending on
subchannel. The \( (\textrm{l}\nu ) \) events have significant amount
of missing energy, so \( \textrm{e}^{+}\textrm{e}^{-}\rightarrow \textrm{qql}\nu  \)
is a major background, where gluon radiation generates the additional
two jets. The \( \textrm{qq}\mu \nu  \) and \( \textrm{qq}\tau \nu  \)
background events are primarily produced from W pairs, but \( \textrm{qqe}\nu  \)
can be produced either from W pairs or from a non-resonant exchange
process producing an electron, a neutrino, and a real W boson which
may then decay to \( \textrm{q}\bar{\textrm{q}}' \). This second
process is known as {}``single-W''. We used a four-fermion generator
named EXCALIBUR \cite{mc:excalibur} that includes both the resonant
and non-resonant diagrams to produce \( \textrm{qqe}\nu  \) events
and used the KORALW \cite{mc:koralw} generator for all other WW decays.

In the \( (\textrm{l}\nu )^{*} \) case, the lepton and neutrino energies
are small, so the major backgrounds are actually the same four-jet
and \( \textrm{q}\bar{\textrm{q}}(\gamma ) \) backgrounds as in the
six-jet case. The leptons arise from the semileptonic decay of quarks
in jets and from the misidentification of low multiplicity jets as
taus. 

\begin{figure}
{\centering \resizebox*{0.8\textwidth}{!}{\includegraphics{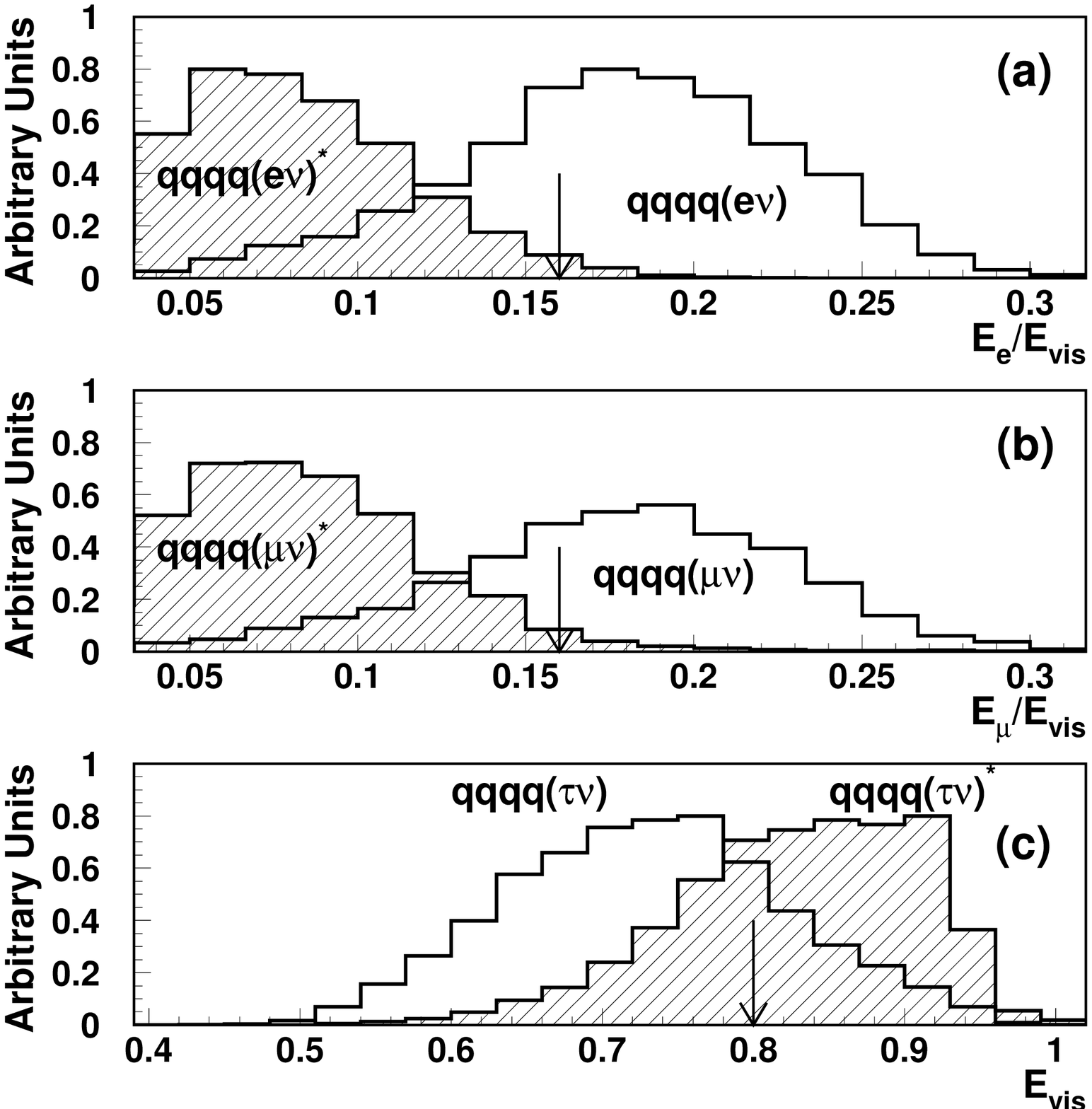}} \par}

\caption{Variables used to define subchannels in the \protect\( \textrm{qqqql}\nu \protect \)
channel.\label{fig:subchan_vars}}

The two \( \textrm{qqqqe}\nu  \) and \( \textrm{qqqq}\mu \nu  \)
subchannels are split at \( \frac{\textrm{E}_{l}}{\Evis }=0.16 \),
while the tau subchannel separation point is \( \Evis =0.8 \).
\end{figure}

We classified each event into a subchannel using the most energetic
identified lepton in the event. For the \( \textrm{qqqqe}\nu  \)
and \( \textrm{qqqq}\mu \nu  \) channels, we separated the two subchannels
using the variable \( \frac{E_{l}}{\Evis } \), as in Figures \ref{fig:subchan_vars}a
and \ref{fig:subchan_vars}b. In the \( \textrm{qqqq}\tau \nu  \)
channels, the initial lepton energy was difficult to reconstruct,
so the subchannels were separated using the visible energy, although
somewhat less efficiently as seen in Figure \ref{fig:subchan_vars}c.
 An event was only considered for identification as a \( \textrm{qqqq}\tau \nu  \)
event if it was not identified as \( \textrm{qqqqe}\nu  \) or \( \textrm{qqqq}\mu \nu  \).
We used Monte Carlo to determine the identification efficiency matrix
given in Table \ref{tab:qqqqlv_id}.
\begin{table}
{\centering \begin{tabular}{|l|c|c|c|c|c|c|c|}
\hline 
MC Type&
\( (\textrm{e}\nu ) \)&
\( (\textrm{e}\nu )^{*} \)&
\( (\mu \nu ) \)&
\( (\mu \nu )^{*} \)&
\( (\tau \nu ) \)&
\( (\tau \nu )^{*} \)&
Total\\
\hline
\hline 
\( \textrm{qqqq}(\textrm{e}\nu ) \)&
45.9&
4.5&
0.9&
1.3&
8.6&
4.0&
65.2\\
\hline 
\( \textrm{qqqq}(\textrm{e}\nu )^{*} \)&
2.9&
42.5&
1.1&
1.8&
1.1&
4.8&
54.2\\
\hline 
\( \textrm{qqqq}(\mu \nu ) \)&
0.2&
0.8&
40.7&
5.7&
12.7&
2.2&
62.3\\
\hline 
\( \textrm{qqqq}(\mu \nu )^{*} \)&
0.2&
1.0&
6.1&
42.8&
2.9&
6.5&
59.5\\
\hline 
\( \textrm{qqqq}(\tau \nu ) \)&
3.7&
7.9&
4.5&
7.6&
29.6&
5.4&
58.7\\
\hline 
\( \textrm{qqqq}(\tau \nu )^{*} \)&
0.6&
5.8&
3.5&
6.9&
7.9&
18.5&
43.2\\
\hline
\end{tabular}\par}

\caption{Identification matrix for \protect\( \textrm{qqqql}\nu \protect \).\label{tab:qqqqlv_id}}

Each row shows the percentages of signal events generated in a specific
subchannel which are identified in each subchannel. Since there is
a finite efficiency for lepton identification, not all events can
be identified since some have no identified lepton.
\end{table}
 To pass preselection, an event must have been identified into one
and only one subchannel. 

\begin{figure}
{\centering \begin{tabular}{cc}
\resizebox*{0.45\textwidth}{!}{\includegraphics{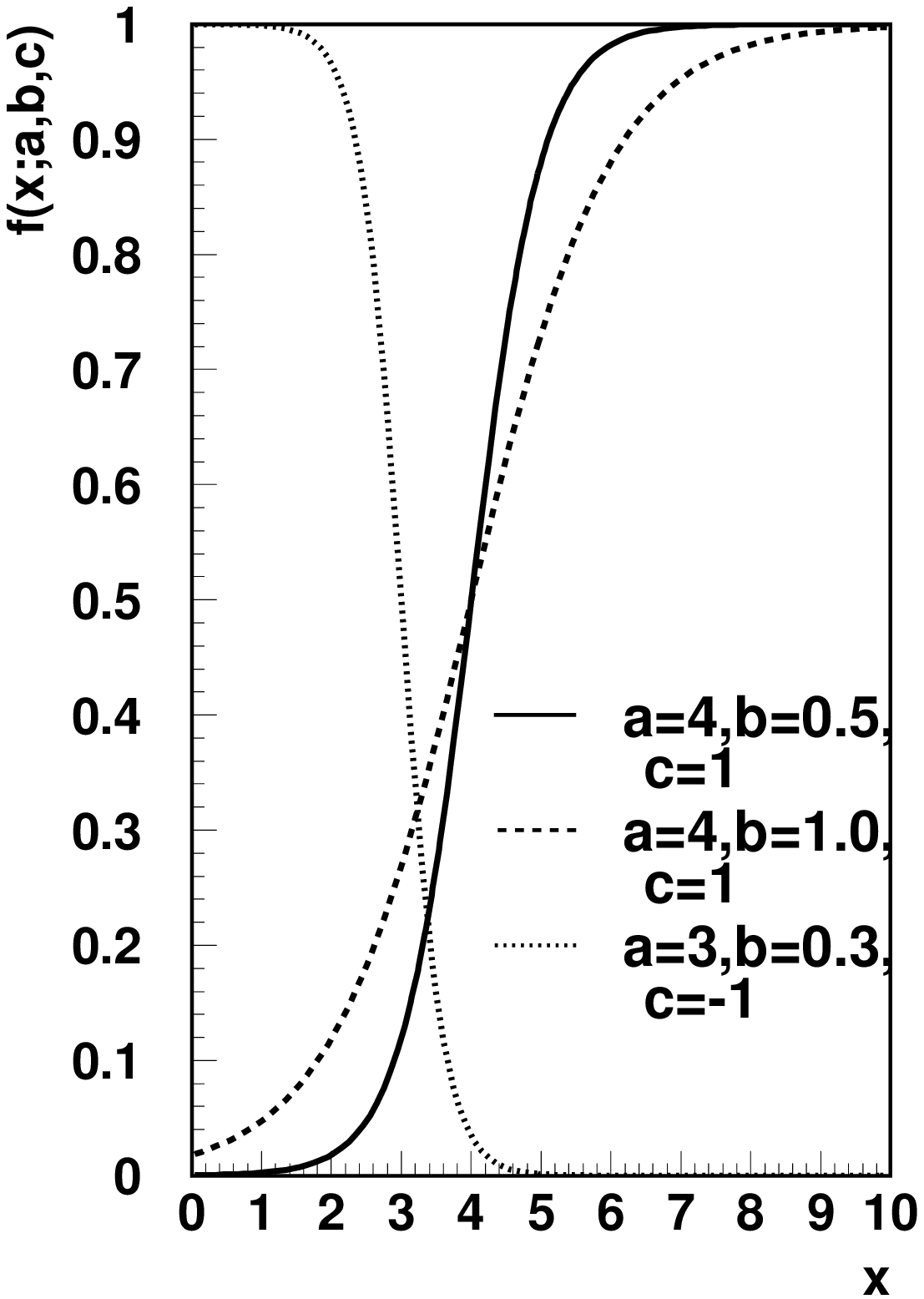}} &
\resizebox*{0.45\textwidth}{!}{\includegraphics{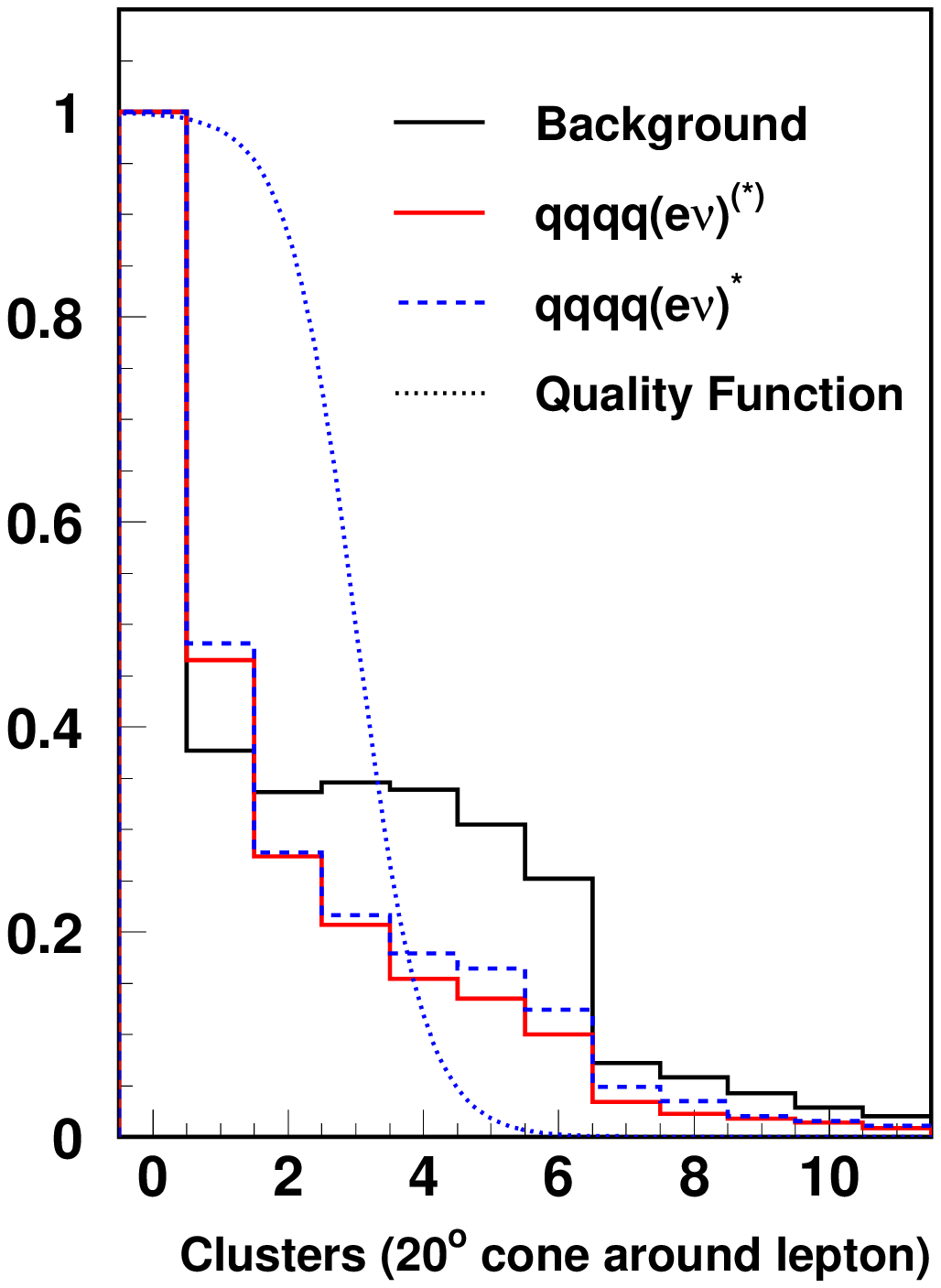}} \\
(a)&
(b)\\
\end{tabular}\par}

\caption{Sigmoid function used for quality calculations and an example quality
variable.\label{fig:sigmoid}}

The sigmoid is parameterized as \( f(x;a,b,c)=\frac{1}{1+e^{\frac{-c(x-a)}{b}}} \).
The parameter \( a \) sets the point at which \( f=\frac{1}{2} \),
\( b \) sets the width of the transition from 0 to 1, and \( c=\pm 1 \)
determines whether the \( f\rightarrow 0 \) or \( f\rightarrow 1 \)
as \( x\rightarrow \infty  \). 

(a) Sigmoid function for several choices of \( a \), \( b \), and
\( c \).

(b) Example quality variable from the \( \textrm{qqqqe}\nu  \) subchannels.
Good isolated electrons should have small numbers of calorimeter clusters
around them, while electrons from semileptonic decays will be close
to jets and their calorimeter clusters.
\end{figure}

The candidate lepton also had to pass certain {}``quality'' requirements.
The purpose of these quality requirements was to remove leptons produced
by semileptonic decay of quarks in jets. To improve the smoothness
of the systematic error calculation, we did not apply hard cuts on
the lepton quality variables. Instead, we fit a sigmoid function by
hand for each variable, as shown in Figure \ref{fig:sigmoid}.
 The parameters were chosen so that \( f_{i}(x;a,b,c) \) is near
one for good signal leptons and drops to zero for more poorly reconstructed
leptons. We multiplied the results for the different variables and
placed a cut at 0.1 on the product.

Besides the subchannel identification and quality requirements, we
applied additional preselection cuts to remove other obvious backgrounds.
To preselect hadronic events, we required 30 calorimetric clusters,
20 charged tracks, 40 GeV of energy in the BGO, and 10 GeV of energy
in the HCAL. To select against single-W and \( \textrm{q}\bar{\textrm{q}}(\gamma ) \)
backgrounds, we required the event thrust to be less than 0.9, the
fraction of visible energy in a \( 30^{\circ } \) cone around the
beampipe to be less than 60\%, and \( \left| \cos \theta _{{\scriptsize \textrm{missing}}}\right| <0.92 \).
We also required the event to contain no photons of greater than 20
GeV. Another major source of background for this channel is \( \textrm{e}^{+}\textrm{e}^{-}\rightarrow \textrm{WW}\rightarrow \textrm{qql}\nu  \),
so we fit the event to two jets after excluding the lepton identified
above and required that the dijet mass be greater than 90 GeV.

After preselection, the major remaining background was \( \textrm{e}^{+}\textrm{e}^{-}\rightarrow \textrm{WW} \),
particularly \( \textrm{WW}\rightarrow \textrm{qqqq} \) events where
one of the quarks decays semileptonically. For each subchannel, we
prepared one network with the ten input variables listed in Table
\ref{tab:qqqqlv_nets}, twenty hidden nodes, and one output node to
remove the WW and \( \textrm{e}\nu \textrm{qq} \) backgrounds.
\begin{table}
\begin{tabular}{|c|p{2.1in}|p{2.5in}|}
\hline 
\emph{Variable}&
\emph{Description}&
\emph{Trend}\\
\hline
\hline 
\( \chi _{\textrm{WW}}^{2} \)&
\( \chi ^{2} \) of a 5C fit to \( \textrm{e}^{+}\textrm{e}^{-}\rightarrow \textrm{WW}\rightarrow \textrm{qqqq} \)&
WW background events should have a good \( \chi ^{2} \) for this
fit, while signal events should not fit as well.\\
\hline 
\( \textrm{E}^{\textrm{max}}_{4\textrm{j}-1\textrm{l}} \)&
Energy of the most energetic jet from a fit to four jets, having removed
the candidate lepton.&
\\
\cline{1-1} \cline{2-2} 
\( \textrm{E}^{\textrm{min}}_{4\textrm{j}-1\textrm{l}} \)&
Energy of the least energetic jet from a fit to four jets, having
removed the candidate lepton.&
\multicolumn{1}{p{2.5in}|}{\vspace{-0.75in}\( (\textrm{l}\nu )^{*} \) signal events should have
four reasonably equal jets, while many backgrounds have several high
energy jets and several very low energy gluon jets. \( (\textrm{l}\nu )^{(*)} \)
signal events will look more like background events.}\\
\hline 
\( \theta ^{\textrm{min}}_{4\textrm{j}-1\textrm{l}} \)&
Minimum angle between any two of the four jets.&
Gluon jets tend to be emitted at small angles relative to the emitting
quark jet.\\
\hline 
\( m^{4c}_{\textrm{Z}} \)&
Mass of the dijet pair after the 4C fit with mass closest to \mZ.&
For signal events, this should be close to \mZ. \\
\hline 
\( m^{4c}_{\textrm{L}} \)&
Mass of the lepton-neutrino system after the 4C fit.&
For \( (\textrm{l}\nu )^{*} \), this should be small, while for \( (\textrm{l}\nu )^{(*)} \)
it should be close to \mW.\\
\hline 
\( m^{4c}_{\textrm{Q}} \)&
Mass of the other two jets after the 4C fit.&
For \( (\textrm{l}\nu )^{*} \), this should be close to \mW, while
for \( (\textrm{l}\nu )^{(*)} \) it should be small.\\
\hline 
\( p^{4c}_{\textrm{L}} \)&
Momentum of the lepton-neutrino system after the 4C fit.&
\multicolumn{1}{b{2.5in}|}{}\\
\cline{1-1} \cline{2-2} 
\( p^{4c}_{\textrm{Q}} \)&
Momentum of the two jet system after the 4C fit.&
\multicolumn{1}{p{2.5in}|}{\vspace{-0.46in}For signal events, the momentum of the decay pairs
should be small and equal.}\\
\hline 
\( \log Y_{34} \)&
Durham Y value where the fit changes from three jets to four.&
Events with gluon jets will tend to have smaller Y values.\\
\hline
\end{tabular}

\caption{Neural network variables for the \protect\( \textrm{qqqql}\nu \protect \)
selection networks.\label{tab:qqqqlv_nets}}
\end{table}
 For the electron and muon subchannels, we cut on the network output
at 0.5, while we cut the tau subchannels at 0.3. The numbers of events
expected and observed in this channel are listed in Table \ref{tab:qqqqlv_numbers},
broken down by subchannel. We produced final discriminant distributions
using the output of the neural network and the reconstructed mass
separately for each subchannel.
\begin{table}
\begin{sideways}
\begin{tabular}{|l|cc|cc||cc|cc|}
\hline 
&
\multicolumn{2}{c|}{1999}&
\multicolumn{2}{c||}{2000}&
\multicolumn{2}{c|}{1999}&
\multicolumn{2}{c|}{2000}\\
&
{\footnotesize Preselection}&
{\footnotesize Selection}&
{\footnotesize Preselection}&
{\footnotesize Selection}&
{\footnotesize Preselection}&
{\footnotesize Selection}&
{\footnotesize Preselection}&
{\footnotesize Selection}\\
\hline
&
\multicolumn{4}{c||}{\( \textrm{qqqqe}\nu  \)}&
\multicolumn{4}{c|}{\( \textrm{qqqq}(\textrm{e}\nu )^{*} \)}\\
\hline
WW background&
0.41&
0.18&
0.75&
0.27&
1.10&
0.37&
1.82&
0.67\\
\hline 
ZZ background&
0.47&
0.33&
0.99&
0.64&
0.39&
0.20&
0.49&
0.25\\
\hline 
\( \textrm{q}\bar{\textrm{q}} \) background&
0.34&
0.17&
0.66&
0.39&
0.78&
0.20&
1.03&
0.38\\
\hline 
\( \textrm{e}\nu \textrm{qq} \) background&
0.68&
0.21&
1.49&
0.43&
0&
0&
0.01&
0\\
\hline 
Zee background&
0.22&
0.11&
0.56&
0.29&
0.11&
0.07&
0.09&
0.03\\
\hline 
Total MC Bkgd&
2.13&
1.01&
4.47&
2.04&
2.39&
0.85&
3.46&
1.35\\
\hline 
Data&
4&
1&
4&
3&
2&
1&
3&
1\\
\hline 
{\footnotesize Signal for \( m_{\textrm{h}}=105 \) GeV}&
0.31&
0.26&
0.91&
0.80&
0.27&
0.23&
0.77&
0.69\\
\hline 
&
\multicolumn{4}{c||}{\( \textrm{qqqq}\mu \nu  \)}&
\multicolumn{4}{c|}{\( \textrm{qqqq}(\mu \nu )^{*} \)}\\
\hline 
WW background&
0.77&
0.32&
1.42&
0.53&
0.88&
0.29&
1.88&
0.51\\
\hline 
ZZ background&
0.31&
0.16&
0.63&
0.39&
0.17&
0.07&
0.32&
0.19\\
\hline 
\( \textrm{q}\bar{\textrm{q}} \) background&
0.06&
0.06&
0.13&
0.1&
0.17&
0.06&
0.49&
0.24\\
\hline 
Total MC Bkgd&
1.16&
0.55&
2.18&
1.03&
1.24&
0.44&
2.73&
0.96\\
\hline 
Data&
1&
0&
1&
0&
0&
0&
1&
1\\
\hline 
{\footnotesize Signal for \( m_{\textrm{h}}=105 \) GeV}&
0.19&
0.16&
0.60&
0.53&
0.24&
0.20&
0.64&
0.56\\
\hline 
&
\multicolumn{4}{c||}{\( \textrm{qqqq}\tau \nu  \)}&
\multicolumn{4}{c|}{\( \textrm{qqqq}(\tau \nu )^{*} \)}\\
\hline 
WW background&
5.7&
2.0&
12.1&
4.3&
41.7&
6.2&
78.0&
11.9\\
\hline 
ZZ background&
0.6&
0.2&
1.2&
0.5&
2.7&
1.1&
5.9&
2.4\\
\hline 
\( \textrm{q}\bar{\textrm{q}} \) background&
2.9&
1.0&
4.9&
1.8&
13.4&
2.0&
21.2&
3.8\\
\hline 
\( \textrm{e}\nu \textrm{qq} \) background&
3.9&
0.7&
8.2&
1.7&
0.6&
0.1&
1.2&
0.31\\
\hline 
Total MC Bkgd&
13.2&
4.0&
26.5&
8.32&
58.5&
9.5&
106.2&
18.4\\
\hline 
Data&
13&
3&
36&
8&
64&
8&
138&
22\\
\hline 
{\footnotesize Signal for \( m_{\textrm{h}}=105 \) GeV}&
0.26&
0.22&
0.63&
0.55&
0.17&
0.14&
0.40&
0.35\\
\hline
\end{tabular}
\end{sideways}

\caption{Preselection and selection for the \protect\( \textrm{qqqql}\nu \protect \)
analysis.\label{tab:qqqqlv_numbers}}
\end{table}

\newpage\begin{minipage}{0.60\linewidth}\vfill\section{The \ensuremath{\nu\nu\mathrm{qqqq}} Channel\label{sec:vvqqqq}}\end{minipage}\hfill\begin{minipage}{0.35\linewidth}

\vspace{0.3cm}
{\centering \resizebox*{0.9\textwidth}{!}{\includegraphics{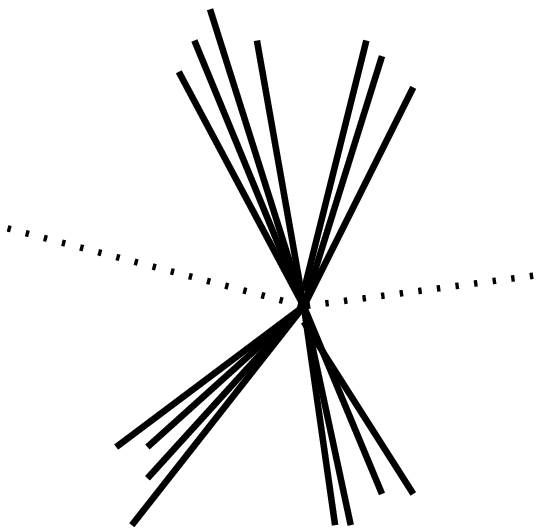}} \par}
\vspace{0.3cm}

\end{minipage}

In this channel, the Z decays to neutrinos and the W's decay hadronically,
so all the visible energy in the event comes from the Higgs. The signature
is two medium-energy jets with the invariant mass of the W, two low-energy
jets with a much smaller invariant mass, and a missing mass of the
Z. The total energy in the event should be twice the beam energy and
the vector sum of all momenta in the event should be zero. Thus the
missing mass is\[
m_{{\scriptsize \textrm{missing}}}=\sqrt{(\sqrt{s}-\Evis )^{2}-p^{2}_{{\scriptsize \textrm{vis}}}}.\]
 In this case, the two neutrinos produced by the Z should have an
invariant mass of \mZ.

The most important background to this channel was the \( \textrm{e}^{+}\textrm{e}^{-}\rightarrow \textrm{WW} \)
process, particularly the case where \( \textrm{WW}\rightarrow \textrm{q}\bar{\textrm{q}}'\tau \nu  \).
A tau decays hadronically 65\% of the time, leaving an event with
two high energy jets from the \( \textrm{q}\bar{\textrm{q}}' \) and
one low energy jet from the tau. The tau decay also involves a neutrino
which contributed to the missing mass. Gluon radiation or jet reconstruction
can easily account for a fourth low energy jet. 

Another very important background was the \( \textrm{e}^{+}\textrm{e}^{-}\rightarrow \textrm{q}\bar{\textrm{q}}(\gamma \gamma ) \)
process, where the both the electron and positron emit a photon before
annihilating, or one emits two photons. After emitting the photons,
the electron and positron interact at smaller effective center-of-mass
energy (\( \sqrt{s}' \)). This {}``double-radiative'' process has
a sharp peak for \( \sqrt{s}'=\mZ  \), where the emission of the
photons effectively returns the process to the huge Z resonance at
91 GeV visible in Figure \ref{fig:xsecs}. We reduced this background
by requiring the event thrust to be less than 0.9, the fraction of
visible energy in a \( 30^{\circ } \) cone around the beampipe to
be less than 60\%, and \( \left| \cos \theta _{{\scriptsize \textrm{missing}}}\right| <0.96 \). 

We preselected events with substantial missing energy by requiring
\( 0.4<\frac{\Evis }{\sqrt{s}}<0.7 \) and chose hadronic events by
requiring 20 calorimeter clusters and 10 charged tracks in the event.
We also required at least 30 GeV of energy in the BGO and 10 GeV in
the HCAL. To select against radiative events, we required no identified
photon with more than 10 GeV of energy, less than 10 GeV of energy
deposited in either the ALR or luminosity monitor%
\footnote{These detectors are very close to the beampipe and can intercept low-angle
photons from the radiative processes. ALR stands for Active Lead Ring,
which is a low-resolution, small angle detector in the forward region
of L3.
}, and that the missing momentum vector not be pointing toward the
EGAP. We forced the event to four jets using the Durham algorithm
and required at least one charged track in each jet and a minimum
jet energy of 6 GeV to select against low energy jets from gluons
or poor jet reconstruction.

After preselection, we prepared three networks with the eight inputs
described in Table \ref{tab:vvqqqq_nets}, twenty hidden nodes, and
one output node.
\begin{table}
\begin{tabular}{|c|p{2.1in}|p{2.5in}|}
\hline 
\emph{Variable}&
\emph{Description}&
\emph{Trend}\\
\hline
\hline 
\( \textrm{E}^{\textrm{max}}_{4\textrm{j}} \)&
Energy of the most energetic jet from a fit to four jets.&
\\
\cline{1-1} \cline{2-2} 
\( \textrm{E}^{\textrm{min}}_{4\textrm{j}} \)&
Energy of the least energetic jet from a fit to four jets.&
\multicolumn{1}{p{2.5in}|}{\vspace{-0.5in}Signal events tend to have two medium-energy and two
low-energy jets, while backgrounds will tend to have higher \( \textrm{E}^{\textrm{max}} \)
values and lower \( \textrm{E}^{\textrm{min}} \) values. }\\
\hline 
\( \theta ^{\textrm{min}}_{4\textrm{j}} \)&
Minimum angle between any two of the four jets.&
Gluon jets tend to be emitted at small angles relative to the emitting
quark jet.\\
\hline 
\( \alpha _{\textrm{W}^{(*)}\textrm{W}^{*}} \)&
Angle between the decay planes of the W candidate and \( \textrm{W}^{*} \)
candidate.&
This angle is likely to be smaller for gluon jets which fake the \( \textrm{W}^{*} \).\\
\hline 
\( m^{5c}_{\textrm{W}} \)&
Mass of the dijet with invarient mass closest to \mW after the 5C
fit.&
For signal events, this mass should be close to \mW.\\
\hline 
\( m_{\textrm{recoil}} \)&
Recoil mass of the event.&
\multicolumn{1}{p{2.5in}|}{For signal events, the recoil mass should be \mZ. Background events
will tend to have smaller recoil mass.}\\
\hline 
\( \log Y_{23} \)&
Durham Y value where the fit changes from two jets to three.&
\\
\cline{1-1} \cline{2-2} 
\( \log Y_{34} \)&
Durham Y value where the fit changes from three jets to four.&
\multicolumn{1}{p{2.5in}|}{\vspace{-0.28in} Events with gluon jets will tend to have smaller
Y values.}\\
\hline
\end{tabular}

\caption{Neural network variables for the \protect\( \nu \nu \textrm{qqqq}\protect \)
selection networks.\label{tab:vvqqqq_nets}}
\end{table}
 One network was trained to reject WW and \( \textrm{e}\nu \textrm{qq} \)
backgrounds, a second to reject ZZ, and a third to remove \( \textrm{q}\bar{\textrm{q}}(\gamma ) \).
We used only the \( \textrm{h}\rightarrow \textrm{WW} \) signal for
training, not the \( \textrm{h}\rightarrow \textrm{ZZ} \). At the
selection stage, we required \( \Nww >0.4 \), \( \Nzz >0.4 \), and
\( \Nqq >0.6 \). The numbers of predicted and observed events after
preselection and selection are given in Table \ref{tab:vvqqqq_numbers}.
\begin{table}
\begin{center}\begin{tabular}{|c|c|c|c|c|}
\hline 
&
\multicolumn{2}{c|}{1999}&
\multicolumn{2}{c|}{2000}\\
&
Preselection&
Selection&
Preselection&
Selection\\
\hline
\hline 
WW background&
94.1&
10.2&
169.1&
21.3\\
\hline 
ZZ background&
8.0&
1.9&
17.3&
3.0\\
\hline 
qq background&
32.3&
2.6&
43.6&
4.3\\
\hline 
\( \textrm{e}\nu \textrm{qq} \) background&
26.1&
1.3&
47.5&
3.1\\
\hline 
Total MC&
161.1&
16.0&
278.6&
31.9\\
\hline 
Data&
147&
13&
304&
28\\
\hline 
Signal for \( m_{\textrm{h}}=105 \) GeV&
1.04&
0.84&
2.93&
2.52\\
\hline
\end{tabular}

\end{center}

\caption{Preselection and selection totals for the \protect\( \nu \nu \textrm{qqqq}\protect \)
channel.\label{tab:vvqqqq_numbers}}
\end{table}

As in the \( \textrm{qqqqqq} \) channel, we can add \( \textrm{hZ}\rightarrow \textrm{Z}^{(*)}\textrm{Z}^{*}\textrm{Z}\rightarrow \nu \nu \textrm{qqqq} \)
events to our base \( \textrm{h}\rightarrow \textrm{WW} \) signal.
Of course, any of the three Z bosons can be the one which decays to
the neutrino pair, not only the Higgsstrahlung Z. Fortunately, the
kinematics of the event minimize the error on the Higgs mass generated
by taking \( \textrm{Z}^{(*)}\rightarrow \nu \nu  \) instead of the
Higgsstrahlung \( \textrm{Z}\rightarrow \nu \nu  \). The selection
accepted events where either the radiated Higgsstrahlung Z or the
\( \textrm{Z}^{(*)} \) from the Higgs decayed to neutrinos, but the
missing energy in the \( \textrm{Z}^{*}\rightarrow \nu \nu  \) case
was too small. The accepted signatures made up 20\% of the total \( \textrm{hZ}\rightarrow \textrm{ZZZ} \)
branching fraction, and including them increased the expected channel
signal rate by 15\%.

\newpage\begin{minipage}{0.60\linewidth}\vfill\section{The \ensuremath{\nu\nu\mathrm{qql}\nu} Channel}\end{minipage}\hfill\begin{minipage}{0.35\linewidth}

\vspace{0.3cm}
{\centering \resizebox*{0.9\textwidth}{!}{\includegraphics{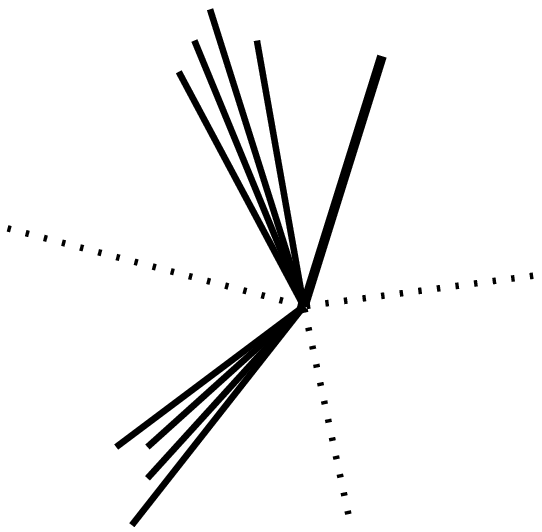}} \par}
\vspace{0.3cm}

\end{minipage}

In the \( \nu \nu \textrm{qql}\nu  \) channel, the Z decays to neutrinos,
while one W decays leptonically and the other into quarks. In this
channel, there are not enough constraints to reconstruct the Higgs
mass directly, so the channel is mostly a counting experiment. As
in the \( \textrm{qqqql}\nu  \) channel, the signal divides into
six subchannels as a function of the lepton flavor and source ( \( \textrm{W}^{(*)} \)
or \( \textrm{W}^{*} \)). We used the same variables to separate
the two channels and similar lepton quality requirements as in the
\( \textrm{qqqql}\nu  \) analysis. We required a minimum quality
value as the basic level of preselection.

The visible energy in this channel is quite small compared to the
other channels, since there are at least three energetic neutrinos
in the event. The low visible energy means that two-photon processes
become important sources of background and several cuts are applied
to eliminate them. A particularly useful variable for reducing the
two-photon and \( \textrm{q}\bar{\textrm{q}}(\gamma ) \) backgrounds
was \( \sin \Psi \equiv \left| \left( \hat{j}_{1}\times \hat{j}_{2}\right) \cdot \hat{z}\right|  \),
where \( \hat{j}_{1} \) and \( \hat{j}_{2} \) are the unit vectors
along the directions of the jets determined from fitting the event
into a two-jet topology. This variable preferentially selects events
in which the jets are at right angles to each other and to the beampipe.
Many background events have back-to-back jets or jets with small angles
relative to the beampipe. We required events to have \( \sin \Psi >0.07 \). 

For preselection, we also required \( \left| \cos \theta _{{\scriptsize \textrm{missing}}}\right| <0.9 \),
the fraction of visible energy in a \( 30^{\circ } \) cone around
the beampipe to be less than 40\%, and that there be less than 7 GeV
of energy in the ALR. We also set cuts on the BGO energy, recoil mass,
and numbers of clusters and tracks on a subchannel-basis as given
in Table \ref{tab:vvqqlv_presel}.
\begin{table}
{\centering \begin{tabular}{|l|c|c|c|}
\hline 
Quantity&
\( \nu \nu \textrm{qq}(\textrm{e}\nu ) \)&
\( \nu \nu \textrm{qq}(\mu \nu ) \)&
\( \nu \nu \textrm{qq}(\tau \nu ) \)\\
\hline
\hline 
BGO Energy/\( \sqrt{s} \)&
0.14 < x < 0.4&
0.02 < x < 0.22&
0.02 < x < 0.3\\
\hline 
Calorimeter clusters&
10 < x < 80&
10 < x < 75&
10 < x < 80\\
\hline 
Charge tracks&
4 < x < 23&
3 < x < 25&
5 < x < 25\\
\hline 
Recoil mass&
> 95 GeV&
> 80 GeV&
> 115 GeV\\
\hline 
\multicolumn{4}{l}{}\\
\hline 
&
\( \nu \nu \textrm{qq}(\textrm{e}\nu )^{*} \)&
\( \nu \nu \textrm{qq}(\mu \nu )^{*} \)&
\( \nu \nu \textrm{qq}(\tau \nu )^{*} \)\\
\hline
\hline 
BGO Energy/\( \sqrt{s} \)&
0.1 < x < 0.42&
0.05 < x < 0.35&
0.05 < x < 0.4\\
\hline
Calorimeter clusters&
15 < x < 80&
10 < x < 80&
15 < x < 85\\
\hline
Charge tracks&
6 < x < 33&
5 < x < 35&
5 < x < 33\\
\hline
Recoil mass&
> 80 GeV&
> 80 GeV&
> 75 GeV\\
\hline
\end{tabular}\par}

\caption{Subchannel-specific preselection cuts for \protect\( \nu \nu \textrm{qql}\nu \protect \).\label{tab:vvqqlv_presel}}
\end{table}

The most important background was \( \textrm{e}^{+}\textrm{e}^{-}\rightarrow \textrm{WW}\rightarrow \textrm{qql}\nu  \),
particularly the \( \textrm{WW}\rightarrow \textrm{qq}\tau \nu  \)
channel. We reconstructed an average W pair mass to help reject this
background. First, we scaled the energies and masses of the two jets
by a common factor until the sum of their energies was \( \frac{\sqrt{s}}{2} \),
as would be the case in a real WW event. Then, we constructed a {}``neutrino''
to balance the event. We calculated \( \mW ^{{\scriptsize \textrm{recon}}} \)
as the average of the invariant mass of the {}``neutrino'' - lepton
system and the scaled di-jet invariant mass. We used this variable
along with several others listed in Table \ref{tab:vvqqlv_nets} to
prepare one \( 8\times 20\times 1 \) neural network for each subchannel.
\begin{table}
\begin{tabular}{|c|p{1.75in}|p{2.8in}|}
\hline 
\emph{Variable}&
\emph{Description}&
\emph{Trend}\\
\hline
\hline 
\( \frac{E_{l}}{\sqrt{s}} \)&
Energy of the candidate lepton normalized to \( \sqrt{s} \).&
Lepton energy for the \( (\textrm{l}\nu )^{*} \)channel will be small,
while in the \( (\textrm{l}\nu )^{(*)} \) channel it will be larger.\\
\hline 
\( \frac{E_{{\scriptsize \textrm{jet}1}}+E_{{\scriptsize \textrm{jet}2}}}{\sqrt{s}} \)&
Jet energy sum scaled by the center-of-mass energy.&
The jet energy is complementary to the lepton energy - larger for
\( (\textrm{l}\nu )^{*} \)and smaller for \( (\textrm{l}\nu )^{(*)} \).\\
\hline 
\( \alpha  \)&
Angle between the plane containing the two jets and the lepton. &
Leptons from quark decay backgrounds will tend to lie in the plane
of the jets.\\
\hline 
\( \theta ^{\textrm{min}}_{\textrm{jl}} \)&
Minimum angle between either jet and the candidate lepton.&
Leptons from quark decay will tend to be close in angle to a jet.\\
\hline 
\( \theta _{\textrm{lm}} \)&
Angle between the lepton and missing momentum vector.&
For semi-leptonic WW events, the direction angle between the lepton
and missing momentum should coorespond to the W mass. For other backgrounds,
this angle should have no particular value. For signal, we expect
some coorelation between the lepton and missing energy.\\
\hline 
\( \mW ^{{\scriptsize \textrm{recon}}} \)&
Average mass for the \( \textrm{WW}\rightarrow \textrm{qql}\nu  \)
reconstruction.&
This mass should be close to \mW\ for semileptonic WW background events.\\
\hline 
\( m_{{\scriptsize \textrm{missing}}} \)&
Recoil mass against the visible part of the event. &
\multicolumn{1}{p{2.8in}|}{For signal events, the recoil mass should be \mZ, but energy losses
from the third neutrino and in the jets mean that the observed missing
mass will be > \mZ. Background events will tend to have a smaller
recoil mass.}\\
\cline{3-3} 
\hline 
\( m_{\textrm{jj}} \)&
Dijet mass.&
\multicolumn{1}{p{2.8in}|}{For \( (\textrm{l}\nu )^{(*)} \) signal events, this will peak around
25 GeV, depending on \mh. This variable is not used for \( (\textrm{l}\nu )^{*} \)networks.}\\
\hline
\end{tabular}

\caption{Neural network variables for the \protect\( \nu \nu \textrm{qql}\nu \protect \)
selection networks.\label{tab:vvqqlv_nets}}
\end{table}

For final selection, we placed cuts on \( {\cal N} \) around \( 0.35 \),
depending on the subchannel. The numbers of events predicted and observed
after selection and preselection are listed in Table \ref{tab:vvqqlv_numbers}.
There were insufficient constraints to fully reconstruct the Higgs
mass, so we used the visible mass as the final variable, as it was
quite correlated with the Higgs mass.
\begin{table}
{\centering \begin{tabular}{|l|c|c|c|c|}
\hline 
&
\multicolumn{2}{c|}{1999}&
\multicolumn{2}{c|}{2000}\\
&
\multicolumn{4}{c|}{ \( \nu \nu \textrm{qq}(\textrm{e}\nu )+\nu \nu \textrm{qq}(\mu \nu )+\nu \nu \textrm{qq}(\tau \nu ) \)}\\
&
Preselection&
Selection&
Preselection&
Selection\\
\hline
\hline 
WW background&
1.9&
0.3&
3.8&
0.7\\
\hline 
ZZ background&
0.5&
<0.1&
0.7&
0.0\\
\hline 
qq background&
0.1&
<0.1&
0.3&
0.0\\
\hline 
\( \textrm{e}\nu \textrm{qq} \) background&
1.2&
0.2&
2.7&
0.5\\
\hline 
Total MC&
3.7&
0.5&
7.6&
1.2\\
\hline 
Data&
6&
0&
15&
2\\
\hline 
{\footnotesize Signal for \( m_{\textrm{h}}=105 \) GeV}&
0.29&
0.24&
0.67&
0.60\\
\hline
\multicolumn{5}{c}{}\\
\hline 
&
\multicolumn{4}{c|}{\( \nu \nu \textrm{qq}(\textrm{e}\nu )^{*}+\nu \nu \textrm{qq}(\mu \nu )^{*} \)}\\
\hline
WW background&
11.2&
1.2&
23.2&
2.7\\
\hline
ZZ background&
1.4&
0.1&
2.2&
0.4\\
\hline
qq background&
0.4&
<0.1&
0.5&
<0.1\\
\hline
\( \textrm{e}\nu \textrm{qq background} \)&
1.6&
0.2&
3.8&
0.7\\
\hline
Total MC&
14.6&
1.5&
29.8&
3.8\\
\hline
Data&
15&
1&
24&
6\\
\hline
{\footnotesize Signal for \( m_{\textrm{h}}=105 \) GeV}&
0.19&
0.17&
0.43&
0.40\\
\hline
\end{tabular}\par}

\caption{Preselection and selection totals for the \protect\( \nu \nu \textrm{qql}\nu \protect \)
channel.\label{tab:vvqqlv_numbers}}
\end{table}

\newpage\begin{minipage}{0.60\linewidth}\vfill\section{The llqqqq Channel}\end{minipage}\hfill\begin{minipage}{0.35\linewidth}

\vspace{0.3cm}
{\centering \resizebox*{0.9\textwidth}{!}{\includegraphics{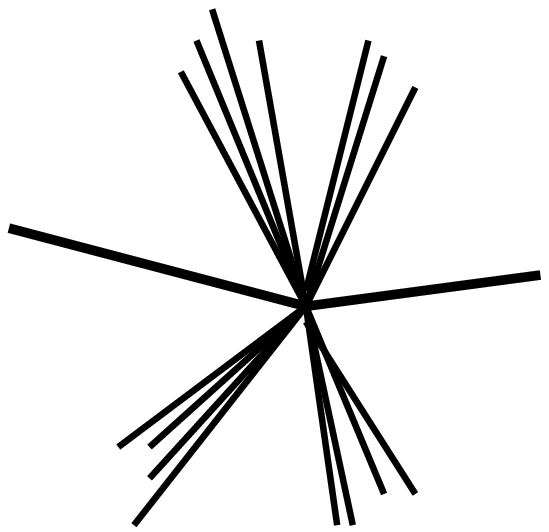}} \par}
\vspace{0.3cm}

\end{minipage}

In this channel, the Z decays to a pair of electrons or muons, while
the W's decay hadronically. The decay of the Z to taus was not considered
in the analysis. The BGO and muon chambers give very good resolution
on the leptons from the Z decay, so the Z was very well reconstructed
and the Higgs mass could be determined from the recoil. The only difficulty
with this channel was its small branching ratio, which implied a small
number of expected signal events. The event signature is two energetic
leptons with the invariant mass of the Z, two energetic jets with
about \mW, and two smaller jets. 

The requirement that the two leptons have an invariant mass close
to \mZ\ removed almost all background events which did not contain
legitimate Z bosons. At least one of the leptons in the rejected events
generally came from semileptonic quark decay in a jet. The most important
remaining background was Z pair production, but good recoil mass resolution
allowed the suppression of this background except for \( \mh \sim \mZ  \).

The first phase of the analysis was identification of the two leptons
considered to be the decay products of the Z. The analysis calculated
quality values for each pair of identified leptons, considering both
the single lepton factors, such as minimum angle to a jet, as well
as di-lepton factors such as \( \left| \mZ -m_{ll}\right|  \). Pairs
such as \( \mu \textrm{e} \) which would not arise from Z decay were
not considered. The pair with the best quality factor was chosen as
the candidate pair. This assignment put each event into either the
\( \textrm{eeqqqq} \) or \( \mu \mu \textrm{qqqq} \) subchannel
or rejected it.

After the identification, several preselection cuts were applied to
select events consistent with slightly more than two jets, while removing
as much of the four-jet background as possible. We required at least
30 calorimeter clusters and between 10 and 40 charged tracks in the
event. We required \( 5\textrm{ GeV}<\Ehcal <65\textrm{ GeV} \),
and \( 0.4<\frac{\Ebgo }{\sqrt{s}}<0.9 \) for \( \textrm{eeqqqq} \)
events and \( 0.1<\frac{\Ebgo }{\sqrt{s}}<0.5 \) for \( \mu \mu \textrm{qqqq} \)%
\footnote{The cut is set much higher for \( \textrm{eeqqqq} \) events since
the whole energy of the electron pair should go into the BGO, while
only a small fraction of the muon pair's energy will be deposited
there.
}. Since there were no neutrinos in our event signature, we required
\( 0.8<\frac{\Evis }{\sqrt{s}}<1.2 \) for \( \textrm{eeqqqq} \)
events and \( 0.7<\frac{\Evis }{\sqrt{s}}<1.2 \) for \( \mu \mu \textrm{qqqq} \)
events. 

After preselection, we applied a 4C fit requiring energy and momentum
conservation and then applied a few selection cuts. We required the
\( \chi ^{2} \) of the 4C fit to be less than 3 for \( \textrm{eeqqqq} \)
and less than 4.5 for \( \mu \mu \textrm{qqqq} \) and that the reconstructed
Z mass was between 60 GeV and 110 GeV. We also forced the event to
a four-jet + two lepton topology and required the \( \log Y_{34} \)
determined from the fit be larger than -7.5. These cuts removed the
remaining {}``obvious'' backgrounds, and then we trained three neural
networks to remove the remaining background: one for the \( \textrm{eeqqqq} \)
subchannel and two for \( \mu \mu \textrm{qqqq} \). The eeqqqq neural
network and one of the \( \mu \mu \textrm{qqqq} \) networks were
trained against the ZZ and Zee backgrounds. The second \( \mu \mu \textrm{qqqq} \)
network was trained against WW and \( \textrm{e}\nu \textrm{qq} \)
backgrounds. The six variables used in these \( 6\times 25\times 1 \)
neural networks are listed in Table \ref{tab:llqqqq_nets}.
\begin{table}
\begin{tabular}{|c|p{2.1in}|p{2.5in}|}
\hline 
\emph{Variable}&
\emph{Description}&
\emph{Trend}\\
\hline
\hline 
\( \textrm{E}^{\textrm{max}}_{4\textrm{j}-2\textrm{l}} \)&
Energy of the most energetic jet from a fit to four jets.&
\\
\cline{1-1} \cline{2-2} 
\( \textrm{E}^{\textrm{min}}_{4\textrm{j}-2\textrm{l}} \)&
Energy of the least energetic jet from a fit to four jets.&
\multicolumn{1}{p{2.5in}|}{\vspace{-0.5in}Signal events tend to have two medium-energy and two
low-energy jets, while backgrounds will tend to have higher \( \textrm{E}^{\textrm{max}} \)
values and lower \( \textrm{E}^{\textrm{min}} \) values. }\\
\hline 
\( \theta ^{\textrm{min}}_{\textrm{jj}} \)&
Minimum angle between any two of the four jets.&
Gluon jets tend to be emitted at small angles relative to the emitting
quark jet.\\
\hline 
\( \theta ^{\textrm{min}}_{\textrm{jl}} \)&
Minimum angle between any jet and any lepton.&
Backgrounds where the lepton is produced from quark decay will tend
to have smaller values of \( \theta ^{\textrm{min}}_{\textrm{jl}} \)
than signal events.\\
\hline 
\( m_{\textrm{ll}} \)&
Di-lepton invarient mass of the event.&
\multicolumn{1}{p{2.5in}|}{For signal events, the dilepton mass should be \mZ. Background events
where the leptons are not from a Z will tend to have a smaller dilepton
mass.}\\
\hline 
\( \log Y_{34} \)&
Durham Y value where the fit changes from three jets to four.&
Events with gluon jets will tend to have smaller Y values.\\
\hline
\end{tabular}

\caption{Neural network variables for the llqqqq selection networks.\label{tab:llqqqq_nets}}
\end{table}
 We required that selected events have \( \Nzz >0.25 \) and \( \Nww >0.2 \)(for
\( \mu \mu \textrm{qqqq} \)). The number of background and signal
events expected and data events observed after preselection and selection
are given in Table \ref{tab:llqqqq}.
\begin{table}
{\centering \begin{tabular}{|c|c|c|c|c|}
\hline 
&
\multicolumn{2}{c|}{1999}&
\multicolumn{2}{c|}{2000}\\
&
Preselection&
Selection&
Preselection&
Selection\\
\hline
&
\multicolumn{4}{c|}{eeqqqq}\\
\hline
\hline 
ZZ background&
2.2&
0.7&
4.6&
1.6\\
\hline 
Zee background&
0.6&
0.1&
1.1&
0.5\\
\hline 
\( \textrm{e}\nu \textrm{qq} \) background&
0&
0&
0.1&
0\\
\hline 
Total MC&
2.8&
1.1&
5.8&
2.1\\
\hline 
Data&
1&
0&
3&
2\\
\hline 
Signal for \( m_{\textrm{h}}=105 \) GeV&
0.11&
0.10&
0.31&
0.29\\
\hline
\multicolumn{5}{c}{}\\
\hline 
&
\multicolumn{4}{c|}{\( \mu \mu \textrm{qqqq} \)}\\
\hline
ZZ background&
2.4&
0.9&
4.5&
1.9\\
\hline
WW background&
3.4&
0.8&
6.1&
1.1\\
\hline
qq background&
0.5&
0&
0.7&
0.1\\
\hline
Total MC&
6.3&
1.7&
11.3&
3.1\\
\hline
Data&
5&
2&
12&
4\\
\hline
Signal for \( m_{\textrm{h}}=105 \) GeV&
0.12&
0.10&
0.29&
0.27\\
\hline
\end{tabular}\par}

\caption{Preselection and selection totals for the \protect\( \textrm{llqqqq}\protect \)
channel.\label{tab:llqqqq}}
\end{table}

The llqqqq channel was also able to benefit from a matching signature
in \( \textrm{h}\rightarrow \textrm{ZZ} \). However, signal Monte
Carlo was not available for this process. Since the efficiency and
shape for the \( \textrm{h}\rightarrow \textrm{ZZ} \) in both the
\( \nu \nu \textrm{qqqq} \) and qqqqqq channels matched the \( \textrm{h}\rightarrow \textrm{WW} \),
we simply increased the cross-section for this channel by the appropriate
factor to account for the case where Higgsstrahlung Z decayed leptonically.
Without signal Monte Carlo, we cannot properly adjust for the case
where one of the Higgs decay Z bosons decays leptonically.

\newpage\begin{minipage}{0.60\linewidth}\vfill\section{The \ensuremath{\mathrm{qql}\nu\mathrm{l}\nu} Channel}\end{minipage}\hfill\begin{minipage}{0.35\linewidth}

\vspace{0.3cm}
{\centering \resizebox*{0.9\textwidth}{!}{\includegraphics{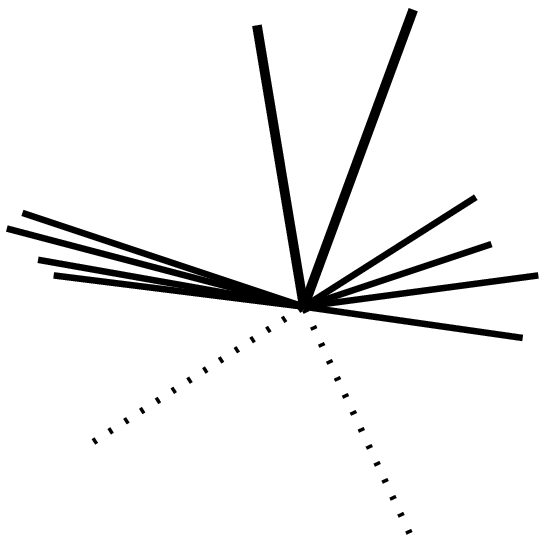}} \par}
\vspace{0.3cm}

\end{minipage}

In the \( \textrm{qql}\nu \textrm{l}\nu  \) channel, the Z decays
hadronically while both W bosons decay to lepton-neutrino pairs. The
analysis was lepton-flavor blind, except for lepton quality cuts which
were flavor-specific. We required two identified leptons, one with
more than 12 GeV of energy and the second with energy greater than
10 GeV. We accepted electrons, muons, or taus, but did not accept
minimum-ionizing particle (MIP) candidates as muons or taus.

At the preselection level, we required \( 0.5<\frac{\Evis }{\sqrt{s}}<0.85 \),
since the neutrinos should not carry away more than approximately
40\% of the center-of-mass energy. To select events with hadronic
content, we required 25 calorimetric clusters, 8 tracks, at least
30 GeV of energy in the BGO, and at least 8 GeV in the HCAL. To remove
two-photon and \( \textrm{q}\bar{\textrm{q}}(\gamma ) \) backgrounds,
we required \( \left| \cos \theta _{{\scriptsize \textrm{missing}}}\right| <0.92 \),
an event thrust of less than 0.93, less than 40\% of the total visible
energy within \( 30^{\circ } \) of the beampipe, and there be no
photon in the event with more than 20 GeV of energy. To remove unmodeled
or poorly modeled backgrounds, we required less than 10 GeV in the
ALR or luminosity monitor.

After preselection, the dominant backgrounds were W pair production
and \( \textrm{e}^{+}\textrm{e}^{-}\rightarrow \textrm{e}\nu \textrm{q}\bar{\textrm{q}} \)
processes. To remove these backgrounds, we constructed an \( 8\times 16\times 1 \)
neural network using the variables listed in Table \ref{tab:qqlvlv_nets}.
\begin{table}
\begin{tabular}{|l|p{2.1in}|p{2.5in}|}
\hline 
\emph{Variable}&
\emph{Description}&
\emph{Trend}\\
\hline
\hline 
\( \textrm{m}_{2\textrm{j}-2\textrm{l}} \)&
Invarient mass of the two jets after the fit removing the identified
leptons.&
For signal events, this should be close to \mZ, while for WW background
it will less than \mW.\\
\hline 
\( \theta ^{\textrm{min}}_{\textrm{jj}-\textrm{L}} \)&
Minimum angle between the more energetic lepton and either of the
two jets.&
\\
\cline{1-1} \cline{2-2} 
\( \theta ^{\textrm{min}}_{\textrm{jj}-\textrm{l}} \)&
Minimum angle between the less energetic lepton and either of the
two jets.&
\multicolumn{1}{p{2.5in}|}{\vspace{-20pt}Leptons from quark decay in jets tend to be emitted
at small angles relative to the emitting jet.}\\
\hline 
\( m^{\textrm{scale}}_{\textrm{jj}} \)&
Dijet mass from the fit to two jets and one lepton after scaling the
energies and masses of the jets by a common factor until \( \sum \textrm{E}_{\textrm{j}}=\sqrt{s} \).&
For WW background events, the dijet mass should be \mW, while it should
be considerably larger for signal events.\\
\hline 
\( m_{\textrm{l}\nu }^{\textrm{scale}} \)&
Invarient mass of the lepton-missing momentum system after scaling
the jets.&
For WW background events, this mass should also be \mW.\\
\hline 
\( \theta _{\textrm{ll}} \)&
Lepton-lepton angle.&
For signal events, both leptons should be in the same hemisphere,
while background leptons tend to be at higher angles.\\
\hline 
\( \log \frac{Y_{23}}{Y_{34}} \)&
Ratio of the Durham Y value where the fit changes from two jets to
three to the Y value for three to four.&
A signal event will appear either three-jet-like or four-jet-like
depending on the direction and energy of the second lepton, while
many background events will be two-jet or four-jet.\\
\hline
\end{tabular}

\caption{Neural network variables for the \protect\( \textrm{qql}\nu \textrm{l}\nu \protect \)
selection networks.\label{tab:qqlvlv_nets}}
\end{table}
We required \( {\cal N}>0.5 \) for selection. The numbers of expected
background and signal and observed data after preselection and selection
are given in Table \ref{tab:qqlvlv}. 
\begin{table}
{\centering \begin{tabular}{|c|c|c|c|c|}
\hline 
&
\multicolumn{2}{c|}{1999}&
\multicolumn{2}{c|}{2000}\\
&
Preselection&
Selection&
Preselection&
Selection\\
\hline
&
\multicolumn{4}{c|}{\( \textrm{qql}\nu \textrm{l}\nu  \) without taus}\\
\hline
\hline 
WW background&
4.26&
0.22&
6.40&
0.32\\
\hline 
\( \textrm{e}\nu \textrm{qq} \) background&
3.11&
0.12&
4.53&
0.09\\
\hline 
ZZ background&
0.56&
0.19&
1.06&
0.36\\
\hline 
qq background&
0.58&
0&
0.55&
0.02\\
\hline 
Total MC&
8.62&
0.55&
12.74&
0.86\\
\hline 
Data&
7&
0&
16&
2\\
\hline 
Signal for \( m_{\textrm{h}}=105 \) GeV&
0.15&
0.09&
0.58&
0.36\\
\hline
\multicolumn{5}{c}{}\\
\hline 
&
\multicolumn{4}{c|}{\( \textrm{qql}\nu \textrm{l}\nu  \) with taus}\\
\hline
WW background&
23.6&
5.6&
33.7&
7.5\\
\hline
\( \textrm{e}\nu \textrm{qq} \) background&
8.6&
2.3&
12.2&
2.3\\
\hline
qq background&
2.7&
0.6&
4.0&
0.9\\
\hline
ZZ background&
2.5&
1.3&
3.9&
2.2\\
\hline
Total MC&
37.6&
10.0&
54.1&
13.1\\
\hline
Data&
34&
9&
62&
19\\
\hline
Signal for \( m_{\textrm{h}}=105 \) GeV&
0.22&
0.15&
0.41&
0.28\\
\hline
\end{tabular}\par}

\caption{Preselection and selection totals for the \protect\( \textrm{qql}\nu \textrm{l}\nu \protect \)
channel.\label{tab:qqlvlv}}
\end{table}

After selection, we separated the events into two groups: events where
neither lepton was identified as a tau and events where at least of
one of the leptons was identified as a tau. The amount of background
in the second group was much larger than in the first. We created
discriminant final variables for these two subchannels separately,
using the reconstructed Higgs mass and the network output. We reconstructed
the Higgs mass by scaling the jet masses and energies by a common
factor until \( m_{jj}=\mZ  \), and then calculating the recoil mass
off the di-jet.

\chapter{Results of the Search \label{chap:results}}

\textit{Attempt the end, and never stand to doubt;}\\
\textit{Nothing's so hard but search will find it out.}

\begin{quotation}
Robert Herrick
\end{quotation}

\section{Search in the Two Photon Channel}

All four LEP collaborations have performed searches for a Higgs decaying
to two photons using the Higgsstrahlung production mode. The details
of each experiment's analysis are given in journal articles\cite{higgs:gg_Aleph,higgs:gg_Delphi,higgs:gg_l3,higgs:gg_opal,higgs:gg_opal2},
but the general strategies will be outlined here.

The decays of the Z define three search channels: \( \textrm{hZ}\rightarrow \gamma \gamma \textrm{q}\bar{\textrm{q}} \),
\( \textrm{hZ}\rightarrow \gamma \gamma \nu \bar{\nu } \), and \( \textrm{hZ}\rightarrow \gamma \gamma \textrm{l}^{+}\textrm{l}^{-} \).
For all the channels, the major background is double-radiative Z production
as described in Section \ref{sec:vvqqqq}. The process tends to produce
on-shell, boosted Z bosons and one or more high energy photons that
mimic the signal. For the signal, the di-photon spectrum peaks at
the Higgs mass, while the background spectrum is fairly flat over
a wide range of invariant masses. 

In general, photons are experimentally identified as isolated clusters
in the electromagnetic calorimeters with no matching track to indicate
an electron. Additional selection criteria are also used to reject
merged \( \pi _{0} \)'s and other possible sources of fakes.

\begin{figure}
{\centering \resizebox*{0.6\textwidth}{!}{\includegraphics{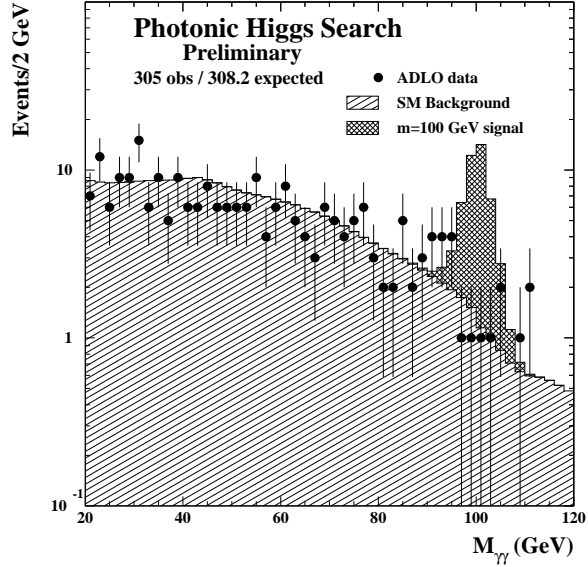}} \par}

\caption{Mass distribution of selected events in the LEP combined \protect\( \textrm{h}\rightarrow \gamma \gamma \protect \)
search.\label{fig:m_gg}}
\end{figure}

DELPHI, L3, and OPAL have developed analyses for each of the three
search channels separately, using similar cut-based approaches to
reduce the background. In contrast, the ALEPH experiment performs
a ``global'' analysis, focusing on the di-photon system and combining
all the Z decay channels together. The results from these selections
are presented in Table \ref{tab:hgg_results}.
\begin{table}
{\centering \begin{tabular}{cccccc}
\hline 
&
Energy&
&
Expected&
Benchmark&
Expected\\
Experiment&
Range&
Candidates&
Bkgd&
Limit (GeV)&
Limit (GeV)\\
\hline
ALEPH&
192-209&
10&
10.8&
104.4&
104.6\\
DELPHI&
189-209&
47&
42.2&
103.6&
105.1\\
L3&
189-209&
64&
69.5&
104.1&
104.9\\
OPAL&
88-209&
184&
185.7&
104.8&
105.2\\
\hline
\hline 
LEP&
&
305&
308.5&
106.5&
109.6\\
\hline
\end{tabular}\par}

\caption{Results of the \protect\( \textrm{h}\rightarrow \gamma \gamma \protect \)
search.\label{tab:hgg_results}}
\end{table}
 Each experimental group provides data, background, and predicted
signal distributions of the reconstructed Higgs mass to the LEP Higgs
Working Group (LHWG). These distributions, added together, are plotted
in Figure \ref{fig:m_gg} for a mass hypothesis of \( m_{\textrm{h}}=100\textrm{ GeV} \).
 The results can also be statistically combined to produce limits
or indicate the presence of signal.

\section{Hypothesis Testing and Limit Setting}

In January of 2000, just before the last year of LEP, CERN hosted
a workshop to discuss the best way to derive confidence level limits
and combine the results from different experiments and search channels
\cite{higgs:limits}. In the Standard Model search, there are four
physical channels and at least nine center of mass energies. Thus,
there are 36 independent channels to combine from each of the four
experiments. To obtain the maximum search power from the analyses,
the goal is to consider the shape of the final variables, rather than
just execute a global counting experiment. The solution is to consider
each bin of each final variable as a separate counting experiment
in the form of a log-likelihood ratio: \[
L_{i}=s_{i}-n_{i}\ln \left( \frac{b_{i}+s_{i}}{b_{i}}\right) =s_{i}-n_{i}\ln \left( 1+\frac{s_{i}}{b_{i}}\right) \]
where \( s_{i} \), \( b_{i} \), and \( n_{i} \) are the number
of expected signal events, expected background events, and observed
events in the \( i\textrm{th} \) bin, respectively. The ratio compares
the data to the hypotheses of background only and signal+background.
We can combine bins, center-of-mass energies, and channels simply
by adding the log-likelihood ratio values for each bin. The specific
form of the statistical estimator chosen by the LHWG is\[
-2\ln Q=2\sum _{i}s_{i}-2\sum _{i}n_{i}\ln \left( 1+\frac{s_{i}}{b_{i}}\right) .\]
A positive value of \( -2\ln Q \) indicates the observed data agrees
better with the background-only hypothesis, while a negative value
favors the signal plus background hypothesis. The expected values
of \( -2\ln Q \) in the presence and absence of signal can be obtained
by replacing \( n_{i} \) with \( s_{i}+b_{i} \) and \( b_{i} \)
respectively. The prefactor was chosen so that in the large number
limit, \( -2\ln Q\rightarrow \chi ^{2} \). Both signal and background
are calculated in terms of efficiencies \( \epsilon ^{s}_{i} \) and
\( \epsilon ^{b}_{i} \) and the number of expected events is extracted
by multiplying by the signal and background cross-sections and the
luminosity. This means that \( -2\ln Q \) scales linearly with luminosity,
so the performance gain expected from additional luminosity is easily
calculated. Also, the significance of a bin is entirely captured by
its signal-to-background ratio, so we can collect bins of similar
signal-to-background ratio together to speed the computation.

A single round of computations generates the observed and expected
estimator values, but to gauge the importance of any excess or deficit,
we determined the range of estimator values consistent with the observation.
To determine this range, we create two new distributions from the
signal and background distributions by replacing the value in each
bin with a value drawn from the Poisson distribution with the expected
value of the original bin value. These new distributions simulate
what might be observed in a real experiment, as a Monte Carlo trial.
We recorded the \( -2\ln Q \) values from these distributions and
repeat the process 50,000 or more times. The result of these trials
is new statistical distributions of \( -2\ln Q \).

\begin{figure}
{\centering \resizebox*{0.8\textwidth}{!}{\includegraphics{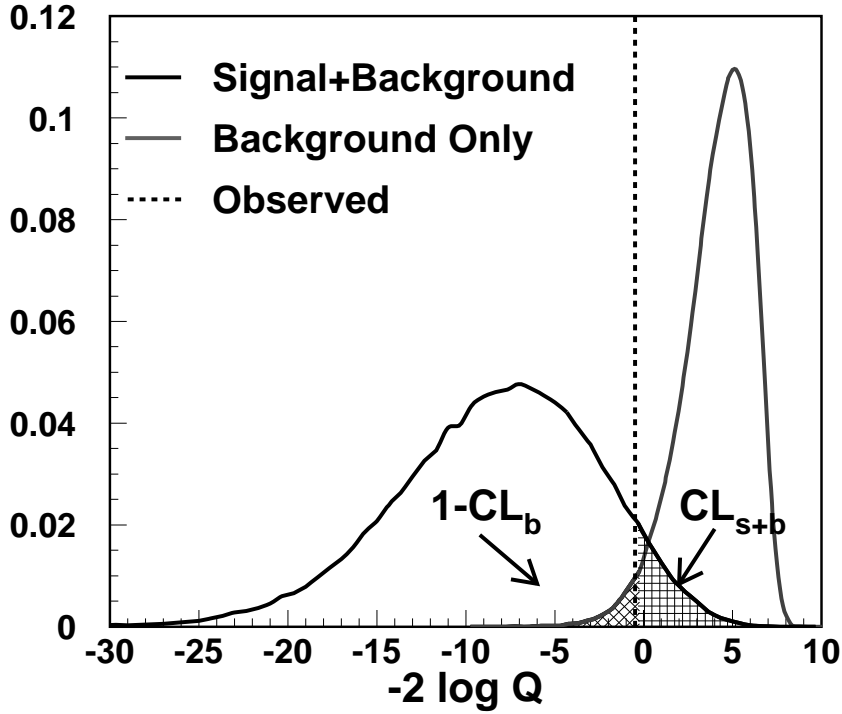}} \par}

\caption{Example of the Background-only and Signal+Background \protect\( -2\ln Q\protect \)
distributions.\label{fig:lnq}}
\end{figure}

In Figure \ref{fig:lnq} are the \( -2\ln Q \) trial distributions
from the \( \textrm{h}\rightarrow \gamma \gamma  \) search for \( m_{\textrm{h}}=106\textrm{ GeV} \).
 The solid black and gray curves are the \( -2\ln Q \) distributions
for the signal+background and background only hypotheses, respectively.
The dashed line indicates the observed value calculated from the data.
Two important confidence levels can be immediately obtained from this
plot. The fractional area of the signal+background curve above the
observed line indicates the observed confidence level for signal.
In the presence of signal, one would expect \( \CLsb \approx \frac{1}{2} \).
Very small values of \CLsb\ statistically exclude the presence of
signal. Underfluctuations in data may cause anomolously small values
of \CLsb, so the LHWG exclusion crition is based on a modified confidence
level \( \CLs \equiv \frac{\CLsb }{\CLb } \). The second confidence
level in the plot is the fractional area of the background-only plot
below the observed line, which gives \( 1-\CLb  \). For data that
contains no signal events, \CLb\ should be near \( \frac{1}{2} \).
However, in the presence of a signal \( 1-\CLb  \) will become very
small. An extremely small \( 1-\CLb  \) (\( <5\times 10^{-7} \))
is the criterion for the discovery of the Higgs boson.

\section{Results of the Two-Photon Search}

\begin{figure}
\begin{tabular}{p{2.5in}p{2.5in}l}
\resizebox*{0.48\textwidth}{!}{\includegraphics{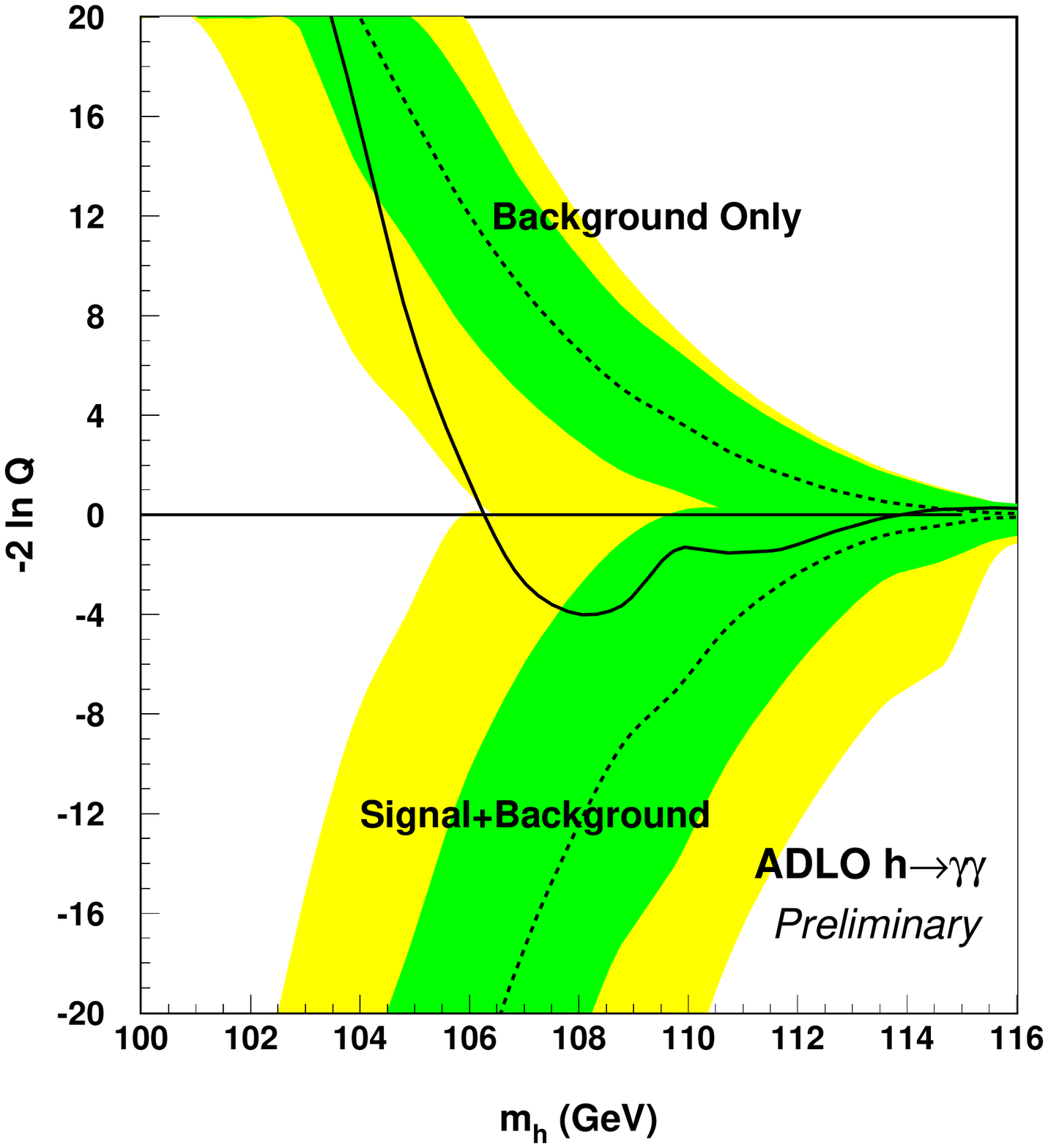}} 

\centering (a) &
\resizebox*{0.48\textwidth}{!}{\includegraphics{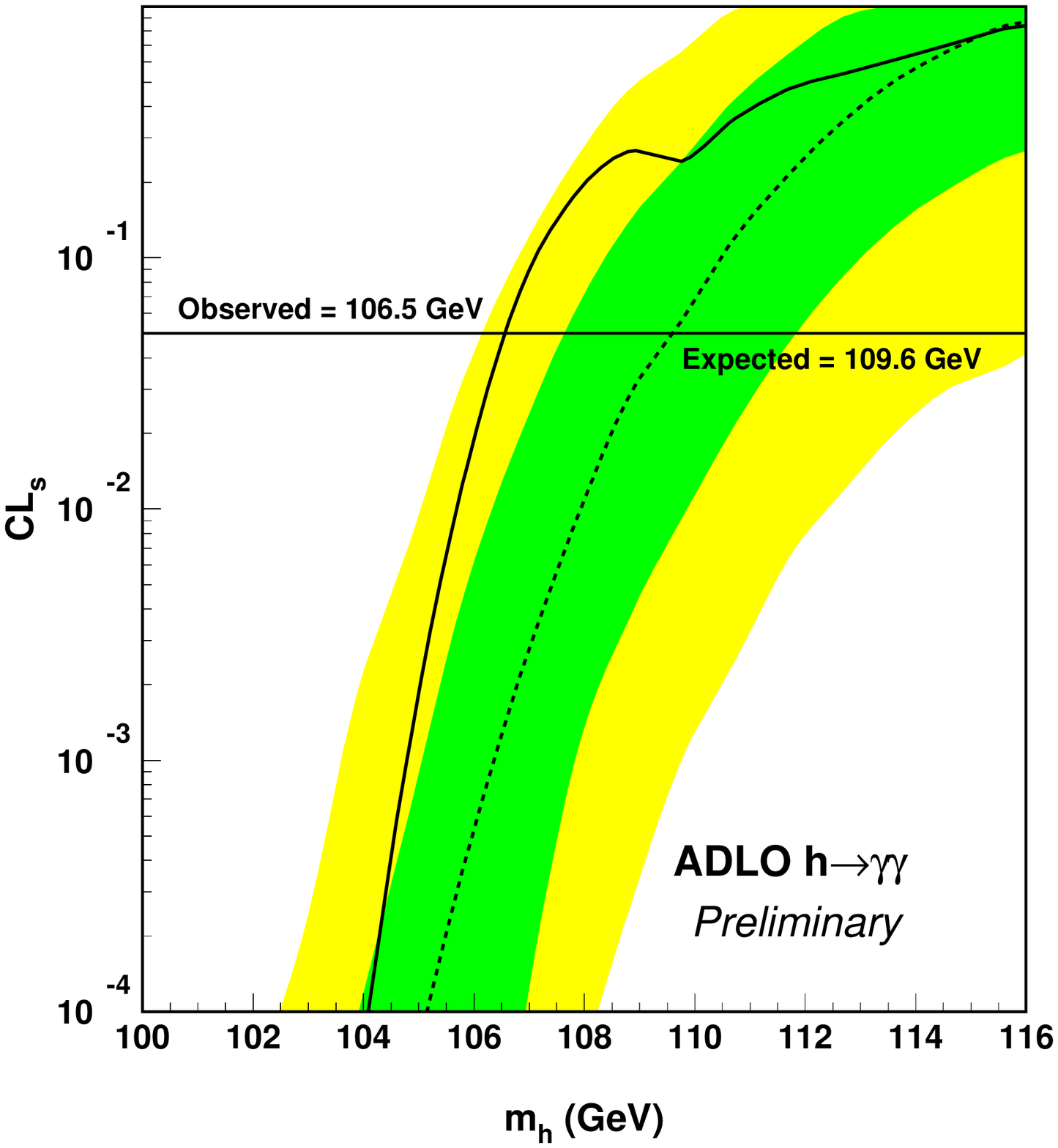}} 

\centering (b)&
\\
\vspace{-2in}

\caption{Results and limits for the LEP combined \protect\( \textrm{h}\rightarrow \gamma \gamma \protect \)
search.\label{fig:gg_results}}
\medskip{}

(a) \( -2\ln Q \) results for the benchmark model
\medskip{}

(b) CL limits for the benchmark model

(c) Branching-Ratio limits \medskip{}
&
\resizebox*{0.48\textwidth}{!}{\includegraphics{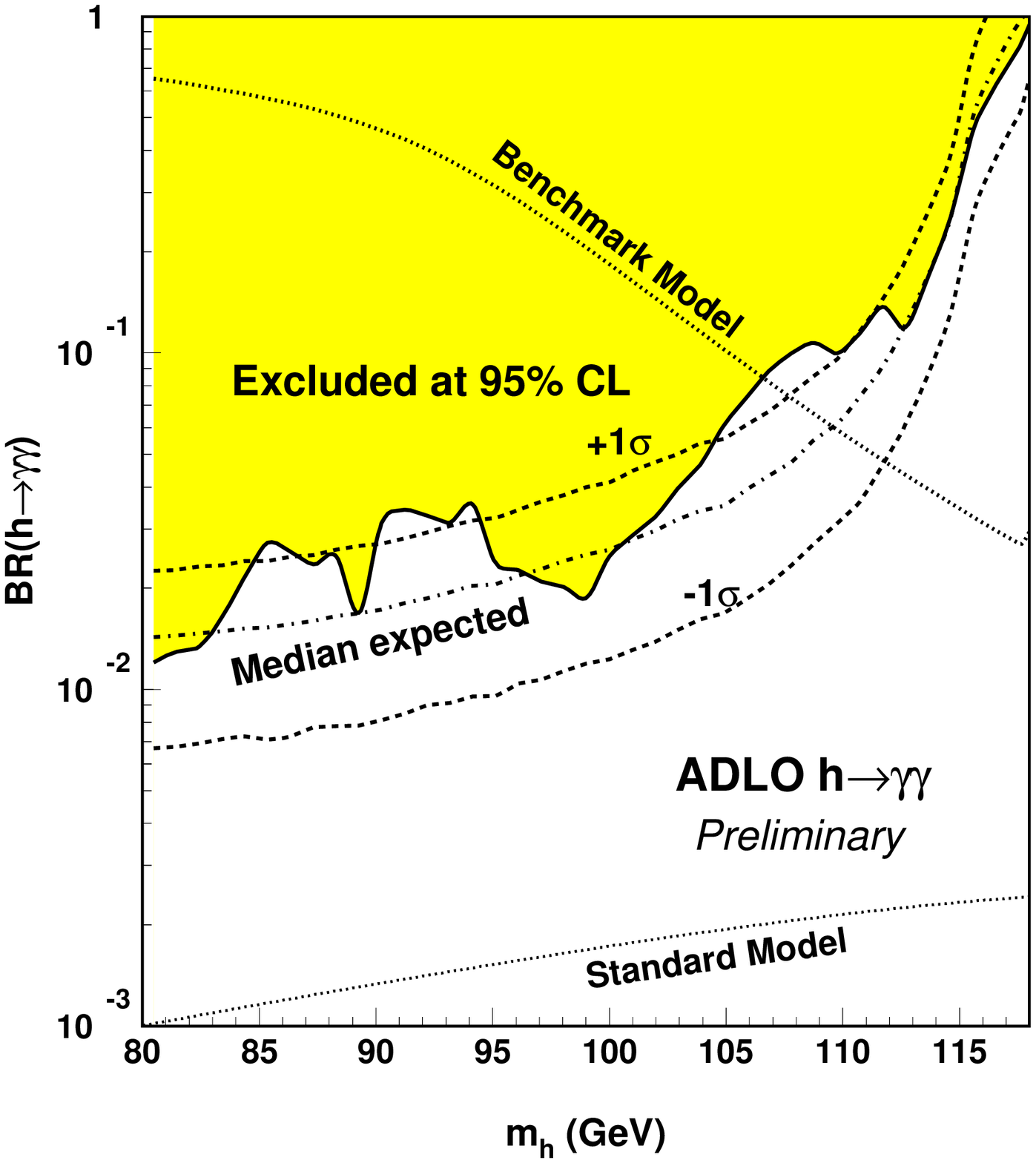}} 

\centering (c)&
\\
\end{tabular}
\end{figure}

In the benchmark model, we can calculate the \( -2\ln Q \) distributions
using inputs provided by all four experiments. To visualize the results
as a function of mass, we plot the central value of the \( -2\ln Q \)
distributions for the signal+background hypothesis and the background-only
hypothesis as a functions of mass using dashed lines. Then we indicate
the \( \pm 1\sigma  \) and \( \pm 2\sigma  \) edges from the Monte
Carlo trials by shaded regions. Finally, we indicate the observed
\( -2\ln Q \) values using a solid line. The results of the \( \textrm{h}\rightarrow \gamma \gamma  \)
search are shown in this form in Figure \ref{fig:gg_results}a. Where
the line is in the upper half of the plot, the observation favors
the background-only hypothesis and where it is in the lower it favors
the signal+background. The shaded regions indicate the level of compatibility
between the observed and either hypothesis. At each mass, a {}``cross-section''
vertically through the plot would reveal distributions similar to
Figure \ref{fig:lnq}. 

To determine the exclusion levels as a function of mass, we can perform
the \CLsb\ integration on each \( -2\ln Q_{{\scriptsize \textrm{s}+\textrm{b}}} \)
distribution. Using the \( -2\ln Q_{{\scriptsize \textrm{b}}} \)
distribution, we can also determine the median expected exclusion
and the \( \pm 1\sigma  \) and \( \pm 2\sigma  \) bands. The results
of this calculation are shown in Figure \ref{fig:gg_results}b. This
plot shows an observed 95\% CL limit on the benchmark fermiophobic
Higgs at \( \textrm{m}_{\textrm{h}}=106.5\textrm{ GeV} \) with a
expected limit of \( \textrm{m}_{\textrm{h}}=109.6\textrm{ GeV} \).
These numbers are not identical to those presented by the LHWG because
of the use of a different statistical method than that usually employed
in the fermiophobic results%
\footnote{These results, like the LHWG Standard Model results, use the frequentist
\CLs\ method while the fermiophobic combination has been traditionally
performed using a Bayesian technique \cite{higgs:lhwg_gg}.
}. There is an excess in signal events observed above background in
the high-mass region, but this excess is well within the \( \pm 2\sigma  \)
band. 

We can derive model-independent results by scaling up and down the
predicted signal cross-section until \CLs\ = 95\%. This provides a
limit on the rate of the production of any Higgs which decays to \( \gamma \gamma  \),
regardless of model. We show the results of this calculation in Figure
\ref{fig:gg_results}c along with the predictions of two models. The
upper dotted line shows the prediction of the benchmark model, while
the lower dotted line shows the prediction of the Standard Model.
The LEP combined \( \textrm{h}\rightarrow \gamma \gamma  \) results
can exclude the benchmark model over a large mass range, but cannot
exclude the Standard Model at any mass.

\section{Results of the \protect\( \textrm{h}\rightarrow \textrm{WW}/\textrm{ZZ}\protect \)
Search}

After all analysis cuts were applied in Chapter 3, the final output
of each analysis was a set of final variable distributions for data,
predicted background, and signal as a function of Higgs mass hypothesis.
These final variable distributions were the {}``inputs'' to the
hypothesis-testing and limit-setting algorithms. These distributions
are quite complicated, so they can be visualized in several different
ways.

The final variable distributions for a given mass are most conveniently
plotted by grouping together bins of the final variable which have
the same signal-to-background ratio, since the log-likelihood ratio
uses only the signal-to-background ratio. To give a sense of the mass-dependence
of the analyses, the final variable distributions for each analysis
are given for two different mass hypotheses: \mh=100 GeV and \mh=110
GeV. Most analyses show better signal-to-background ratio for the
higher mass hypothesis even though the Higgsstrahlung cross-section
is larger for the lower masses. This behavior arises because background
events tend to have low reconstructed {}``Higgs masses,'' usually
close to \mW\ or \mZ. 

The final results of the analysis can also be viewed using a plot
of reconstructed Higgs mass directly. In such a plot, any additional
information about the {}``signal-ness'' of an event is lost, so
the search sensitivity of the plot is limited. Each channel with a
reconstructed mass variable (all the channels except \( \nu \nu \textrm{qql}\nu  \))
has a plot of the reconstructed mass for all data and background events
along with the signal distribution for a single mass. 

The plotting technique with the highest information density is the
\( -2\ln Q \) plot. These plots show the expected and observed results
as a continuous function of Higgs mass hypothesis. Each channel has
a \( -2\ln Q \) plot, but the scale used is channel-dependent. Each
of \( -2\ln Q \) plots also has a dark band surrounding the observed
line which indicates the magnitude of the systematic error estimated
for the channel. We determine the systematic errors by shifting variables
in the the signal and background events while keeping the discriminant
probability density functions constant. The change in \( -2\ln Q \)
from the normal results indicates the magnitude of the systematic
error. More details of the calculation as well as tables of the major
systematic errors channel are given in Appendix \ref{sec:systematics}.

We can compare the relative search power of the channels in various
ways. In the log-likelihood plot, the \( \Delta (-2\ln Q) \) distance
between the background-only and signal+background hypotheses is larger
for channels and mass hypotheses which have greater search power.
Another is a table of the signal-to-background ratio for given number
of expected signal events, as in Table \ref{tab:tully}. For each
channel, we integrate the final variable distribution from the largest
signal-to-background ratio until we reach 0.1, 0.5, and 1.0 expected
events. For each expectation we report the signal-to-background ratio.
Some channels never expect as many as 0.5 or 1.0 events at \mh=110
GeV for any signal-to-background ratio.
\begin{table}
{\centering \begin{tabular}{|c|c|c|c|c|}
\hline 
&
\multicolumn{3}{c|}{s/b ratio}&
Total expected\\
Channel&
1.0 signal events&
0.5 signal events&
0.1 signal events&
signal events\\
\hline
\hline 
qqqqqq&
0.14&
0.22&
0.24&
8.05\\
\hline 
\( \textrm{qqqql}\nu  \)&
0.75&
1.8&
3.7&
2.71\\
\hline 
\( \nu \nu \textrm{qqqq} \)&
0.14&
0.43&
1.1&
1.47\\
\hline 
\( \nu \nu \textrm{qql}\nu  \)&
-&
0.12&
1.1&
0.79\\
\hline 
llqqqq&
-&
-&
4.7&
0.38\\
\hline 
\( \textrm{qql}\nu \textrm{l}\nu  \)&
-&
-&
5.1&
0.47\\
\hline
\end{tabular}\par}

\caption{Comparison of the channels by signal to background ratio for \mh=110
GeV.\label{tab:tully}}

Cases where a channel expects less than 1.0 or 0.5 events total are
indicated by a dash.
\end{table}

\subsection{Results by Channel}

\begin{figure}
\begin{tabular}{p{2.5in}p{2.5in}l}
\resizebox*{0.48\textwidth}{!}{\includegraphics{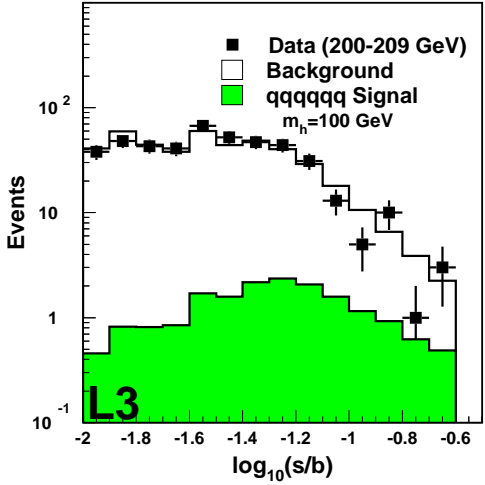}} 

\centering (a) Final variable rebinned in s/b for \( m_{\textrm{h}}=100 \)
GeV&
\resizebox*{0.48\textwidth}{!}{\includegraphics{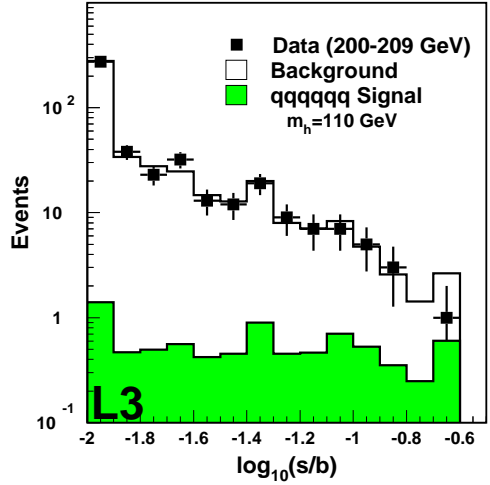}} 

\centering (b) Final variable rebinned in s/b for \( m_{\textrm{h}}=110 \)
GeV&
\\
\resizebox*{0.48\textwidth}{!}{\includegraphics{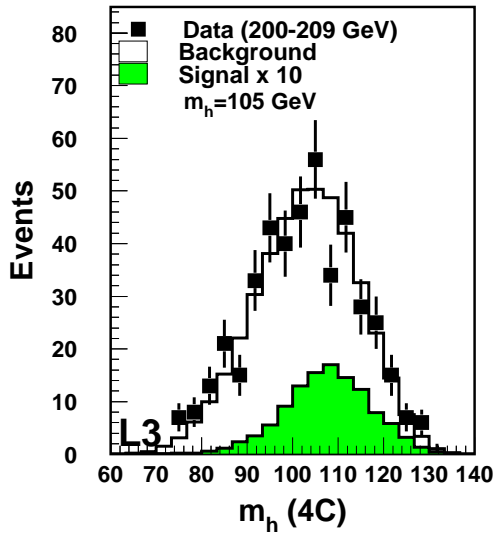}} 

(c) Mass distribution of selected events\medskip{}
&
\resizebox*{0.48\textwidth}{!}{\includegraphics{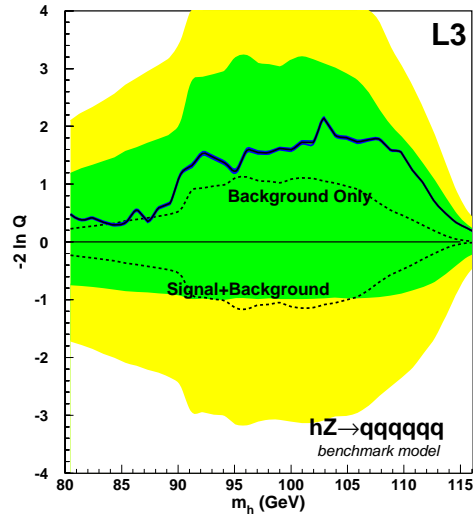}} 

\centering (d) \( -2\ln Q \) curves for qqqqqq&
\\
\end{tabular}

\caption{Results for the \protect\( \textrm{hZ}\rightarrow \textrm{qqqqqq}\protect \)
search.\label{fig:6j_results}}
\medskip{}
\end{figure}

The largest total number of signal events is expected in the qqqqqq
channel (Figure \ref{fig:6j_results}). However, the analysis has
no bin with a signal-to-background ratio greater than \( \frac{1}{4} \),
which is quite low compared to the other channels. The six broad jets
of the fully hadronic events are difficult to reconstruct in the L3
detector, which leads to a very poor resolution on the reconstructed
Higgs mass. As seen in Figure \ref{fig:6j_results}d, the channel
exhibits a small constant deficit in data relative to the Standard
Model background expectation, which follows directly from the small
observed counting deficit (443 with 446 expected) and the wide resolution.

The \( \textrm{qqqql}\nu  \) channel has the highest total separation
power of any of the channels. Its branching fraction is nearly as
large as the qqqqqq branching fraction. The lepton and neutrino in
the event improve the event identification and mass resolution. The
observed \( -2\ln Q \) curve lies within the \( \pm 1\sigma  \)
region for most of the mass range, but the channel indicates an excess
over the background-only prediction at masses above 110 GeV.
\begin{figure}
\begin{tabular}{p{2.5in}p{2.5in}l}
\resizebox*{0.48\textwidth}{!}{\includegraphics{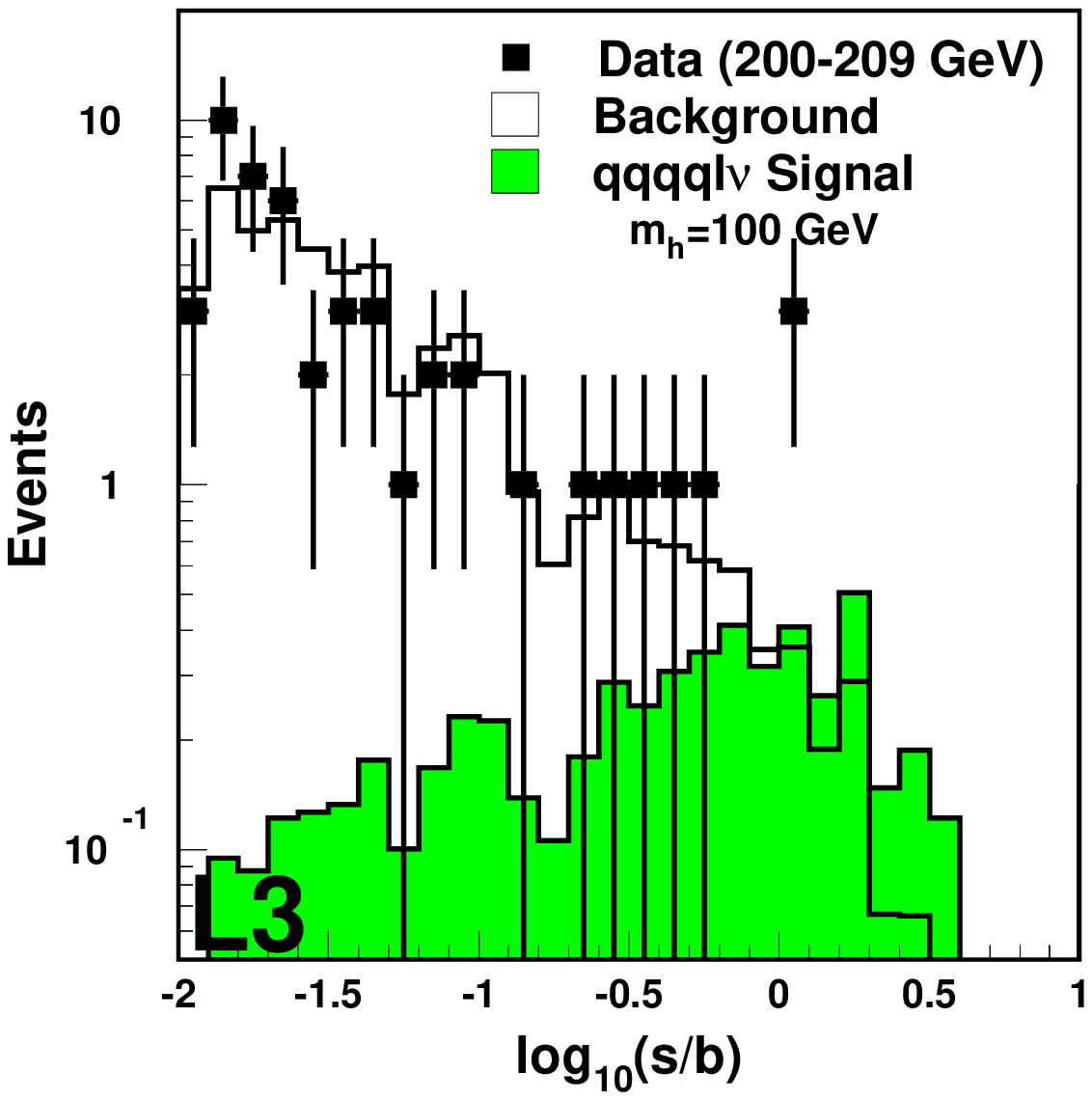}} 

\centering (a) Final variable rebinned in s/b for \( m_{\textrm{h}}=100 \)
GeV&
\resizebox*{0.48\textwidth}{!}{\includegraphics{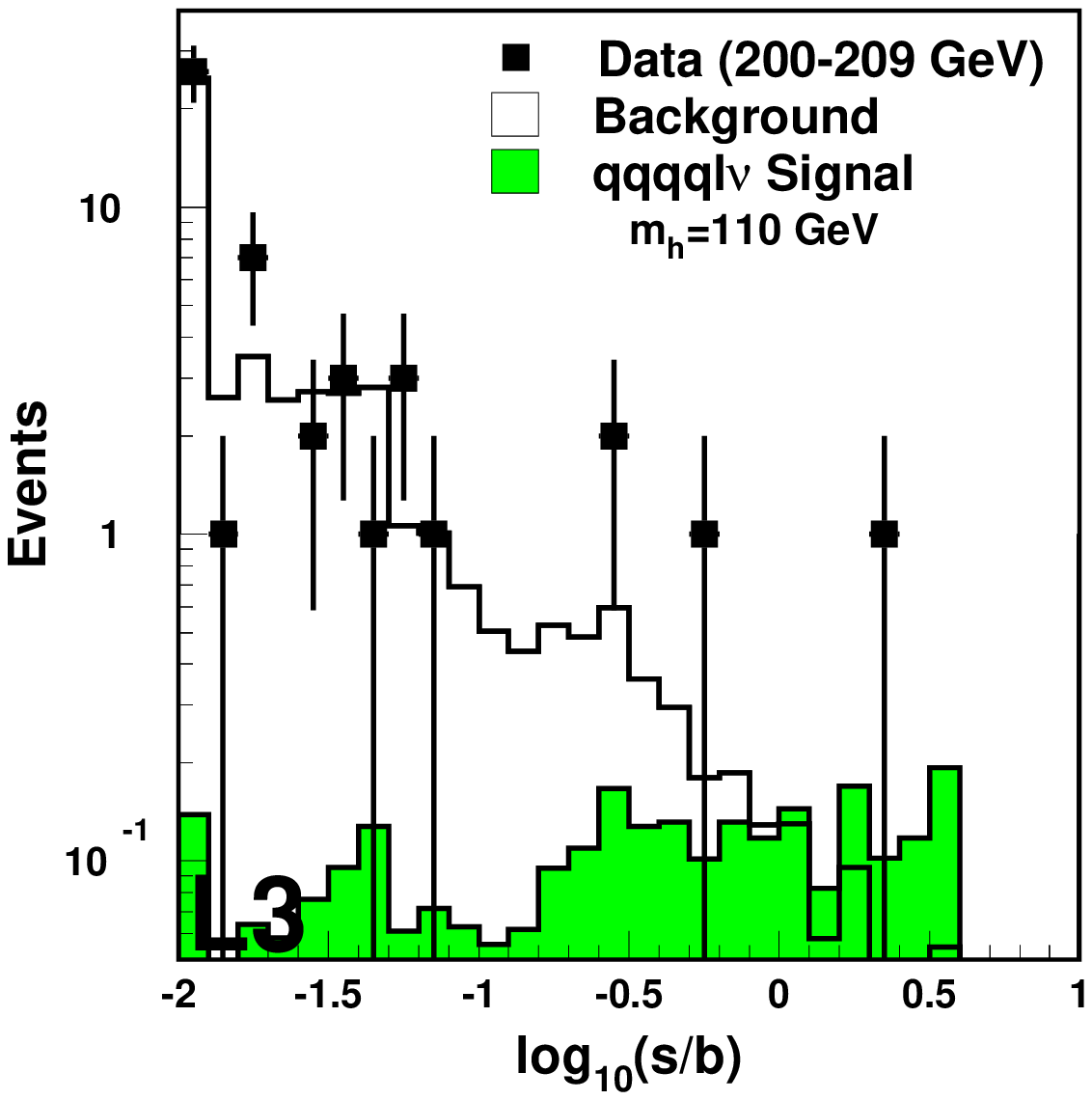}} 

\centering (b) Final variable rebinned in s/b for \( m_{\textrm{h}}=110 \)
GeV&
\\
\resizebox*{0.48\textwidth}{!}{\includegraphics{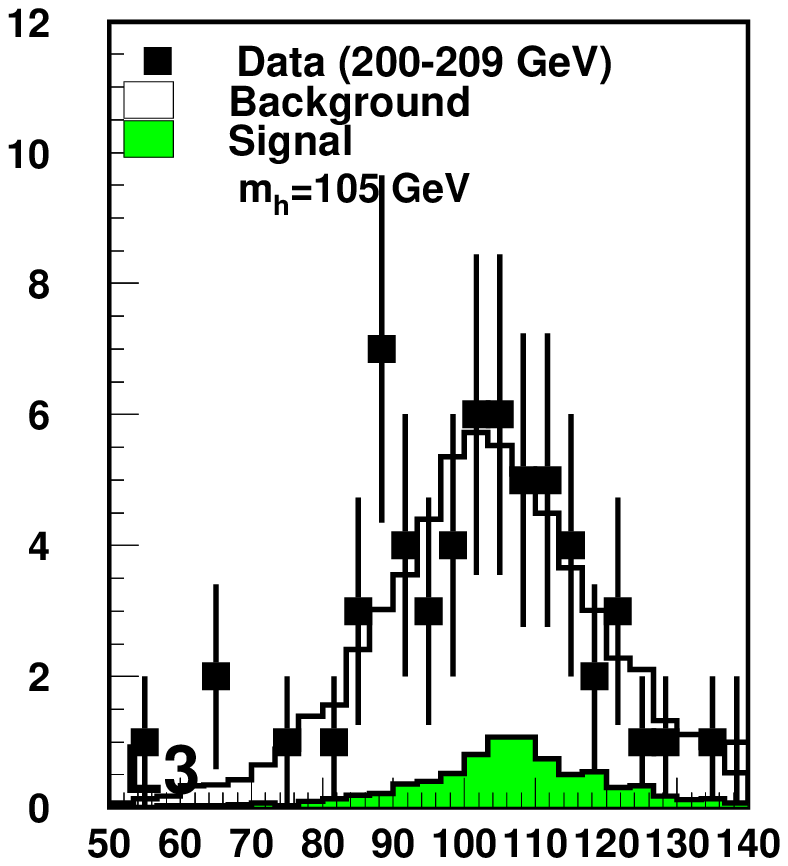}} 
\medskip{}

(c) Mass distribution of selected events\medskip{}
&
\resizebox*{0.48\textwidth}{!}{\includegraphics{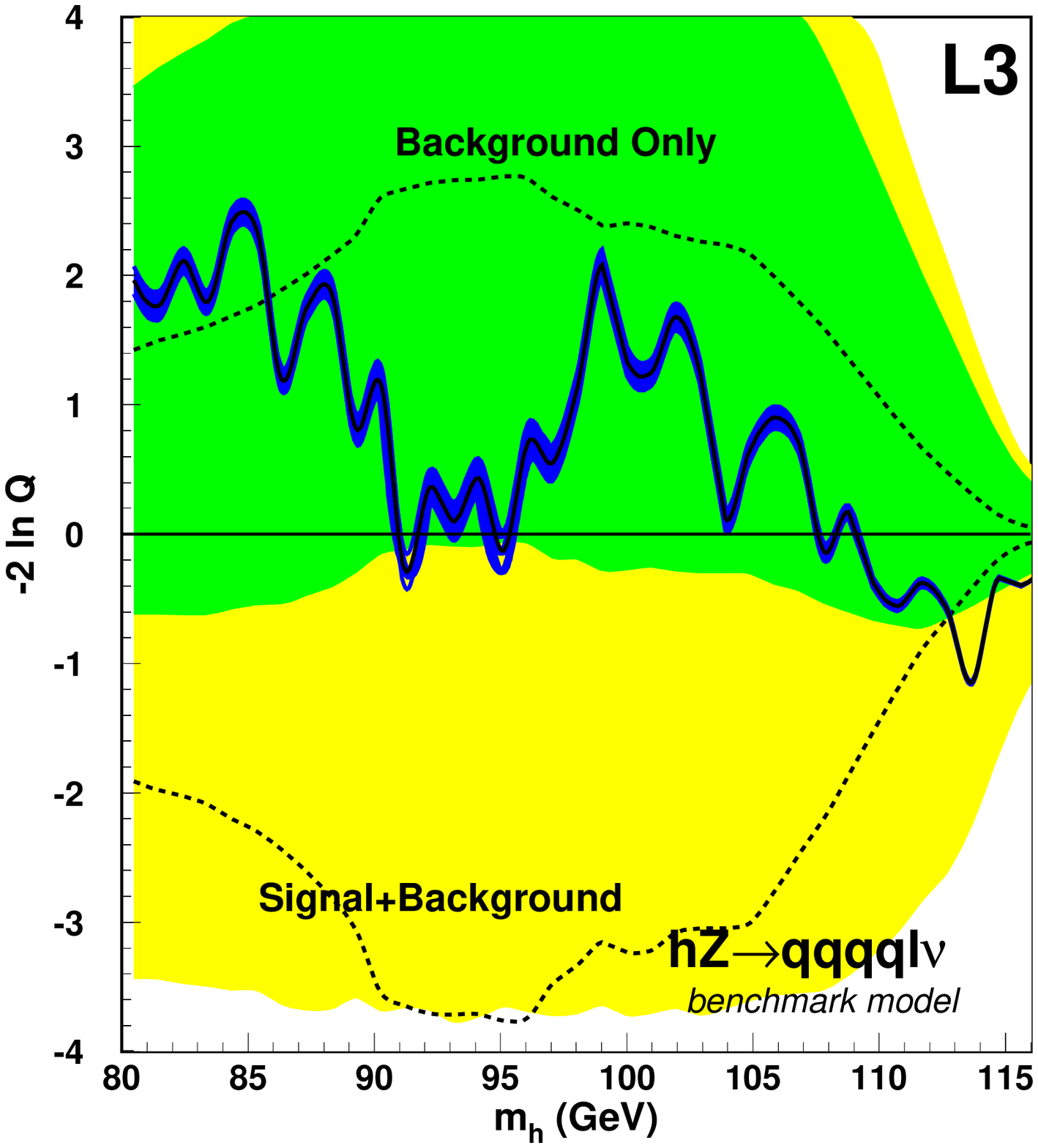}} 

\centering (d) \( -2\ln Q \) curves &
\\
\end{tabular}

\caption{Results for the \protect\( \textrm{hZ}\rightarrow \textrm{qqqql}\nu \protect \)
search.\label{fig:qqqqlv_results}}
\medskip{}
\end{figure}

\begin{figure}
\begin{tabular}{p{2.5in}p{2.5in}l}
\resizebox*{0.48\textwidth}{!}{\includegraphics{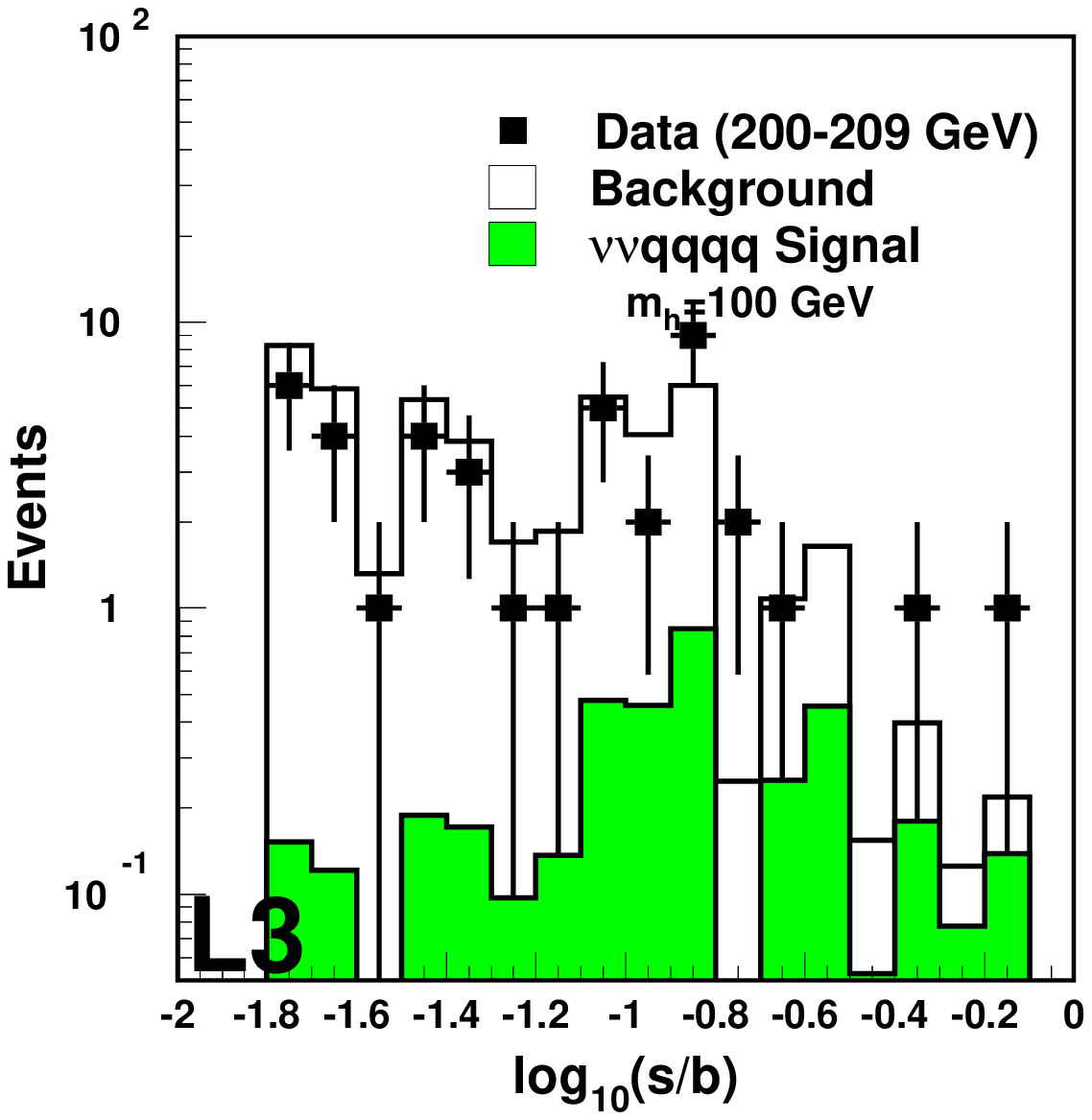}} 

\centering (a) Final variable rebinned in s/b for \( m_{\textrm{h}}=100 \)
GeV &
\resizebox*{0.48\textwidth}{!}{\includegraphics{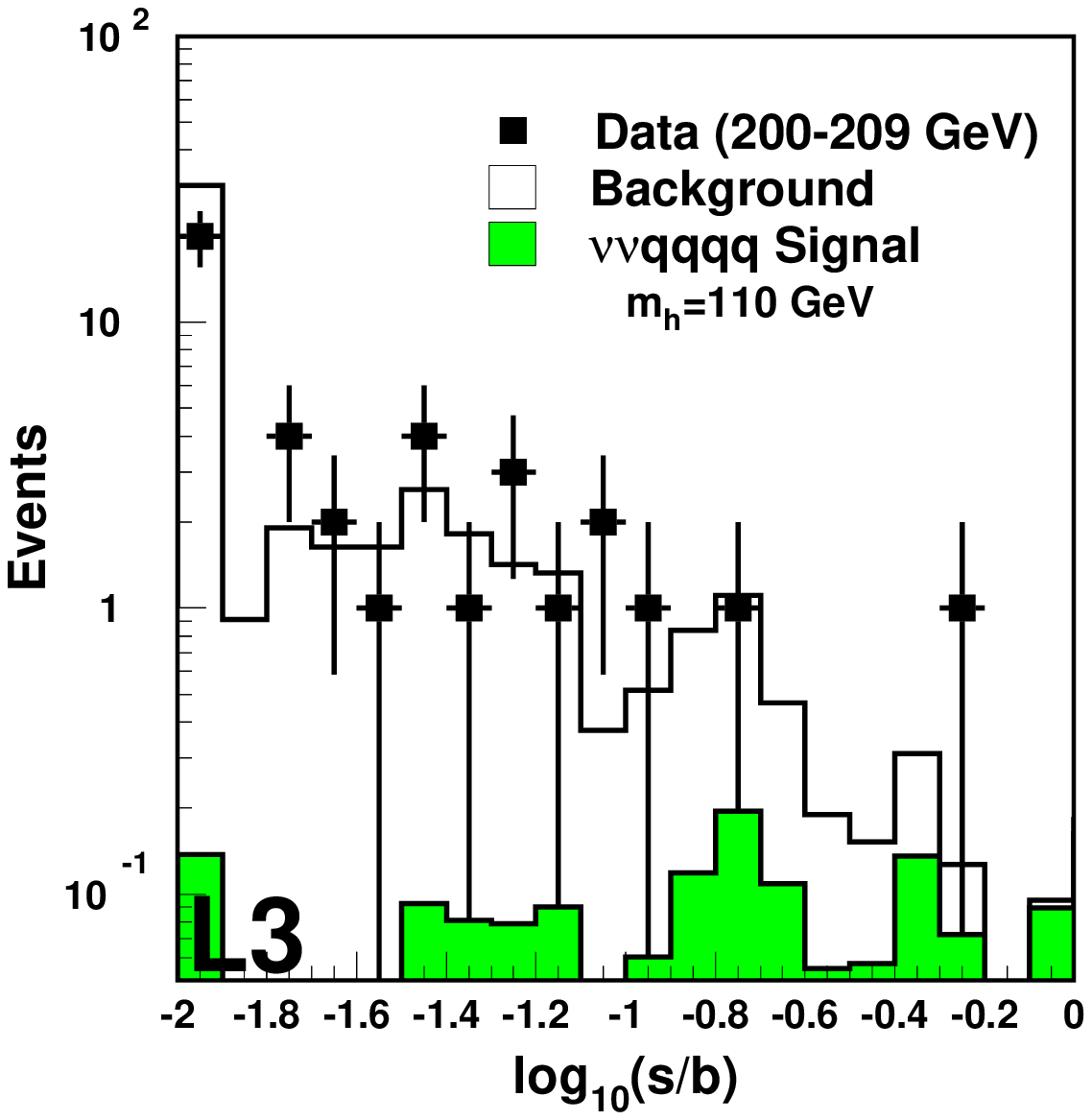}} 

\centering (b) Final variable rebinned in s/b for \( m_{\textrm{h}}=110 \)
GeV&
\\
\resizebox*{0.48\textwidth}{!}{\includegraphics{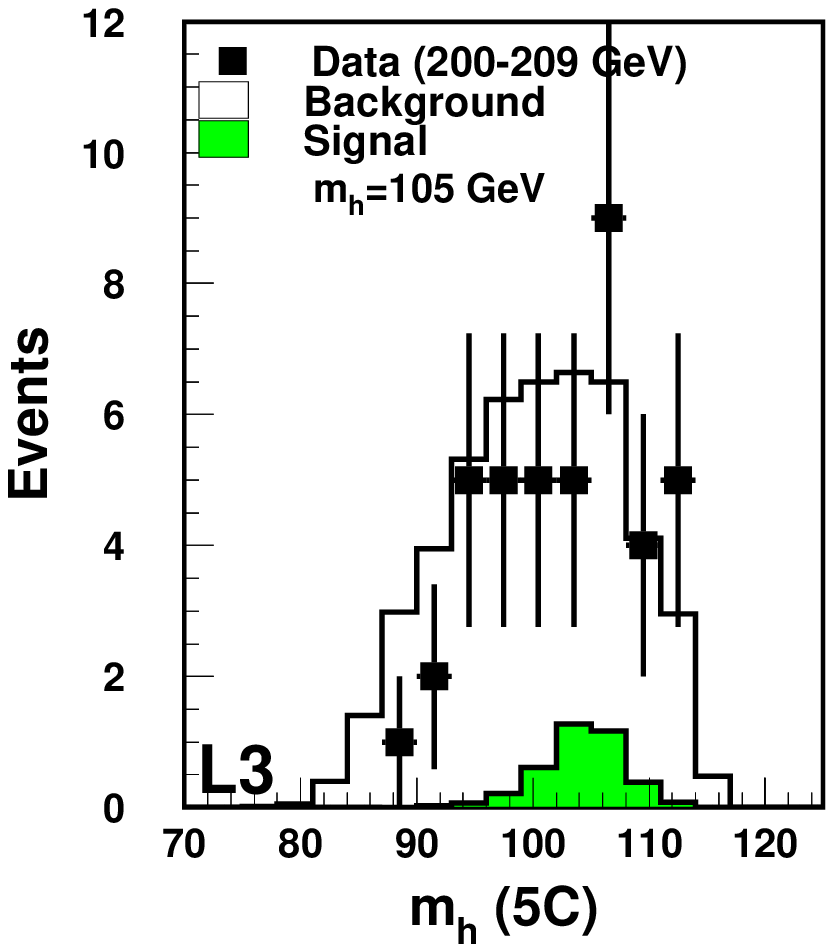}} 

(c) Mass distribution of selected events\medskip{}
&
\resizebox*{0.48\textwidth}{!}{\includegraphics{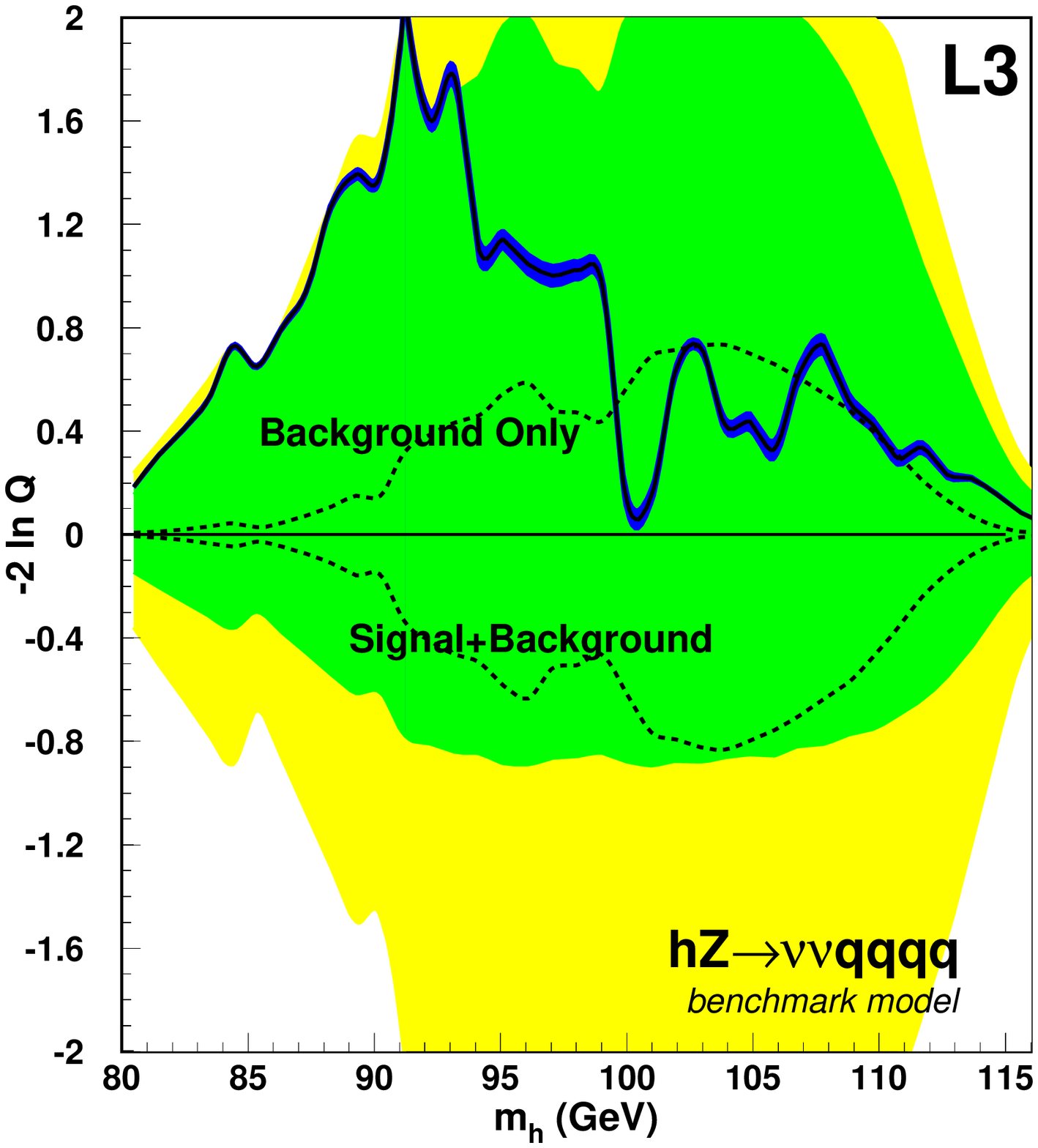}} 

\centering (d) \( -2\ln Q \) curves&
\\
\end{tabular}

\caption{Results for the \protect\( \textrm{hZ}\rightarrow \nu \nu \textrm{qqqq}\protect \)
search.\label{fig:vv4j_results}}
\medskip{}
\end{figure}

As shown in Figure \ref{fig:vv4j_results}d, the \( \nu \nu \textrm{qqqq} \)
channel is significantly deficit in data at masses below \mh=90 GeV,
where the W pair-production background dominates. At higher masses,
the results closely follow the background-only expectation except
for a candidate event near 100 GeV. The analysis is only effective
for masses above \mZ. The separation between the signal+background
and background-only hypotheses for lower masses is quite small.

\begin{figure}
\begin{tabular}{p{2.5in}p{2.5in}l}
\resizebox*{0.48\textwidth}{!}{\includegraphics{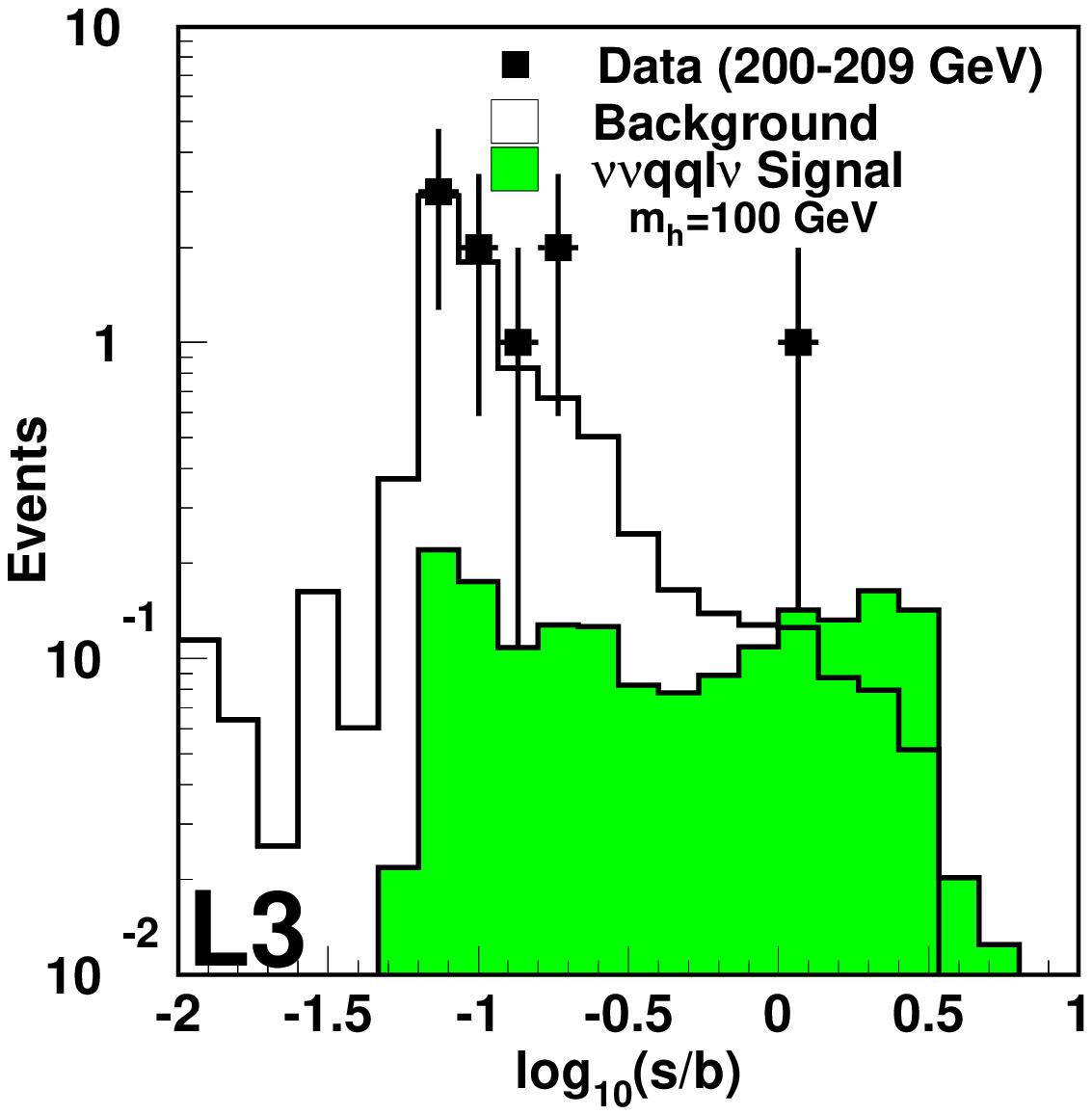}} 

\centering (a) Final variable rebinned in s/b for \( m_{\textrm{h}}=100 \)
GeV &
\resizebox*{0.48\textwidth}{!}{\includegraphics{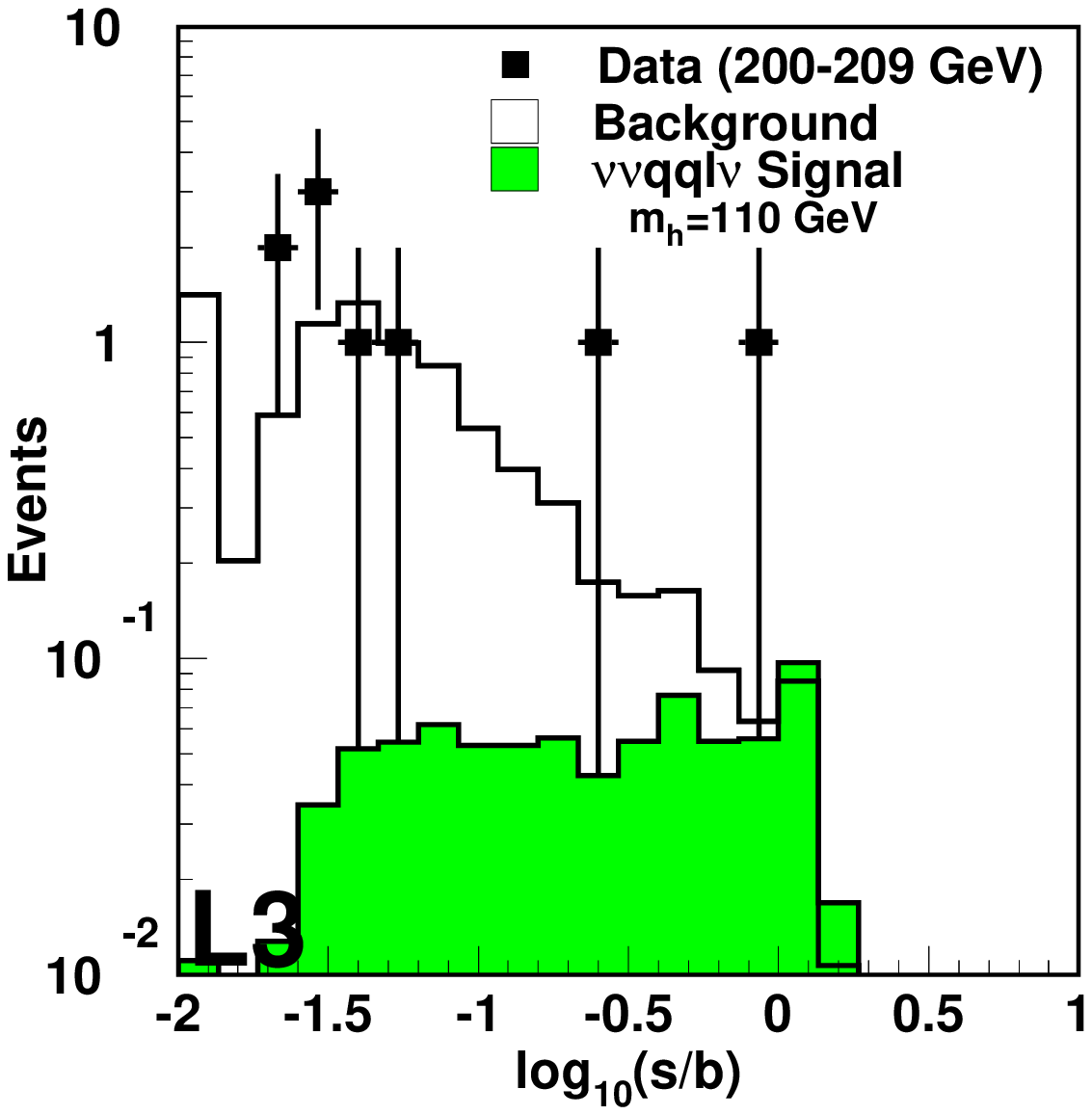}} 

\centering (b) Final variable rebinned in s/b for \( m_{\textrm{h}}=110 \)
GeV&
\\
\multicolumn{2}{p{5in}}{{\centering \resizebox*{0.48\textwidth}{!}{\includegraphics{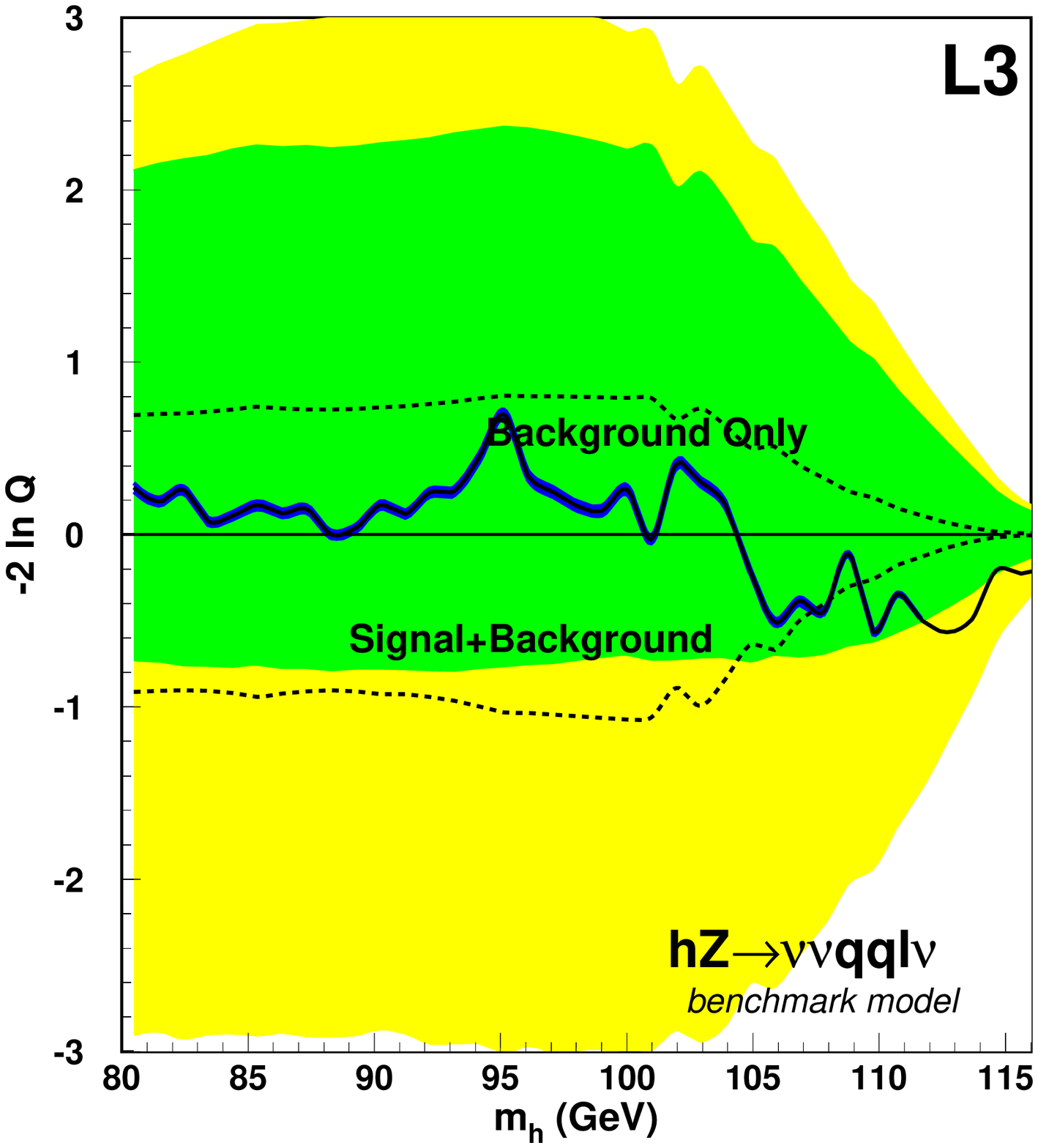}} \par}

\centering (c) \( -2\ln Q \) curves}&
\\
\end{tabular}

\caption{Results for the \protect\( \textrm{hZ}\rightarrow \nu \nu \textrm{qql}\nu \protect \)
search.\label{fig:vvqqlv_results}}
\medskip{}
\end{figure}

The \( \nu \nu \textrm{qql}\nu  \) analysis has remarkably flat performance
for lower masses, as shown in Figure \ref{fig:vvqqlv_results}c. Most
channels suffer from low-mass backgrounds and the \( \textrm{W}^{*} \)
and \( \textrm{W}^{(*)} \) becoming quite low-mass for \( \mh \simeq \mZ  \).
The visible energy in \( \nu \nu \textrm{qql}\nu  \) is always small
and the analysis is less sensitive to the masses of the Higgs decay
bosons since they are hard to reconstruct at all. The observed results
have a small constant excess for lower masses, increasing to a slightly
larger excess for masses above 110 GeV. There is no mass plot for
this channel, since the Higgs mass cannot be reconstructed.

\begin{figure}
\begin{tabular}{p{2.5in}p{2.5in}l}
\resizebox*{0.48\textwidth}{!}{\includegraphics{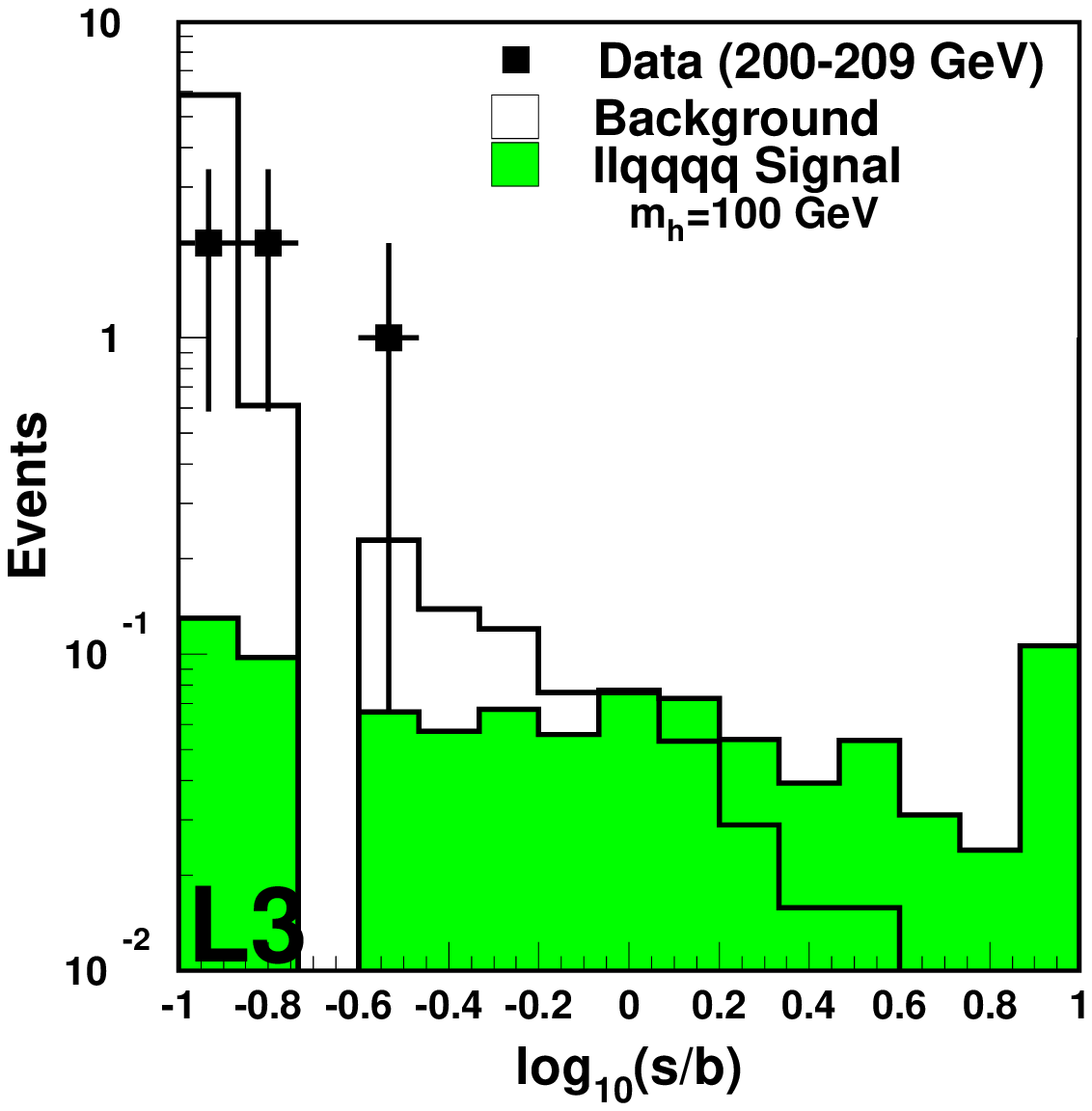}} 

\centering (a) Final variable rebinned in s/b for \( m_{\textrm{h}}=100 \)
GeV &
\resizebox*{0.48\textwidth}{!}{\includegraphics{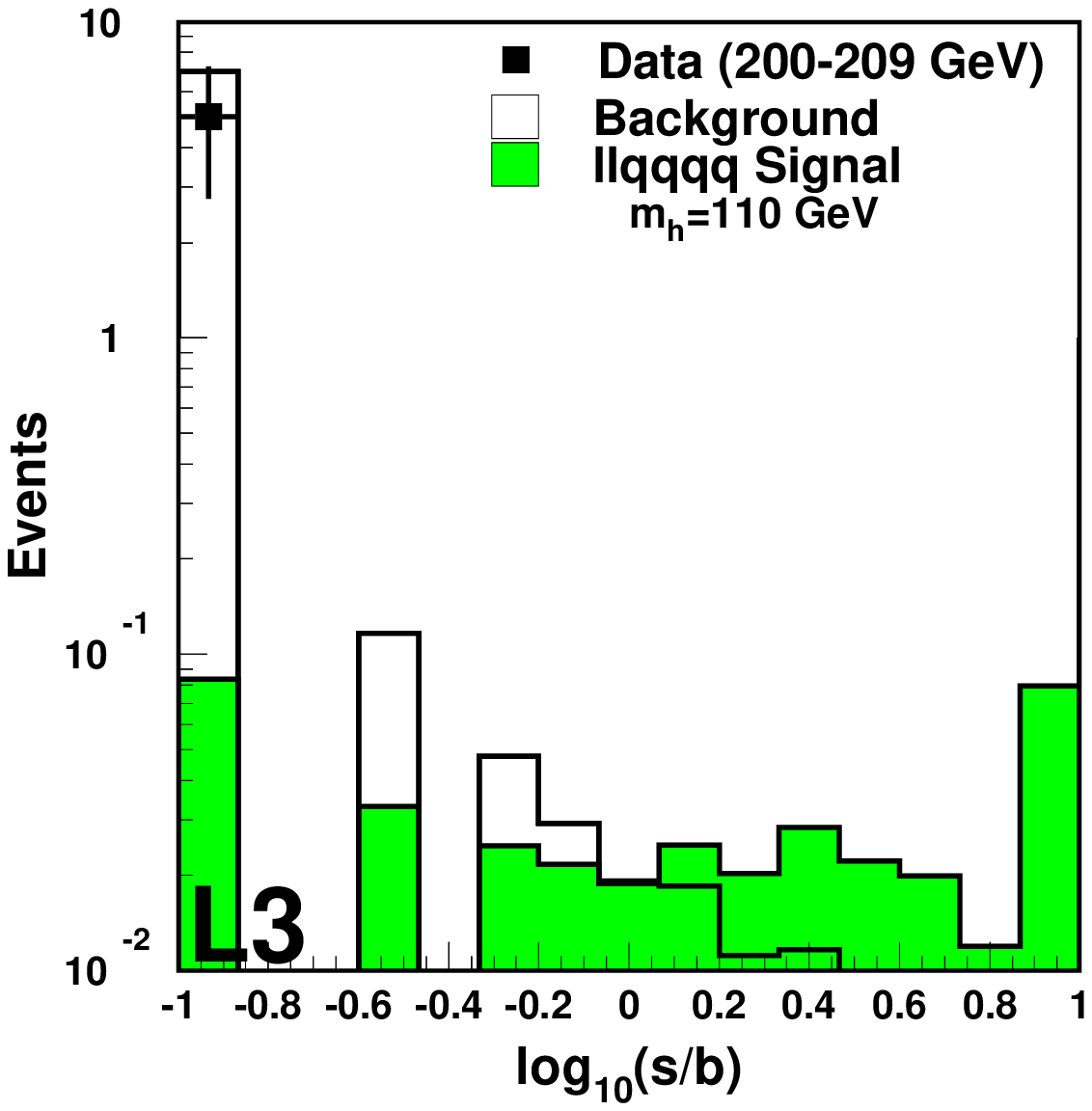}} 

\centering (b) Final variable rebinned in s/b for \( m_{\textrm{h}}=110 \)
GeV&
\\
\resizebox*{0.48\textwidth}{!}{\includegraphics{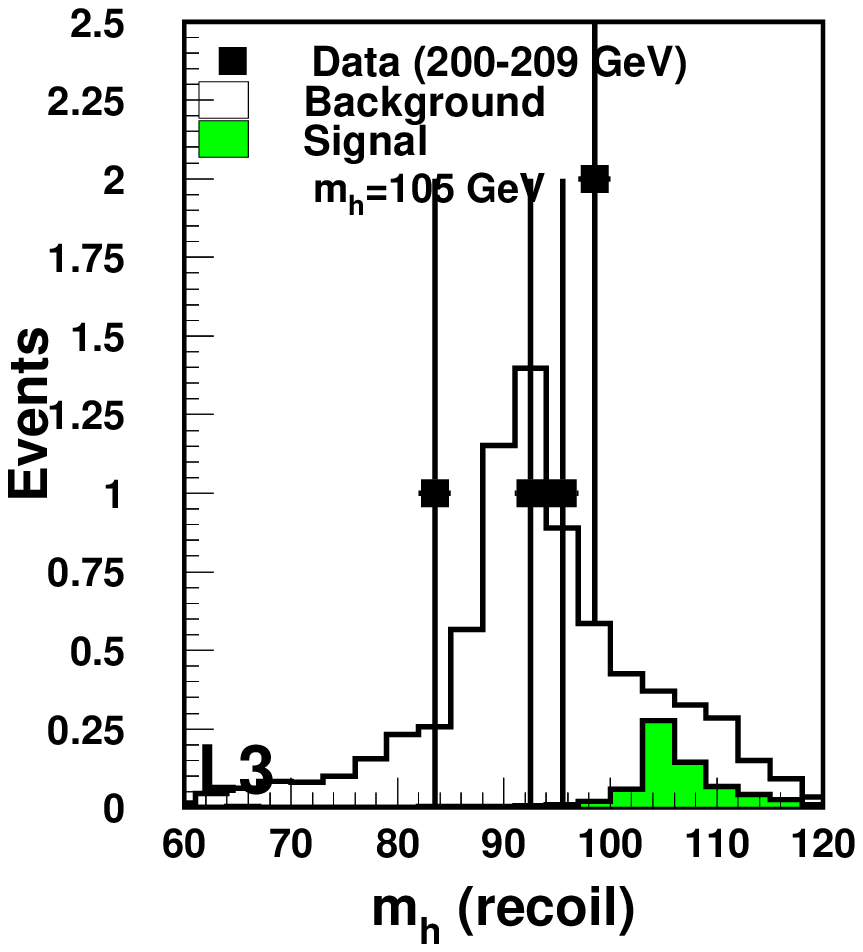}} 

(c) Mass distribution of selected events\medskip{}
&
\resizebox*{0.48\textwidth}{!}{\includegraphics{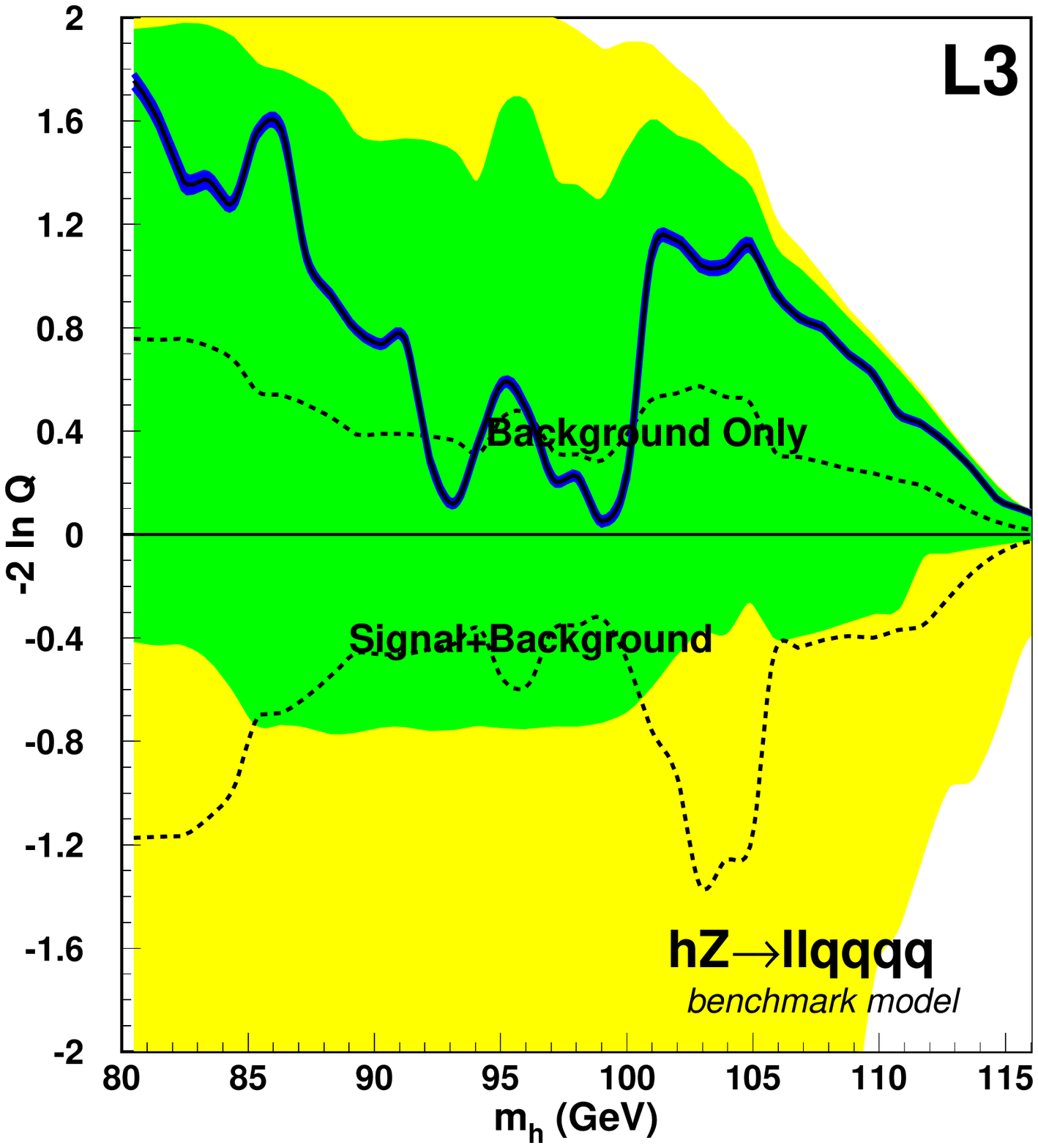}} 

\centering (d) \( -2\ln Q \) curves&
\\
\end{tabular}

\caption{Results for the \protect\( \textrm{hZ}\rightarrow \textrm{llqqqq}\protect \)
search.\label{fig:ll4j_results}}
\medskip{}
\end{figure}

In the llqqqq channel, the effect of the ZZ background can be very
clearly seen in \ref{fig:ll4j_results}d. The separation between background-only
and signal+background has a pronounced narrowing near \mZ. The channel
selects a very small number of data events which makes the upper side
of the \( 2\sigma  \) band extremely narrow. The width is derived
from a series of Poisson trials, and there are very few integer values
available for a down-fluctuation from an observation of five events.
The llqqqq channel is mildly deficit for small mass and high mass
and matches the background-only expectation near and slightly above
\mZ.

\begin{figure}
\begin{tabular}{p{2.5in}p{2.5in}l}
\resizebox*{0.48\textwidth}{!}{\includegraphics{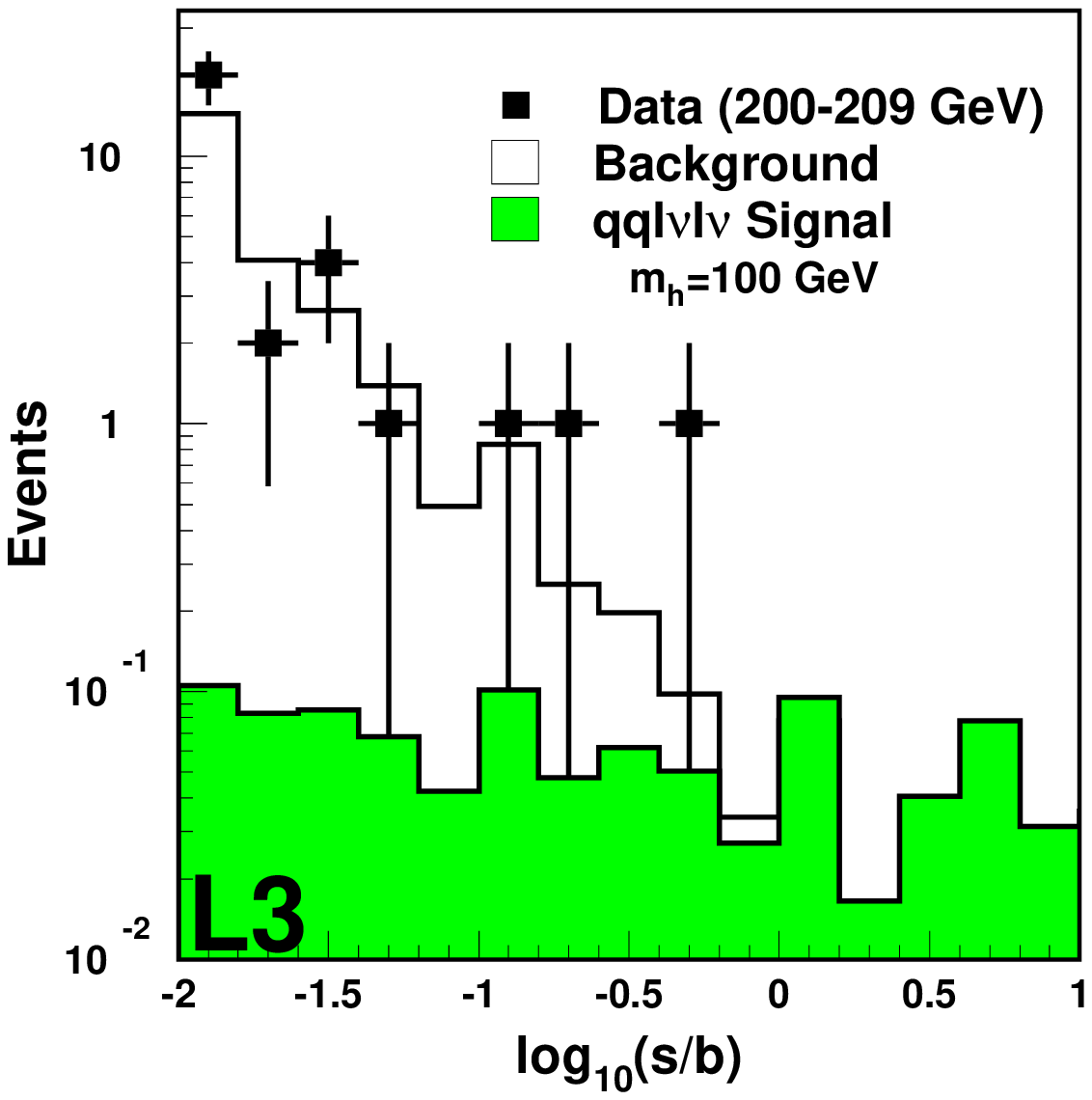}} 

\centering (a) Final variable rebinned in s/b for \( m_{\textrm{h}}=100 \)
GeV &
\resizebox*{0.48\textwidth}{!}{\includegraphics{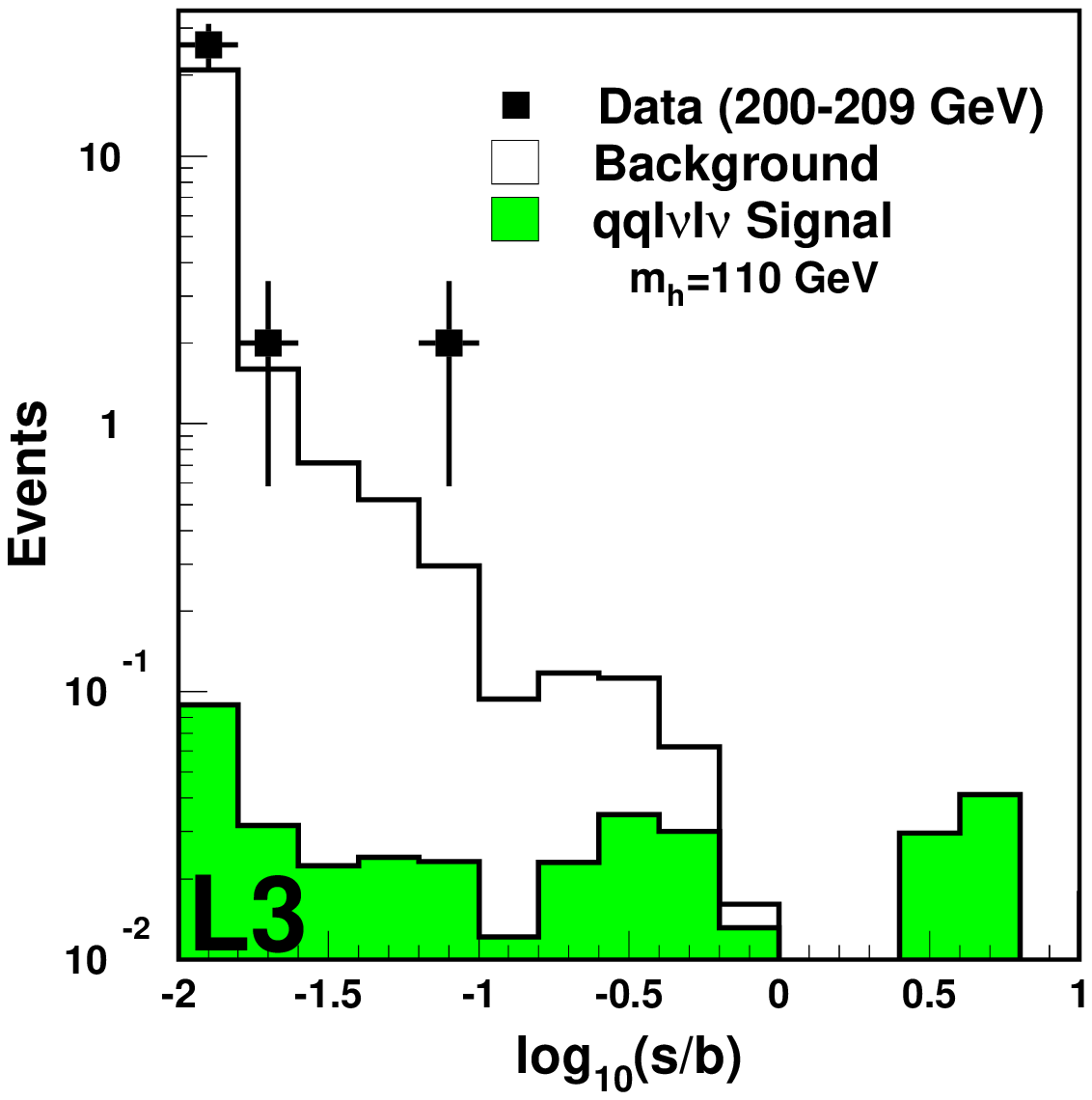}} 

\centering (b) Final variable rebinned in s/b for \( m_{\textrm{h}}=110 \)
GeV&
\\
\resizebox*{0.48\textwidth}{!}{\includegraphics{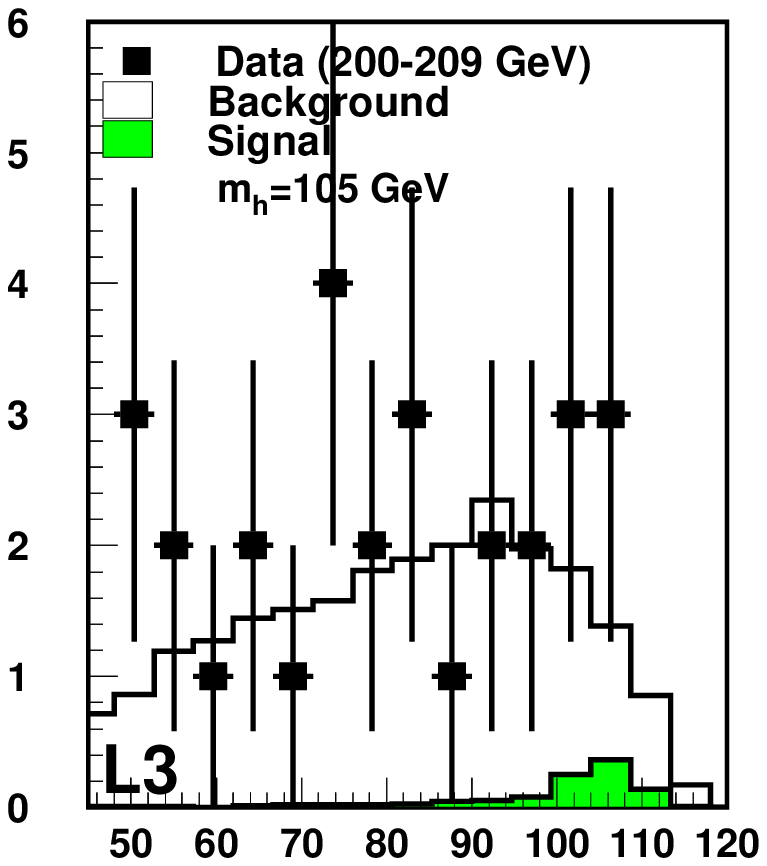}} 

(c) Mass distribution of selected events\medskip{}
&
\resizebox*{0.48\textwidth}{!}{\includegraphics{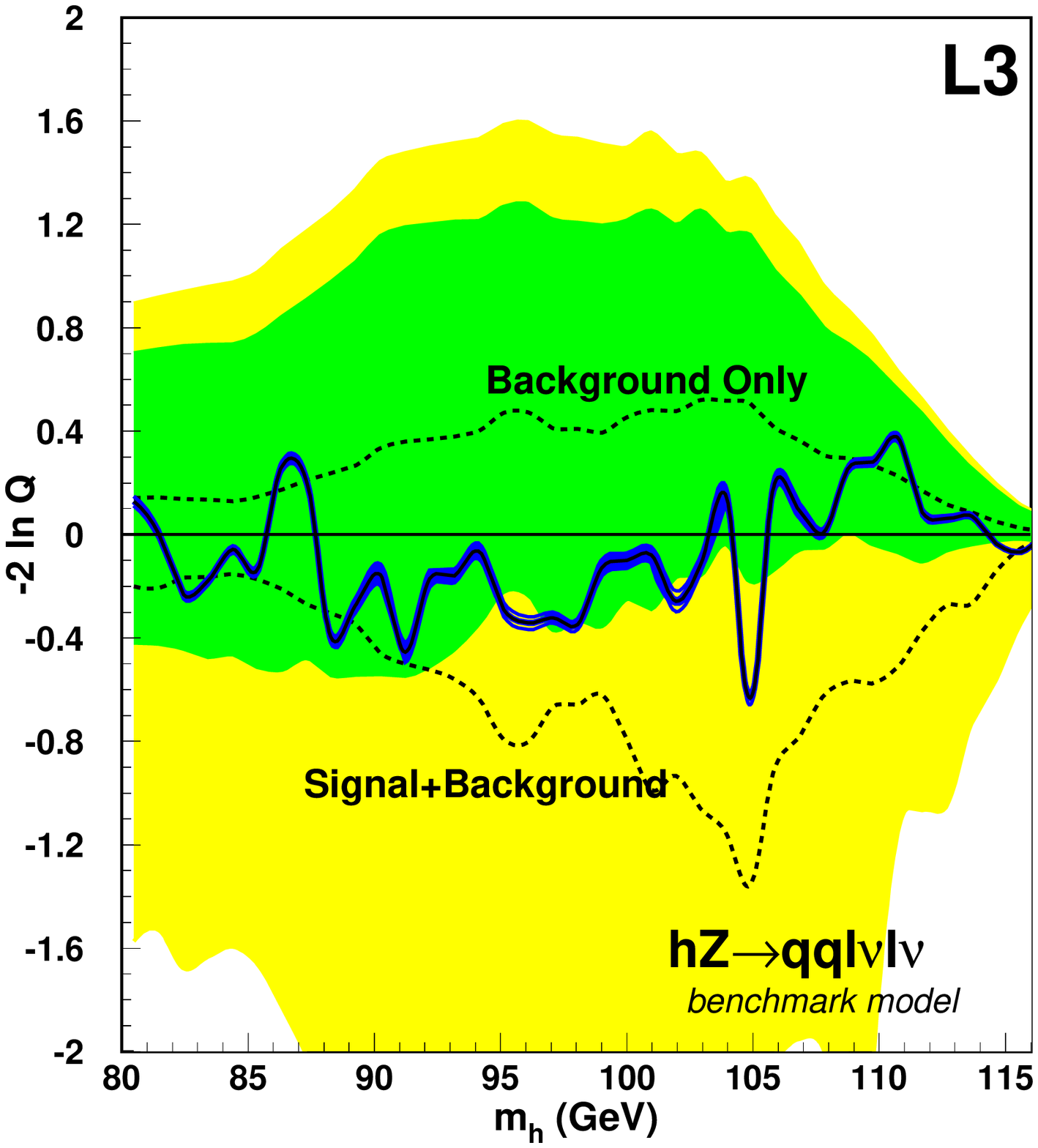}} 

\centering (d) \( -2\ln Q \) curves&
\\
\end{tabular}

\caption{Results for the \protect\( \textrm{hZ}\rightarrow \textrm{qql}\nu \textrm{l}\nu \protect \)
search.\label{fig:qqlvlv_results}}
\medskip{}
\end{figure}

The \( \textrm{qql}\nu \textrm{l}\nu  \) channel has the smallest
total search power of the analyzed channels, but it has some very
high signal-to-background regions in the analysis. Figure \ref{fig:qqlvlv_results}d
shows an excess at lower Higgs mass hypotheses, with a particular
spike at 104 GeV. The analysis shows a small deficit for higher mass
hypotheses.

\newpage

\subsection{Combined \protect\( \textrm{h}\rightarrow \textrm{WW}/\textrm{ZZ}\protect \)
Results}

\begin{figure}
{\centering \resizebox*{0.9\textwidth}{!}{\includegraphics{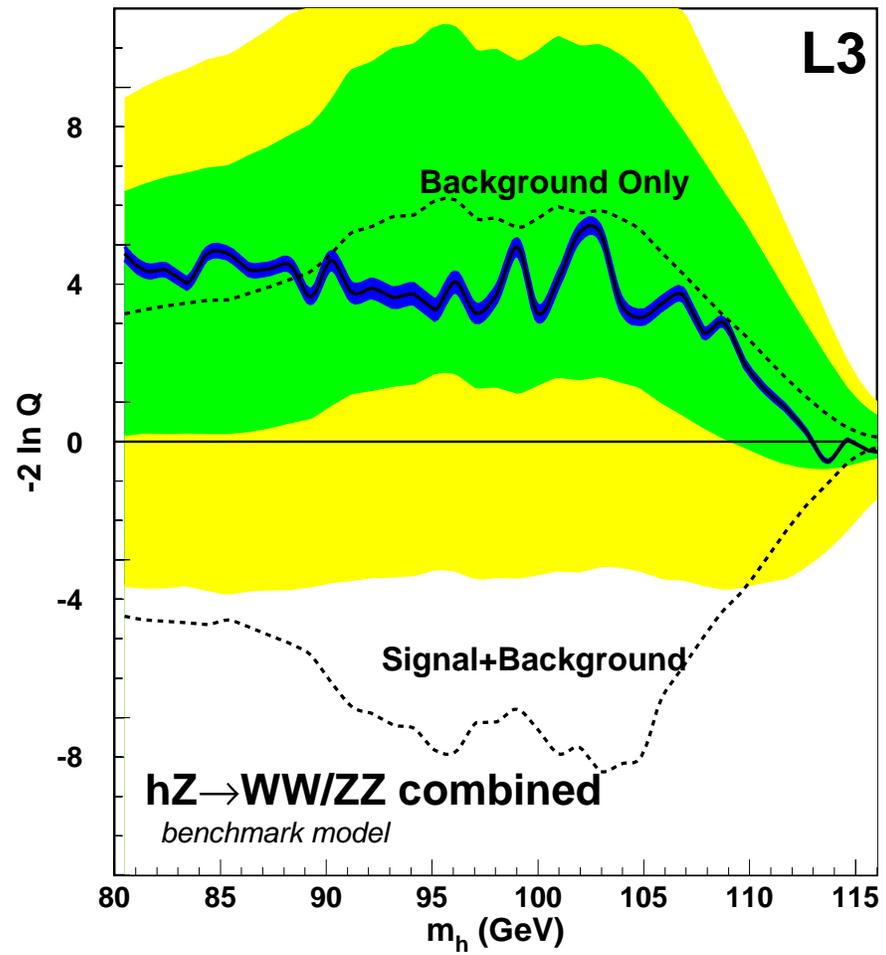}} \par}

\caption{\protect\( -2\ln Q\protect \) plot for the combined \protect\( \textrm{h}\rightarrow \textrm{WW}/ZZ\protect \)
search.\label{fig:llr_combo}}
\end{figure}

The full search power of the analyses is obtained when all the analyses
are combined together into a single analysis, as shown in Figure \ref{fig:llr_combo}.

\begin{figure}
{\centering \resizebox*{0.9\textwidth}{!}{\includegraphics{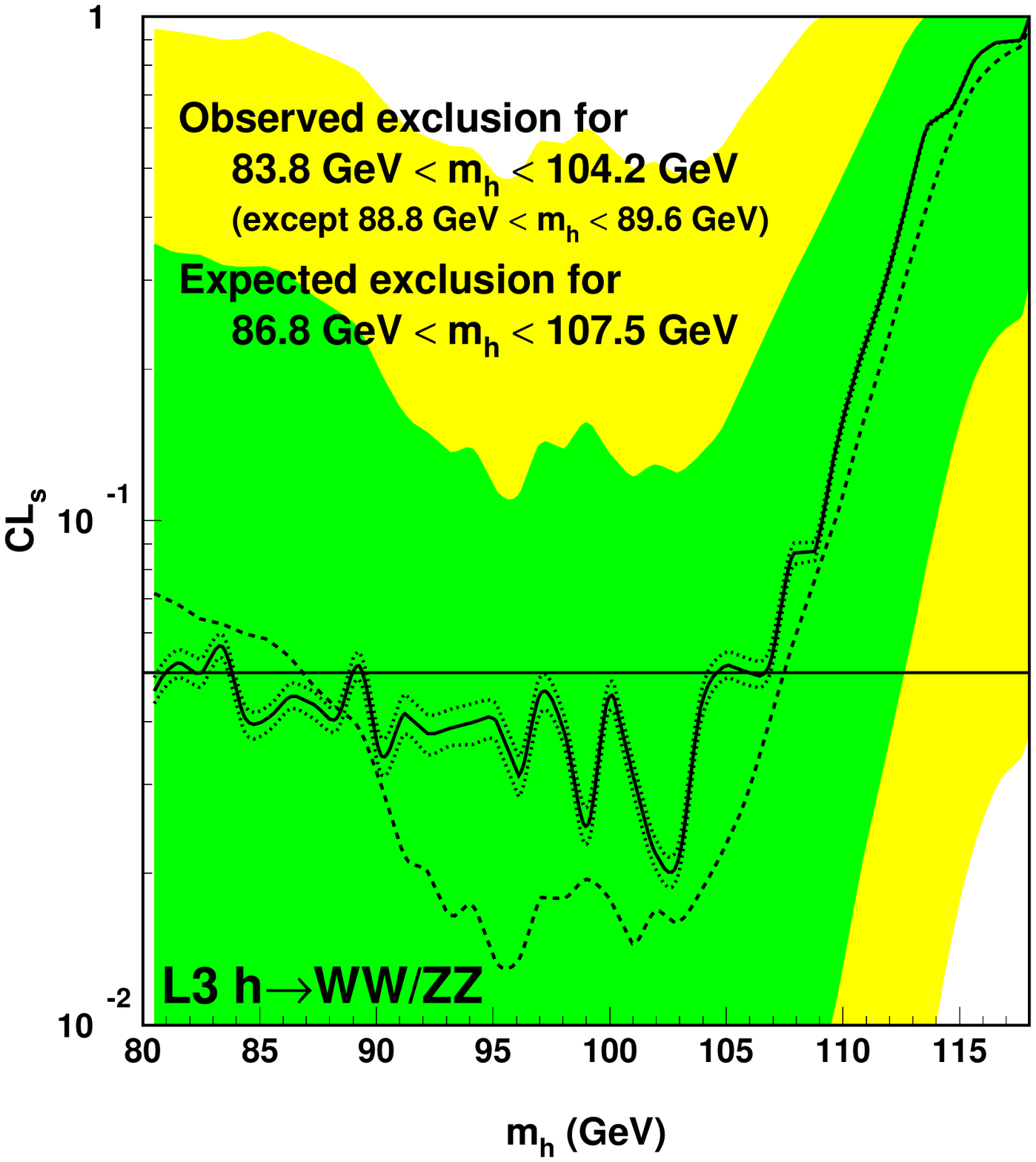}} \par}

\caption{Confidence level limits in the benchmark fermiophobic model.\label{fig:hww_cls}}

The dotted lines indicate the shifts from the observed confidence
levels when systematic errors are taken into account.
\end{figure}

With all the channels combined, the \( \textrm{h}\rightarrow \textrm{WW}/\textrm{ZZ} \)
results can set limits on the production of a fermiophobic Higgs boson
as a function of mass. Figure \ref{fig:hww_cls} shows the exclusion
confidence levels under the benchmark fermiophobic model. The dashed
line indicates the expected exclusion confidence level and the dark
and light areas around the dashed line indicate the \( \pm 1\sigma  \)
and \( \pm 2\sigma  \) expected regions respectively. The solid line
indicates the observed exclusion confidence level, which exhibits
the same deficit at low mass and slight excess at higher masses which
is clear in the \( -2\ln Q \) plot. Ignoring systematic errors, the
observed exclusion region would be \( 83.7\textrm{ GeV }<\mh <104.6\textrm{ GeV} \)
with an unexcluded region between \( 88.9\textrm{ GeV}<\mh <89.4\textrm{ GeV} \).
The dotted lines around the solid curve indicate the effect of taking
the systematic errors into account. The effect of the systematic errors
is small : including errors, the exclusion region is \( 83.8\textrm{ GeV }<\mh <104.2\textrm{ GeV} \)
with an unexcluded region between \( 88.8\textrm{ GeV}<\mh <89.6\textrm{ GeV} \).
 Since the systematic errors are small, we neglect them for the more
computationally difficult branching ratio limits.

\begin{figure}
{\centering \resizebox*{0.9\textwidth}{!}{\includegraphics{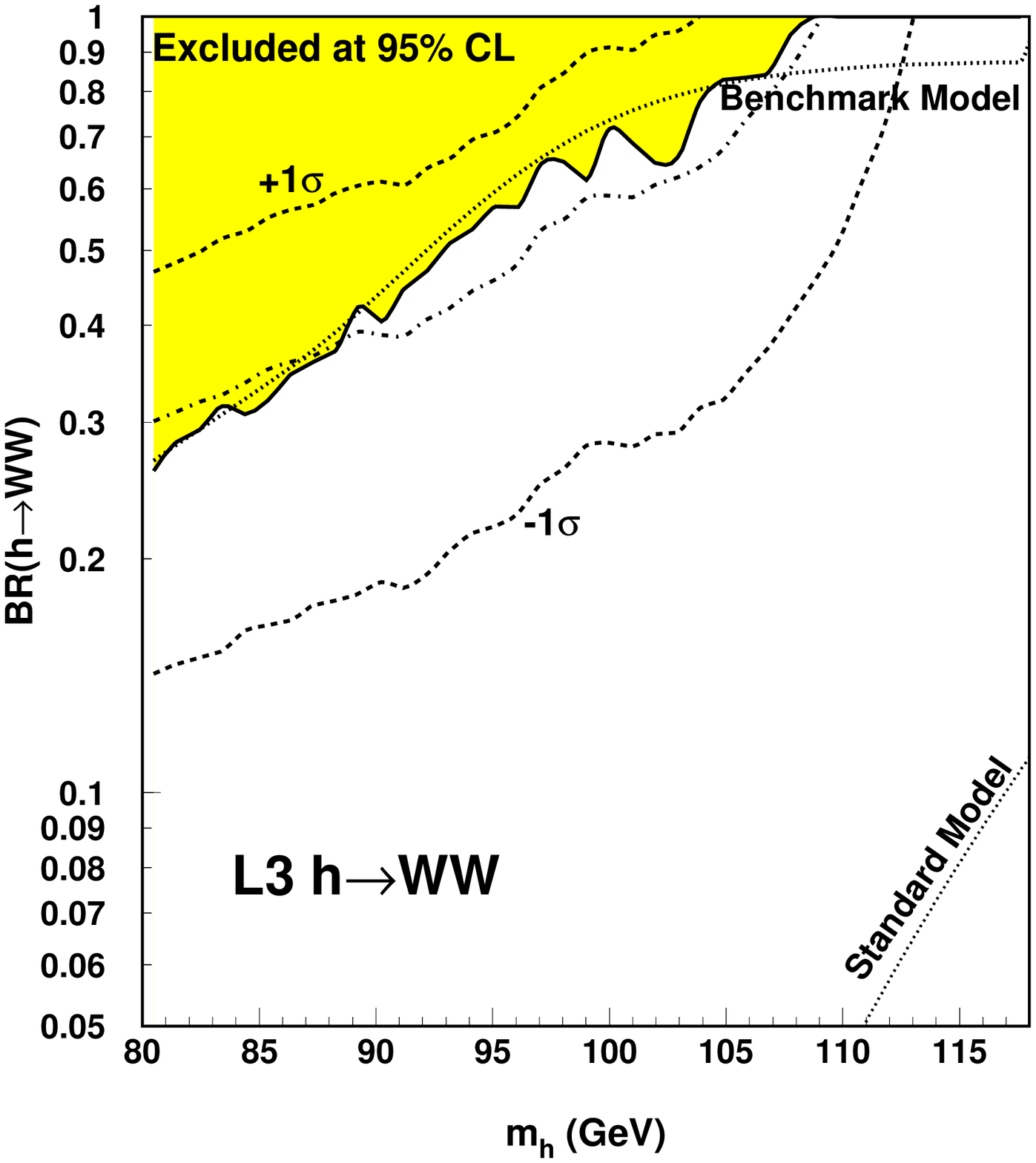}} \par}

\caption{Branching ratio limits for the \protect\( \textrm{h}\rightarrow \textrm{WW}/\textrm{ZZ}\protect \)
search.\label{fig:br_ww}}
\end{figure}

As in the \( \textrm{h}\rightarrow \gamma \gamma  \) search, we obtain
branching ratio limits by scaling the branching ratio to the value
where the observed and expected confidence levels are 95\%. This process
produces the more model-independent result of Figure \ref{fig:br_ww}.
 The predicted branching ratios of both the benchmark model and the
Standard Model are given on the plot as dotted lines. The Standard
Model \( \textrm{H}\rightarrow \textrm{WW} \) branching ratio is
too low by a factor of \( \sim 30 \) at 110 GeV for these analyses
to exclude it, compared to a factor of \( \sim 50 \) for the LEP
combined \( \textrm{H}\rightarrow \gamma \gamma  \) at the same mass.

\section{Combined Fermiophobic Results}

The results of the massive boson search can be combined with the LEP
\( \textrm{h}\rightarrow \gamma \gamma  \) search results to give
wider limits on fermiophobic models. For a single Higgs decay type,
model-independent results can be derived by scaling the branching
ratio of the Higgs to the given decay type as a function of mass to
obtain a 95\% CL limit. A general fermiophobic search with both \( \textrm{h}\rightarrow \gamma \gamma  \)
and \( \textrm{h}\rightarrow \textrm{VV} \) must consider both branching
ratios separately as functions of the mass of the Higgs. A useful
choice of parameterization is \begin{eqnarray*}
\BRphobic  & = & \textrm{BR}(\textrm{h}\rightarrow \gamma \gamma )+\textrm{BR}(\textrm{h}\rightarrow \textrm{WW})+\textrm{BR}(\textrm{h}\rightarrow \textrm{ZZ})\\
\BRgg  & = & \frac{\textrm{BR}(\textrm{h}\rightarrow \gamma \gamma )}{\BRphobic }.
\end{eqnarray*}
 Thus \BRgg\ represents the fraction of fermiophobic decays which
are to \( \gamma \gamma  \) and can range from zero to one, while
\BRphobic\ represents the total Higgs branching fraction to pairs
of gauge bosons. The search algorithm scans values of both \mh\ and
\BRgg\ and determine the values of \BRphobic\ for 95\% CL expected
and observed exclusion. The resulting plane is the most general result
of the LEP fermiophobic search. The process is computationally intensive,
since it involves millions of Monte Carlo trials for each scan point
in the \( 40\times 40 \) plane.

\begin{figure}
{\centering \resizebox*{1\textwidth}{!}{\includegraphics{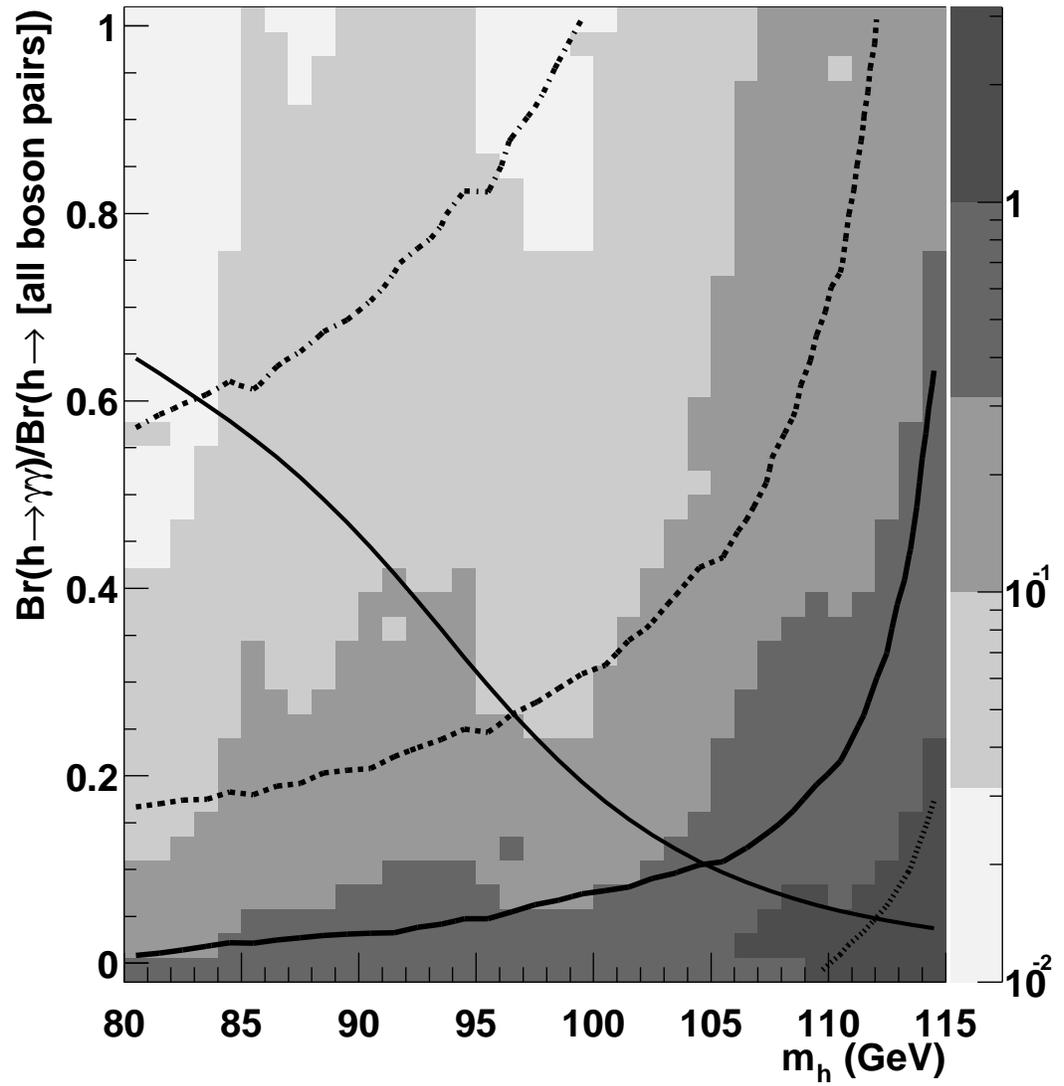}} \par}

\caption{Results of the fermiophobic scan. \label{fig:scan}}
\end{figure}

The results of the scan are plotted in Figure \ref{fig:scan}. The
shades of gray indicate the value of \BRphobic\ corresponding to an
observed 95\% confidence level. The dashed lines give the expected
boundary locations for the color transitions. There is a an excess
observed just above the Z mass independent of \BRgg\, which can be
seen by the darker shades of color extending above the line which
should be the color boundary. There are observed deficits for \( \BRgg >0.1 \)
in the low-mass region as well as the 97 GeV mass region. The line
crossing the plot from upper left to lower right is the benchmark
value of \BRgg. The point where this line crosses the \( \BRphobic =1.0 \)
gives the model-dependent limit. Using a fine-grained search, we obtain
a limit of \mh > 108.1 GeV with the expected limit of \mh > 111.5
GeV, compared to \( \mh >106.5\textrm{ GeV} \) observed and \( \mh >109.6\textrm{ GeV} \)
expected from the \( \textrm{h}\rightarrow \gamma \gamma  \) alone. 

\section{Conclusions}

The first search for a Higgs boson decaying to massive vector bosons
pairs has been remarkablely successful, though it did not discover
the Higgs. The search established the first 95\% confidence level
limits for a Higgs decaying to massive vector boson pairs. Over the
next decade, searches at the Tevatron, LHC, and a linear \( \textrm{e}^{+}\textrm{e}^{-} \)
collider will focus on the WW and ZZ channels for discovering and
studying the Higgs \cite{higgs:rainwater}, and this work should be
a resource for these future studies.

\appendix

\chapter{Final Variable Construction Techniques}

\textit{Man is a tool-using animal. Without tools, he is nothing,}\\
\textit{with tools he is all.}

\begin{quotation}
Thomas Carlyle
\end{quotation}

\section{KEYS}

The final Higgs search results are performed by comparing signal and
background distributions. Choppiness will result in numerical difficulties
for the limit-setting algorithm so these distributions need to be
fairly smooth. However, the high performance of the Higgs search analyses
means that very few Monte Carlo background events remain after selection
to estimate the background. The solution adopted by the LEP Higgs
Working Group is use the KEYS algorithm developed by Kyle Cranmer
to smooth the final distributions.

\begin{figure}
{\centering \resizebox*{0.8\textwidth}{!}{\includegraphics{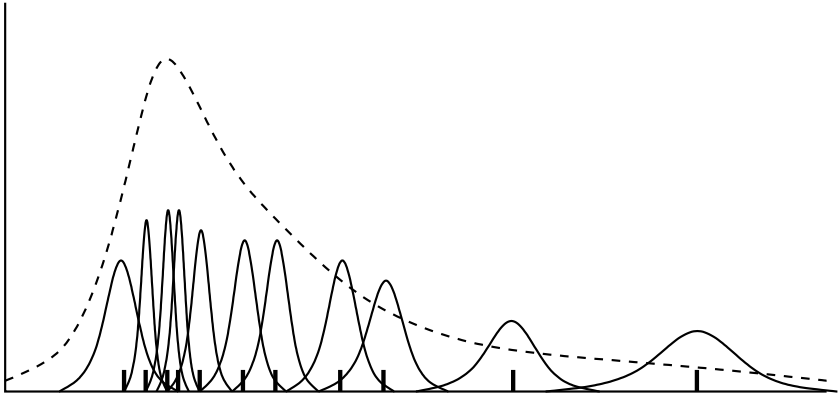}} \par}

\caption{The KEYS kernel-estimation smoothing process.\label{fig:keys}}

The actual value of each event in the distribution is given by the
vertical lines. We replace each point by a Gaussian whose width depends
on the density of points in the immediate neighborhood. Adding many
Gaussians would make the distribution approach the dashed-line curve.
\end{figure}

The entire KEYS algorithm is fairly complicated and is fully described
in \cite{algo:keys}, but the basic concept is straightforward. Each
event is added to the distribution not as a spike at a given value,
but as a Gaussian centered on that value. This technique, known as
{}``kernel estimation,'' is widely used. The key question is the
width of the Gaussian used. KEYS calculates this width in two steps.
First, it constructs an intermediate distribution using fixed widths
dependent on the variance of the source data. Then, this intermediate
distribution is used to determine the widths for the final smoothed
result. Where the intermediate distribution is large, the Gaussians
in the final distribution will be narrow. Where the intermediate distribution
is small, the Gaussians will be wide. This process is shown in cartoon
form in Figure \ref{fig:keys}.
 This technique is very effective at smoothing many distributions,
including neural network distributions. In the case of neural network
distributions where the distribution peaks at the edge and values
above 1.0 and below 0.0 are not possible, we fold the Gaussians back
across the y-axes.

In these analyses, we use our own implementation of the algorithm
which uses a many-bin histogram rather than an event list for calculations.
This modification greatly speeds the algorithm.

\section{Neural Networks\label{sec:neuralnets}}

In many analyses, there are multiple discriminating variables but
usually only one variable can be used in the final fit or search.
Traditionally, one applies hard cuts on all but one variables and
uses the last as the {}``final variable''. In some circumstances,
this technique limits the performance of the analysis since a simple
cut does not extract all the distinguishing power available in a combination
of the variables. One technique for combining variables is the neural
network. The term {}``neural network'' conjures up many images,
but the networks used for analysis purposes are simply non-linear
functions constructed in a certain manner with many parameters, where
the parameters are determined in an iterative process from examples.
There are many possible neural network structures and choices for
combination and activation functions, but we will discuss only those
used here and leave the general discussion to other texts \cite{algo:nn}.

\begin{figure}
{\centering \resizebox*{0.4\textwidth}{!}{\includegraphics{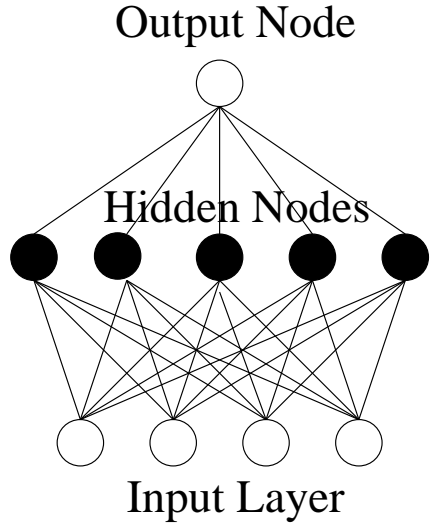}} \par}

\caption{Diagram of a neural network.\label{fig:nnet}}

This network has four input nodes, five hidden nodes, and one output
node.
\end{figure}
Consider a neural network as set of nodes, each with many inputs and
one output. The output of the node can be connected to other nodes
to construct a network, as in Figure \ref{fig:nnet}. The output of
each node is a nonlinear function of its inputs. For the analysis
networks, we form a weighted sum of all the inputs to the node and
use the result as the argument of a sigmoid function to calculate
the node output \( O_{j} \):\[
O_{j}=\frac{1}{2}\left( \tanh (\sum _{i}w_{ij}I_{ij}+\beta _{j})+1\right) ,\]
where the \( I_{ij} \) are the inputs to the \( j \)-th node and
the \( w_{ij} \) are the weights. Thus, for a network with \( n \)
inputs, a single hidden layer of \( m \) nodes and a single output
we can write the output as\[
O=\frac{1}{2}\left[ \tanh \left( \sum ^{m}_{i=1}w_{iO}\frac{1}{2}\left[ \tanh (\sum ^{n}_{j=1}w_{ji}I_{j}+\beta _{i})+1\right] +\beta _{O}\right) +1\right] .\]
The only question is the determination of \( w_{ij} \) and \( \beta _{j} \)
; once these are set, a neural network is a straightforward function.

The weights used in the network are determined through a learning
process. The network is first initialized with random weights. Then
the trainer gives a pattern to the network and the network produces
some output. The trainer then compares the desired output with the
actual output and adjusts the weights by a small increment to make
the output closer to the desired output, propogating the changes back
through the network. The trainer repeats this process many times with
different patterns until the network converges to a set of weights.
There are several mathematical propogation techniques which have been
developed for networks. We use the RPROP technique, which is somewhat
faster than the classic backpropogation technique but otherwise similar
\cite{algo:rprop}. All of our networks were trained using the Stuttgart
Neural Network Simulator (SNNS) package \cite{algo:snns}. SNNS features
a batch training program and a conversion tool to make a trained network
into simple and fast C code.

For high-energy physics use, we compose our patterns out of physically
relevant variables using Monte Carlo signal and background to provide
simulated patterns. Typically, the patterns specify that the network
should output zero for background events and one for signal events.
A fairly large number of patterns is required for training. The exact
number required is dependent on the number of inputs, hidden nodes,
and the level of corellation between inputs, but certainly more than
1000 events each of signal and background should be used if possible.
During the training process, the training software presents each pattern
in the set in random order and then the process repeats.

The training algorithm seeks to minimize the average error over all
the patterns in the set, while the physics goal is generally to minimize
the number of background events in the best signal region. In addition,
a given pattern set will contain non-physical coorelations from sampling
errors, so it is useful to check the performance of the network against
a separate set which is not used for training. Between every second
training cycle, we determine the network outputs for a test pattern
set. From these, we determine the cut position which would leave 70\%
or 80\% of the signal events. Then we count the number of background
events which pass this cut. If this number has decreased from the
last cycle, we save the network. Thus, the final network selected
from the training is the network which has the lowest number of background
events in the signal-pure region from the testing, rather than training
set. This technique is somewhat complicated, but provides significantly
better results than simply accepting the last network from the training.

\section{Discriminant Final Variable\label{sec:discriminant}}

The neural network as described in Section \ref{sec:neuralnets} is
one way to combine multiple variables. However, training a network
requires large numbers of signal events, which can obtained by combining
many generated mass values together. This means that the mass itself
cannot be used as a variable and some other technique is needed to
include this important variable. The technique used in our analyses
to combine such variables is to produce probability distributions
of each of the final variables for each background type and signal
mass hypothesis separately and combine them in a discriminant final
variables for each event.

Let us take \( s_{i} \) to be the probability distribution for variable
\( i \) for a given signal mass hypothesis, and \( b^{j}_{i} \)
to be the probability distribution for variable \( i \) for background
\( j. \) We smooth these distributions using the modified version
of the KEYS algorithm discussed above. If we consider the values of
each variable for a given event as a vector \( \overrightarrow{x} \)
with components \( x_{1},x_{2},\ldots  \), then we form the quantity
\( p_{i} \) which represents probability that event with this value
is signal as:\[
p_{i}=\frac{s_{i}(x_{i})}{s_{i}(x_{i})+\sum _{j}b^{j}_{i}(x_{i})}\]
for each variable \( i \). Similarly, we can construct \( q_{i}^{j} \)
which represents the probability that an event with this value \( x_{i} \)
is of background type \( j \):\[
q^{j}_{i}=\frac{b^{j}_{i}(x_{i})}{s_{i}(x_{i})+\sum _{k}b_{i}^{k}(x_{i})}.\]
We combine the values from each variable by multiplying, yielding
the discriminant\[
f(\overrightarrow{x})=\frac{\prod _{i}p_{i}}{\prod _{i}p_{i}+\sum _{j}\prod _{i}q^{j}_{i}}.\]

\begin{figure}
\begin{center}

{\centering \resizebox*{0.7\textwidth}{!}{\includegraphics{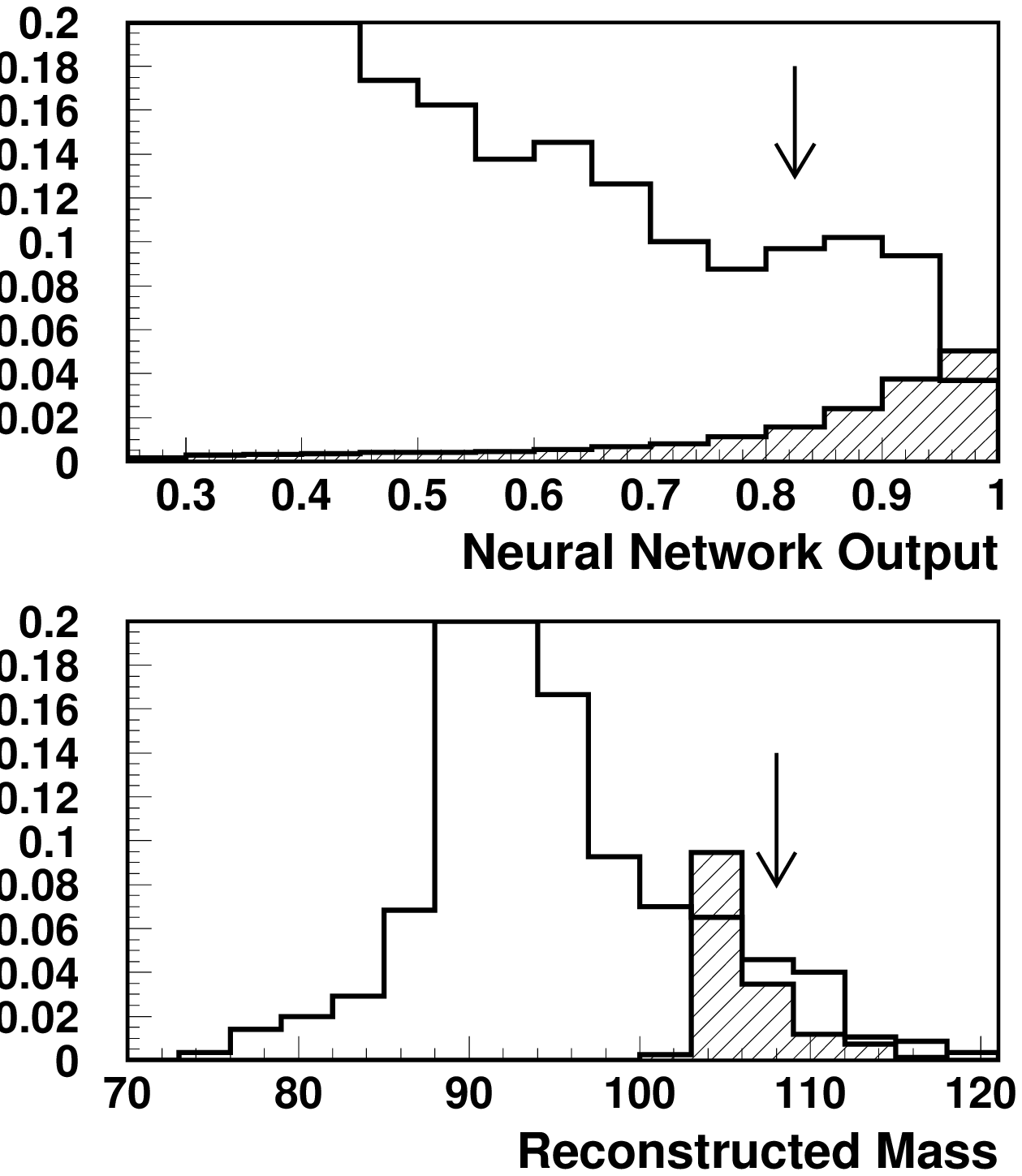}} \par}

\end{center}

\caption{Example of a two-variable discriminant calculation.\label{fig:discrim}}

In this example, there are two variables. The background distribution
is given by the empty curve and the signal distribution is given by
the hatched curve. Consider an event with neural network ouput 0.82
and reconstructed mass of 108 GeV.\begin{eqnarray*}
p_{1}=\frac{0.016}{0.016+0.097}=\frac{1.6}{11.3} &  & p_{2}=\frac{0.036}{0.036+0.047}=\frac{3.6}{8.3}\\
q_{1}=\frac{0.097}{0.016+0.097}=\frac{9.7}{11.3} &  & q_{2}=\frac{0.047}{0.036+0.047}=\frac{4.7}{8.3}\\
 &  & 
\end{eqnarray*}
\[
f=\frac{\frac{1.6}{11.3}\cdot \frac{3.6}{8.3}}{\frac{1.6}{11.3}\cdot \frac{3.6}{8.3}+\frac{9.7}{11.3}\cdot \frac{4.7}{8.3}}=\frac{5.76}{51.35}\approx 0.11\]
For a more signal-like event, \( f \) would be closer to \( 1.0 \),
while for a very background-like event, \( f\rightarrow 0.0 \).
\end{figure}

The discriminant calculation may be more clear in the example in Figure
\ref{fig:discrim}. Of course, in the actual calculation we use hundreds
of bins to improve the smoothness of the distributions rather than
the twenty of the example, but the principle is the same.

\chapter{Systematic Errors\label{sec:systematics}}

\begin{quotation}
\textit{The goal of systematic errors is quantify how much we don't
know about our data.}
\end{quotation}

\section{Strategy for Systematic Error Estimation}

In the calculation of systematic errors, we want to determine what
experimental or other processes might contribute to an error in our
result. Given a reasonable estimate of the scale of these effects
from inspection of large data and Monte Carlo samples, we determine
how our results might be affected. The most important source of systematic
errors is disagreements between the background Monte Carlo and data.
A great deal of experimental effort each year was put into calibrating
the detector and applying measured detector efficiencies to Monte
Carlo. After applying these calibrations and measured efficiencies,
the ensemble of events in the Monte Carlo match the actual conditions
under which the data was taken as closely as possible. At the end
of the process, disagreements between Monte Carlo and data may still
occur, particularly in selected subsamples of the data. 

We estimate the effect of possible shifts in the data using Monte
Carlo. Such shifts are likely to be coorlated across variables. For
example, one of the most important sources of systematic error is
miscalibrations in the energy scale of the calorimeters. A shift in
the energy scale would affect jet energies, visible energy, reconstructed
masses, recoil masses, and many other quantities. To estimate the
effect of an energy scale shift, we shift all the relavent variables
at once in the approproriate directions for each event. We produce
a set of final input files using the shifted events but keeping the
probability density functions constant. We determine the shift in
\( \textrm{LLR}_{\textrm{b}} \) values compared to the unshifted
inputs as a function of mass hypothesis. This process propogates the
effect of uncertainities in physical quantities from the detector
into the final results.

There are several other important sources of potential error besides
the energy scale. Several analyses have significant dependency on
the number of charged tracks overall and in jets. These analyses are
sensitive to changes in the tracking efficiency. The lepton-containing
analyses are sensitive to shifts in the variables used for lepton
identification, but the use of smooth lepton quality variables makes
the error reasonably well-behaved. Several analyses use event shape
variables which depend mostly on cluster counting and can be a significant
source of error since the largest known disagreement between L3 Monte
Carlo and data is the number of clusters. Finally, there are errors
due to the finite numbers of Monte Carlo events available for the
analyses, particularly in the channels which are divided into six
subchannels. We estimate this error by using the standard formula
\( \sqrt{\frac{\epsilon (1-\epsilon )}{N_{{\scriptsize \textrm{MC}}}}} \).
We consider all errors of a given type to be corrolated between the
channels and the center-of-mass energies except for Monte Carlo statistics.

\section{Systematic Errors by Channel}

Although we use the mass-hypothesis-dependent errors for the actual
limit calculations, we give mass-averaged values of the systematic
error coming from various sources in the tables below as a general
guide to the relative importance of the different error sources in
each channel.\\
\begin{tabular}{|l|c|}
\multicolumn{2}{c}{qqqqqq}\\
\hline
Systematic Error Source&
Error\\
\hline
\hline 
Tracking Efficiency&
2.1\%\\
\hline
Event Shape&
1.2\%\\
\hline 
Energy Scale&
1.3\%\\
\hline 
Monte Carlo Statistics&
2\%\\
\hline
\hline 
Total&
3.4\%\\
\hline
\end{tabular}\begin{tabular}{|l|c|}
\multicolumn{2}{c}{\( \textrm{qqqql}\nu  \)}\\
\hline
Systematic Error Source&
Error\\
\hline
\hline 
Lepton Identification&
3.8\%\\
\hline 
Monte Carlo Statistics&
3\%\\
\hline 
Energy Scale&
2.9\%\\
\hline 
Tracking Efficiency&
1.7\%\\
\hline
\hline 
Total&
5.9\%\\
\hline
\end{tabular} \\
\begin{tabular}{|l|c|}
\multicolumn{2}{c}{\( \nu \nu \textrm{qqqq} \)}\\
\hline
Systematic Error Source&
Error\\
\hline
\hline 
Energy Scale&
6.0\%\\
\hline 
Event Shape&
2.6\%\\
\hline 
Monte Carlo Statistics&
2\%\\
\hline
Tracking Efficiency&
0.2\%\\
\hline
\hline 
Total&
6.8\%\\
\hline
\end{tabular}\begin{tabular}{|l|c|}
\multicolumn{2}{c}{\( \nu \nu \textrm{qql}\nu  \)}\\
\hline
Systematic Error Source&
Error\\
\hline
\hline 
Monte Carlo Statistics&
4\%\\
\hline 
Lepton Identification&
2.8\%\\
\hline 
Energy Scale&
2.0\%\\
\hline 
Tracking Efficiency&
0.4\%\\
\hline
\hline 
Total&
5.3\%\\
\hline
\end{tabular}\\
\begin{tabular}{|l|c|}
\multicolumn{2}{c}{\( \textrm{llqqqq} \)}\\
\hline
Systematic Error Source&
Error\\
\hline
\hline 
Energy Scale&
2.2\%\\
\hline 
Tracking Efficiency&
2.1\%\\
\hline 
Monte Carlo Statistics&
2\%\\
\hline
Lepton Identification&
1.8\%\\
\hline
\hline 
Total&
4.1\%\\
\hline
\end{tabular}\begin{tabular}{|l|c|}
\multicolumn{2}{c}{\( \textrm{qql}\nu \textrm{l}\nu  \)}\\
\hline
Systematic Error Source&
Error\\
\hline
\hline 
Energy Scale&
5.3\%\\
\hline 
Lepton Identification&
4.4\%\\
\hline 
Tracking Efficiency&
1.8\%\\
\hline
Monte Carlo Statistics&
2\%\\
\hline
\hline 
Total&
7.4\%\\
\hline
\end{tabular}

\bibliographystyle{unsrt}
\bibliography{l3,algorithms,higgs,hep,hep_discovery,fermiophobic}
Thanks are due to Rene Brun \emph{et al.} for a fine data analysis
package, subject of course to the proviso: {}``PAW can do everything,
but \emph{you} cannot do \emph{anything} with PAW.''
\end{document}